\documentclass[aps,prl,reprint,10pt,superscriptaddress,nofootinbib,longbibliography]{revtex4-2}
\usepackage{amsmath,amssymb,amsfonts,bm,graphicx}
\usepackage[colorlinks=true,linkcolor=blue,citecolor=blue,urlcolor=blue]{hyperref}
\newcommand{\doilink}[2]{\href{https://doi.org/#1}{#2}}
\hypersetup{pdftitle={Dephasing-Enhanced Response and Phase Transitions in Quantum Boltzmann Samplers},pdfauthor={Yu-Xuan Zhang, Jing-Ling Chen, Leong-Chuan Kwek, Peng Wang}}

\usepackage{tikz,pgfplots}
\usetikzlibrary{arrows.meta,calc,patterns,decorations.markings}
\usepgfplotslibrary{fillbetween,groupplots}
\pgfplotsset{compat=1.18}
\definecolor{qblue}{HTML}{0072B2}
\definecolor{qred}{HTML}{CC5A35}
\definecolor{qteal}{HTML}{00876C}
\definecolor{qgray}{HTML}{63666A}
\definecolor{qpurple}{HTML}{775A9A}
\pgfplotsset{paper/.style={
  scale only axis,axis lines*=left,axis line style={black!70,line width=.4pt},
  tick align=outside,tick style={black!70,line width=.4pt},
  major tick length=2.0pt,minor tick length=1.3pt,
  tick label style={font=\fontsize{8}{9}\selectfont,/pgf/number format/fixed,/pgf/number format/precision=3},
  label style={font=\fontsize{9}{10}\selectfont},
  xlabel style={yshift=1pt},ylabel style={yshift=-1pt},
  scaled ticks=false,unbounded coords=jump,clip marker paths=true,
  legend style={draw=none,fill=none,font=\fontsize{8}{9}\selectfont,inner sep=1pt,cells={anchor=west}},
  every axis plot/.append style={line width=.8pt,mark size=1.6pt},
  /pgf/number format/1000 sep={}
}}
\newcommand{\plabel}[3]{\node[anchor=north west,font=\bfseries\fontsize{10}{11}\selectfont] at (#1,#2) {(#3)};}
\tikzset{flow/.style={-{Stealth[length=2.0mm,width=1.25mm]},line width=.75pt,black!65}}

\input{figure_data_main.tex}
\newcommand{\Jcritical}{0.25312\pm0.00016}
\newcommand{\NuExponent}{0.96\pm0.05}
\newcommand{\BetaExponent}{0.123\pm0.003}
\newcommand{\GammaExponent}{1.75\pm0.04}
\newcommand{\CriticalJ}{0.253123518580997}

\newcommand{\ket}[1]{|#1\rangle}\newcommand{\bra}[1]{\langle#1|}

\newcommand{\sech}{\operatorname{sech}}
\newcommand{\dd}{\mathsf d}

\begin{document}
\raggedbottom
\title{Dephasing-Enhanced Response and Phase Transitions in Quantum Boltzmann Samplers}
\author{Yu-Xuan Zhang}
\affiliation{School of Physics, Nankai University, Tianjin 300071, People's Republic of China}
\affiliation{Centre for Quantum Technologies, National University of Singapore, 3 Science Drive 2, Singapore 117543, Singapore}

\author{Jing-Ling Chen}
\email{chenjl@nankai.edu.cn}
\affiliation{Theoretical Physics Division, Chern Institute of Mathematics, Nankai University, Tianjin 300071, People's Republic of China}

\author{Leong-Chuan Kwek}
\email{kwekleongchuan@nus.edu.sg}
\affiliation{Centre for Quantum Technologies, National University of Singapore, 3 Science Drive 2, Singapore 117543, Singapore}
\affiliation{National Institute of Education, Nanyang Technological University, 1 Nanyang Walk, Singapore 637616, Singapore}

\author{Peng Wang}
\email{wp002005@163.com}
\affiliation{Centre for Quantum Technologies, National University of Singapore, 3 Science Drive 2, Singapore 117543, Singapore}
\affiliation{School of Computer and Artificial Intelligence, Southwest Minzu University, Chengdu 610225, China}

\date{September 13, 2026}
\begin{abstract}
Dephasing can drive a quantum Boltzmann sampler into an ordered phase while its programmed probability distribution remains disordered. We trace this transition to an enhanced conditional response: coherence loss changes how each spin responds to its neighbors, and interactions amplify the change. For all-to-all coupling, we establish the stationary transition at finite dephasing. Near criticality, the magnetization distribution depends on the ratio of dephasing strength to the square root of system size, retaining a finite imprint of coherence even as the local coherent correction vanishes. In the strong-dephasing limit on a square lattice, the dynamics generates effective multispin interactions and probability currents, with Ising-like critical statistics. Adjusting the local dissipative update restores the target probability process on arbitrary graphs.
\end{abstract}
\maketitle

In a Boltzmann machine, fields and pair couplings specify a probability distribution, and learning adjusts these parameters using sampled correlations~\cite{Ackley1985,Benedetti2017}. Classical conditional updates realize the target law through detailed balance~\cite{Glauber1963}. A quantum route encodes the square roots of its probabilities as many-body amplitudes~\cite{Somma2007,Wild2021}. Engineered dissipation provides a way to stabilize this state~\cite{Kraus2008,Diehl2008,Verstraete2009}.

Many quantum updates can share the same target state. In interacting spin networks, this freedom can alter relaxation and accelerate convergence~\cite{Fiorelli2019}. During continuous operation, environmental noise acts alongside the programmed updates. Classical nonequilibrium systems show that noise and interactions can cooperate to produce order~\cite{VandenBroeck1994}. For a quantum sampler, the programmed weights specify a reference distribution. The question is which statistical model the running device realizes at those fixed weights.

Dephasing provides a direct way to address this question. It suppresses coherence between spin configurations but leaves their instantaneous measurement probabilities unchanged. The programmed dissipative updates turn coherence loss into changes in the measurement probabilities. Interactions carry these changes back to each spin through the fields produced by its neighbors. Near an ordering threshold, even a small shift of the local response can compete with the collective restoring force. The noise needed to suppress residual coherent feedback may then depend on system size.

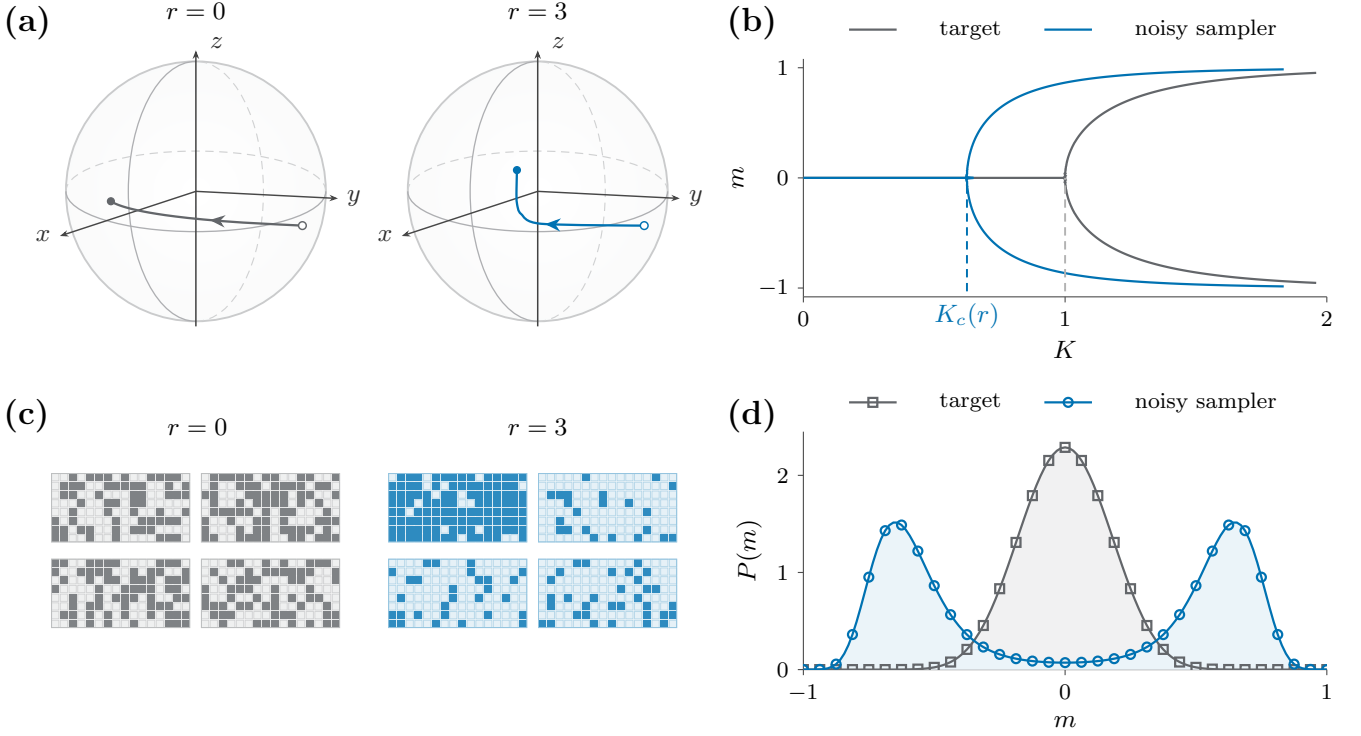
\begin{figure*}[t]
\centering

\pgfmathsetmacro{\sval}{0.8}
\pgfmathsetmacro{\rval}{3.0}
\pgfmathsetmacro{\alphaval}{\sval*\rval/(1+\rval)}
\pgfmathsetmacro{\Kcrit}{1/(1+\alphaval)}

\pgfmathdeclarefunction{ktarget}{1}{\pgfmathparse{ln((1+#1)/(1-#1))/(2*#1)}}
\pgfmathdeclarefunction{tnoisy}{1}{\pgfmathparse{2*(#1)/((1+\alphaval)+sqrt((1+\alphaval)^2-4*\alphaval*(#1)^2))}}
\pgfmathdeclarefunction{knoisy}{1}{\pgfmathparse{ln((1+tnoisy(#1))/(1-tnoisy(#1)))/(2*#1)}}

\newcommand{\blochframe}[2]{
\begin{scope}[shift={(#1,#2)},scale=1.28]
\pgfmathsetmacro{\R}{1.28}

\shade[inner color=white,outer color=qgray!3] (0,0) circle (\R);
\draw[qgray!34,line width=.68pt] (0,0) circle (\R);

\draw[qgray!30,densely dashed,line width=.42pt]
(\R,0) arc[start angle=0,end angle=180,x radius=\R,y radius=.31*\R];

\draw[qgray!52,line width=.48pt]
(-\R,0) arc[start angle=180,end angle=360,x radius=\R,y radius=.31*\R];

\draw[qgray!52,line width=.47pt]
(0,\R) arc[start angle=90,end angle=270,x radius=.52*\R,y radius=\R];

\draw[qgray!30,densely dashed,line width=.40pt]
(0,-\R) arc[start angle=270,end angle=450,x radius=.52*\R,y radius=\R];

\draw[-{Stealth[length=1.08mm]},black!72,line width=.52pt]
(0,0)--(-1.34,-.44) node[left=0pt,text=black!82] {$x$};

\draw[-{Stealth[length=1.08mm]},black!72,line width=.52pt]
(0,0)--(1.40,-.07) node[right=0pt,text=black!82] {$y$};

\draw[-{Stealth[length=1.10mm]},black!74,line width=.54pt]
(0,-1.33)--(0,1.39);

\node[anchor=south west,text=black!84] at (.05,1.32) {$z$};

\end{scope}
}

\def\InitPt{(1.0493,-0.3374)}

\def\TrajZero{
(1.0493,-0.3374)
(0.4657,-0.3077)
(0.0700,-0.2768)
(-0.2000,-0.2477)
(-0.3857,-0.2216)
(-0.5143,-0.1991)
(-0.6042,-0.1800)
(-0.6675,-0.1641)
(-0.7126,-0.1510)
(-0.7449,-0.1403)
(-0.7683,-0.1317)
(-0.7854,-0.1247)
(-0.7980,-0.1192)
(-0.8074,-0.1147)
(-0.8144,-0.1112)
(-0.8197,-0.1083)
(-0.8237,-0.1061)
(-0.8268,-0.1043)
(-0.8291,-0.1029)
(-0.8309,-0.1018)
(-0.8323,-0.1009)
(-0.8334,-0.1002)
(-0.8342,-0.0997)
(-0.8349,-0.0992)
}

\def\TrajNoise{
(1.0493,-0.3374)
(0.0336,-0.3194)
(-0.1648,-0.2218)
(-0.2026,-0.1309)
(-0.2088,-0.0573)
(-0.2091,0.0007)
(-0.2083,0.0462)
(-0.2075,0.0817)
(-0.2069,0.1095)
(-0.2064,0.1313)
(-0.2060,0.1483)
(-0.2057,0.1616)
(-0.2054,0.1720)
(-0.2052,0.1802)
(-0.2051,0.1866)
(-0.2049,0.1915)
(-0.2049,0.1954)
(-0.2048,0.1985)
(-0.2047,0.2009)
(-0.2047,0.2027)
(-0.2046,0.2042)
(-0.2046,0.2053)
(-0.2046,0.2062)
(-0.2046,0.2069)
}

\def\SteadyZero{(-0.8372,-0.0976)}
\def\SteadyNoise{(-0.2045,0.2094)}

\def\CfgTone{0/2,0/4,0/5,0/6,0/8,0/11,0/12,0/13,0/14,1/1,1/2,1/3,1/6,1/7,1/8,1/9,1/10,1/13,1/15,2/1,2/2,2/5,2/9,2/10,2/13,2/14,2/15,3/3,3/6,3/7,3/9,3/10,4/2,4/6,4/9,4/12,4/15,5/0,5/2,5/7,5/10,5/11,5/12,5/13,5/14,6/0,6/1,6/3,6/4,6/7,6/9,6/10,6/11,7/0,7/1,7/4,7/7,7/12,7/13,7/14}
\def\CfgTtwo{0/1,0/2,0/3,0/4,0/5,0/7,0/8,0/11,0/12,0/15,1/1,1/4,1/5,1/6,1/8,1/9,1/10,1/11,1/13,1/14,2/0,2/1,2/5,2/6,2/7,2/8,2/9,2/11,2/12,2/15,3/1,3/4,3/6,3/7,3/8,3/9,3/11,3/13,4/1,4/2,4/3,4/4,4/5,4/7,4/9,4/10,4/14,5/0,5/3,5/6,5/7,5/9,5/13,5/14,5/15,6/0,6/4,6/6,6/7,6/9,6/10,6/11,6/12,6/14,7/0,7/1,7/4,7/5,7/6,7/7,7/10,7/11,7/13,7/14,7/15}
\def\CfgTthree{0/0,0/1,0/4,0/5,0/6,0/7,0/12,0/13,0/14,1/0,1/3,1/4,1/5,1/6,1/9,1/11,1/12,1/15,2/1,2/4,2/6,2/7,2/9,2/10,2/11,2/13,2/14,2/15,3/0,3/5,3/8,3/9,3/10,3/11,3/13,3/14,4/2,4/5,4/7,4/8,4/9,4/11,4/14,4/15,5/2,5/4,5/5,5/8,5/11,5/12,5/13,6/1,6/2,6/4,6/6,6/8,6/11,6/13,6/14,6/15,7/2,7/4,7/6,7/9,7/13,7/14,7/15}
\def\CfgTfour{0/3,0/4,0/7,0/9,0/12,0/13,0/14,1/1,1/5,1/7,1/8,1/9,1/10,1/11,1/12,1/15,2/2,2/4,2/5,2/6,2/10,2/11,2/15,3/0,3/3,3/4,3/5,3/8,3/12,4/0,4/3,4/6,4/8,4/10,4/12,4/13,5/0,5/1,5/2,5/4,5/5,5/7,5/8,5/11,5/14,6/1,6/2,6/3,6/8,6/9,6/10,6/14,6/15,7/2,7/6,7/7,7/8,7/10,7/12,7/14,7/15}

\def\CfgNone{0/1,0/2,0/3,0/5,0/6,0/7,0/8,0/9,0/11,0/12,0/13,0/14,0/15,1/0,1/1,1/2,1/3,1/5,1/7,1/8,1/9,1/10,1/11,1/12,1/13,1/14,2/0,2/1,2/2,2/3,2/4,2/6,2/7,2/9,2/10,2/11,2/12,2/13,2/14,2/15,3/0,3/1,3/2,3/4,3/5,3/6,3/7,3/10,3/11,3/12,3/13,3/14,3/15,4/0,4/1,4/3,4/4,4/5,4/6,4/7,4/8,4/9,4/10,4/11,4/12,4/13,4/14,4/15,5/0,5/1,5/2,5/3,5/4,5/5,5/6,5/7,5/8,5/9,5/10,5/11,5/12,5/13,5/14,5/15,6/0,6/1,6/2,6/3,6/4,6/5,6/7,6/8,6/9,6/11,6/12,6/13,6/14,6/15,7/0,7/1,7/2,7/3,7/4,7/5,7/6,7/7,7/9,7/10,7/11,7/12,7/15}
\def\CfgNtwo{0/5,0/13,1/15,2/1,2/2,2/3,2/7,3/3,3/7,3/10,4/4,4/12,5/2,6/2,6/3,6/5,6/12,7/6,7/8,7/11,7/14,7/15}
\def\CfgNthree{0/4,0/5,0/9,1/0,1/9,1/15,2/2,2/3,2/10,2/11,2/14,3/7,3/11,4/1,4/7,4/10,4/13,5/6,6/0,6/1,6/3,6/6,6/11,6/15,7/0,7/1,7/5,7/8,7/15}
\def\CfgNfour{0/4,0/5,0/11,1/2,1/8,1/10,1/12,2/1,2/11,3/9,3/12,3/13,4/1,4/3,4/6,4/11,5/1,5/5,5/8,5/9,5/12,5/15,6/4,6/5,6/12,6/14,6/15,7/0,7/1,7/6,7/8,7/12,7/14}

\newcommand{\drawconfig}[4]{
\begin{scope}[shift={(#1,#2)}]
\foreach \rr in {0,...,7}{
  \foreach \cc in {0,...,15}{
    \filldraw[fill=#3!10,draw=#3!28,line width=.10pt] ({\cc*0.110},{(7-\rr)*0.110}) rectangle ++(0.092,0.092);
  }
}
\foreach \rr/\cc in #4{
  \fill[#3!82] ({\cc*0.110},{(7-\rr)*0.110}) rectangle ++(0.092,0.092);
}
\draw[#3!42,line width=.35pt] (0,0) rectangle (1.742,0.862);
\end{scope}
}

\resizebox{\textwidth}{!}{%
\begin{tikzpicture}[x=1cm,y=1cm]

\node[anchor=north west,font=\bfseries\large] at (0,9.0) {(a)};

\blochframe{2.54}{6.55}
\blochframe{6.85}{6.55}

\node at (2.54,8.83) {$r=0$};
\node at (6.85,8.83) {$r=3$};

\begin{scope}[shift={(2.54,6.55)},scale=1.28]

\draw[qgray,line width=.82pt,postaction={decorate},
decoration={markings,mark=at position 0.48 with {\arrow{Stealth[length=1.95mm,width=1.55mm]}}}]
plot[smooth] coordinates \TrajZero;

\filldraw[fill=white,draw=qgray,line width=.52pt]
\InitPt circle (.038);

\filldraw[fill=qgray,draw=qgray,line width=.45pt]
\SteadyZero circle (.034);

\end{scope}

\begin{scope}[shift={(6.85,6.55)},scale=1.28]

\draw[qblue,line width=.82pt,postaction={decorate},
decoration={markings,mark=at position 0.58 with {\arrow{Stealth[length=1.95mm,width=1.65mm]}}}]
plot[smooth] coordinates \TrajNoise;

\filldraw[fill=white,draw=qblue,line width=.52pt]
\InitPt circle (.038);

\filldraw[fill=qblue,draw=qblue,line width=.45pt]
\SteadyNoise circle (.034);

\end{scope}

\node[anchor=north west,font=\bfseries\large] at (0,4.05) {(c)};
\node at (2.56,3.58) {$r=0$};
\node at (6.85,3.58) {$r=3$};

\drawconfig{0.72}{2.13}{qgray}{\CfgTone}
\drawconfig{2.61}{2.13}{qgray}{\CfgTtwo}
\drawconfig{0.72}{1.05}{qgray}{\CfgTthree}
\drawconfig{2.61}{1.05}{qgray}{\CfgTfour}

\drawconfig{4.97}{2.13}{qblue}{\CfgNone}
\drawconfig{6.86}{2.13}{qblue}{\CfgNtwo}
\drawconfig{4.97}{1.05}{qblue}{\CfgNthree}
\drawconfig{6.86}{1.05}{qblue}{\CfgNfour}

\node[anchor=north west,font=\bfseries\large] at (9.12,9.0) {(b)};

\begin{axis}[
paper,
at={(10.20cm,5.22cm)},
anchor=south west,
width=6.6cm,
height=3.00cm,
xmin=0,xmax=2,
ymin=-1.08,ymax=1.08,
xtick={0,1,2},
ytick={-1,0,1},
xlabel={$K$},
ylabel={$m$},
legend columns=2,
legend style={at={(.5,1.045)},anchor=south,draw=none,fill=none,column sep=13pt},
clip=false
]

\addplot[qgray,domain=0:1,samples=2] {0};
\addlegendentry{target}

\addplot[qgray,domain=.001:.999,samples=220,variable=\m,restrict x to domain=0:2,forget plot]
({ktarget(\m)},{\m});

\addplot[qgray,domain=.001:.999,samples=220,variable=\m,restrict x to domain=0:2,forget plot]
({ktarget(\m)},{-\m});

\addplot[qblue,domain=0:\Kcrit,samples=2] {0};
\addlegendentry{noisy sampler}

\addplot[qblue,domain=.001:.999,samples=220,variable=\m,restrict x to domain=0:2,forget plot]
({knoisy(\m)},{\m});

\addplot[qblue,domain=.001:.999,samples=220,variable=\m,restrict x to domain=0:2,forget plot]
({knoisy(\m)},{-\m});

\addplot[black!30,densely dashed,line width=.7pt,forget plot]
coordinates {(1,-1.03)(1,0)};

\addplot[qblue,densely dashed,line width=.7pt,forget plot]
coordinates {(\Kcrit,-1.03)(\Kcrit,0)};

\node[text=qblue,font=\small,anchor=north,yshift=-.8mm]
at (axis cs:\Kcrit,-1.00) {$K_c(r)$};

\end{axis}

\node[anchor=north west,font=\bfseries\large] at (9.12,4.05) {(d)};

\begin{axis}[
paper,
at={(10.20cm,.52cm)},
anchor=south west,
width=6.6cm,
height=3.00cm,
xmin=-1,xmax=1,
ymin=0,ymax=2.45,
xtick={-1,0,1},
ytick={0,1,2},
xlabel={$m$},
ylabel={$P(m)$},
legend columns=2,
legend style={at={(.5,1.045)},anchor=south,draw=none,fill=none,column sep=13pt}
]

\addplot[draw=none,fill=qgray!9,forget plot]
table[x=m,y=target]{\DataDistribution}\closedcycle;

\addplot[draw=none,fill=qblue!8,forget plot]
table[x=m,y=p]{\DataDistribution}\closedcycle;

\addplot[qgray,mark=square,mark options={fill=white},mark repeat=4]
table[x=m,y=target]{\DataDistribution};

\addlegendentry{target}

\addplot[qblue,mark=o,mark options={fill=white},mark repeat=4]
table[x=m,y=p]{\DataDistribution};

\addlegendentry{noisy sampler}

\end{axis}

\end{tikzpicture}
}
\caption{Dephasing-enhanced response and stationary ordering at fixed programmed weights.
(a) Single-spin Bloch trajectories at a fixed conditional field, showing the dephasing-induced change in the stationary polarization.
(b) Stationary Curie--Weiss branches, $m=m_*(Km)$. The target orders at $K_c=1$, whereas dephasing shifts the sampler threshold to $K_c(r)<1$.
(c) Four computational-basis samples at $N=128$ and $K=0.75$, showing disordered configurations without dephasing and ordered configurations with dephasing.
(d) Stationary magnetization distributions at the same $N$ and $K$: the target remains unimodal around $m=0$, whereas the dephased sampler is bimodal.
Throughout, the dephased sampler uses $r=3$ and $s=0.8$, with $r=0$ as the no-dephasing reference; $P(m)$ is the probability mass divided by $2/N$.}
\label{fig:mechanism}
\end{figure*}

Here dephasing produces a stationary ordering transition at fixed programmed weights [Fig.~\ref{fig:mechanism}]: an enhanced local response shifts the collective threshold, creating a regime in which the sampler is ordered while the target remains disordered. We establish the thermodynamic stationary law at finite dephasing. Near criticality, locally vanishing coherent feedback still changes the rescaled magnetization distribution. Its effect depends on dephasing strength divided by $\sqrt N$. On a square lattice, the strong-dephasing limit generates multispin interactions and probability currents alongside enhanced order. The same response calculation yields a local update that restores the target probability process.

\textit{Conditional feedback and stationary order.---}Consider $N$ spins $z_i=\pm1$ with target distribution and coherent encoding
\begin{equation}
 \begin{aligned}
 p_\theta(z)&=Z_\theta^{-1}
 e^{\sum_i h_i z_i+\sum_{i<j}J_{ij}z_i z_j},\\
 \ket{\Psi_\theta}&=\sum_z\sqrt{p_\theta(z)}\ket z .
 \end{aligned}
 \label{eq:target}
\end{equation}
Here $\theta=\{h_i,J_{ij}\}$ contains the finite, dimensionless fields and symmetric couplings, and $Z_\theta$ normalizes the distribution. At fixed neighbors, spin $i$ sees $a_i=h_i+\sum_{j\ne i}J_{ij}z_j$ and has conditional polarization $u_i=\tanh a_i$. With Pauli operators $X_i,Z_i$ and the convention $Z_i\ket0_i=\ket0_i$, its conditional target state is
\[
 \ket{\nu_i}=\sqrt{\frac{1+u_i}{2}}\ket0_i+\sqrt{\frac{1-u_i}{2}}\ket1_i.
\]
To construct an update that fixes this state, set $P_i=\ket{\nu_i}\bra{\nu_i}$ and $Q_i=\openone-P_i$ and choose
\begin{equation}
 A_{i0}=P_i+sQ_i,\qquad
 A_{i1}=\sqrt{1-s^2}\ket{\nu_i}\bra{\nu_i^\perp},
 \label{eq:kraus}
\end{equation}
where $0\le s<1$ and $\ket{\nu_i^\perp}$ is orthogonal to $\ket{\nu_i}$. The update retains amplitude $s$ in this orthogonal component and recycles the remainder into $\ket{\nu_i}$. In the network, each neighboring $Z$ configuration coherently controls the corresponding update on spin $i$. This action fixes every conditional branch of $\ket{\Psi_\theta}$, making it a common fixed state of the channels $\Phi_i(\rho)=\sum_{\mu=0,1}A_{i\mu}\rho A_{i\mu}^\dagger$.

Let $\rho_N$ be the density matrix of the entire network. We study its continuous evolution under these updates at rate $\nu$ and phase noise at rate $\gamma$:
\begin{equation}
 \frac{d\rho_N}{dt}
 =\nu\sum_i[\Phi_i(\rho_N)-\rho_N]
 +\frac{\gamma}{2}\sum_i(Z_i\rho_NZ_i-\rho_N).
 \label{eq:generator}
\end{equation}
At $\gamma=0$ the engineered dynamics preserves the coherent target. The noise term damps off-diagonal matrix elements and has zero diagonal contribution. We measure dephasing relative to the programmed relaxation scale by $r=\gamma/[\nu(1-s^2)]$ and use time $\tau=\nu(1-s)t$.

The effect of phase noise on the measured response can first be isolated at a fixed neighboring configuration. Write $m=\langle Z_i\rangle$, $x=\langle X_i\rangle$, $u=\tanh a$, and $q=\sech a$. The local contribution to Eq.~\eqref{eq:generator} gives
\[
 \frac{dm}{d\tau}=(1+s)u-(1+su^2)m-su qx.
\]
The last term couples transverse coherence to the population drift. At stationarity, eliminating $x$ with the transverse equation yields the conditional response
\begin{equation}
 m_*(a)=\frac{(1+\alpha)\tanh a}{1+\alpha\tanh^2a},
 \qquad \alpha=\frac{sr}{1+r}.
 \label{eq:response}
\end{equation}
At zero field, the surviving coherence is $x_0=(1+r)^{-1}$ and the response slope is $g(r)=m_*'(0)=1+s(1-x_0)$. Coherence supplies negative feedback to the longitudinal response. Dephasing weakens this feedback, increasing the response to the same conditional field, as illustrated by the Bloch trajectories in Fig.~\ref{fig:mechanism}(a). The full reset $s=0$ has $g=1$; for $s>0$, the slope increases continuously with noise. The Supplemental Material (SM) derives this response and the network equations from the same local update~\cite{SM}.

Interactions feed the enhanced response back into the conditional field. For Curie--Weiss coupling $J_{ij}=K/N$, each spin couples equally to all the others. At zero field and $0<K<1$, the target is disordered. Define $M_z=N^{-1}\sum_i Z_i$ and its mean $m_N=\langle M_z\rangle$. Taking the magnetization expectation in Eq.~\eqref{eq:generator} gives
\begin{equation}
 \begin{aligned}
 \frac{dm_N}{d\tau}=\frac1N\sum_i\Bigl\langle &(1+s)u_i\\
 &-(1+su_i^2)Z_i-su_iq_iX_i\Bigr\rangle.
 \end{aligned}
 \label{eq:magnetization}
\end{equation}
Here $a_i=(K/N)\sum_{j\ne i}Z_j$ is the neighbor-field operator, with $u_i=\tanh a_i$ and $q_i=\sech a_i$. Within a thermodynamic phase of magnetization $m$, this field concentrates at $Km$. The coupled longitudinal and transverse averages then determine the collective flow~\cite{Fiorelli2023,Carollo2024}. The full stationary state can contain phases of opposite magnetization. Linearizing the magnetization equation about the disordered fixed point uses $u\simeq Km$, $q\simeq1$, and the stationary coherence $x_0$, giving
\begin{equation}
 \frac{d\,\delta m}{d\tau}=[Kg(r)-1]\delta m,
 \qquad
 K_c(r)=\frac{1+r}{1+(1+s)r}.
 \label{eq:threshold}
\end{equation}
The threshold marks the point at which a small magnetization reinforces itself through the conditional response. On a stationary branch, the transverse variables take their fixed-field values and the magnetization satisfies $m=m_*(Km)$, giving the branches shown in Fig.~\ref{fig:mechanism}(b). For $K_c(r)<K<1$, this equation has a unique positive solution $m_r$ and its inverted partner, forming the two stable ordered branches $\pm m_r$.

To determine the long-time sampling statistics, let $P_{N,r}$ be the measurement law of $M_z$ in the finite-system steady state. For fixed $0<s<1$ and $r>0$, its thermodynamic limit is
\begin{equation}
 P_{N,r}\Longrightarrow
 \begin{cases}
 \delta_0,&K\le K_c(r),\\[2pt]
 \tfrac12\delta_{m_r}+\tfrac12\delta_{-m_r},
    &K_c(r)<K<1
 \end{cases},
 \label{eq:stationary-limit}
\end{equation}
where $\delta_m$ is a point mass at $m$. At each finite $N$ the steady state is unique and inversion symmetric, so its first moment vanishes. In the ordered regime, the two peaks give $\langle M_z^2\rangle\to m_r^2$; the target's second moment tends to zero. At $N=128$ and $K=0.75$, Fig.~\ref{fig:mechanism}(c,d) shows the finite-size contrast: at $r=0$ the samples are disordered and $P(m)$ is unimodal, whereas at $r=3$ the samples are ordered and $P(m)$ is bimodal.

To derive this limiting law, we combine the collective flow with the symmetries of the finite-system steady state. Uniqueness and permutation invariance of the generator make that state symmetric under spin permutations. Its reduced thermodynamic limits are mixtures of product states~\cite{ZhangBarthel2024,HudsonMoody1976}, whose single-spin averages follow the collective flow. A stationary drift argument confines the mixture to its fixed points. Above threshold, the leading finite-size magnetization fluctuations drive weight away from the unstable center, and inversion symmetry fixes equal weights on the two surviving branches. The SM develops this argument and details the finite-size quantum steady-state calculations~\cite{Shammah2018}. The~finite-size spectrum also resolves relaxation within an ordered branch and the slower loss of its sign, a signature of quantum metastability~\cite{Macieszczak2016,SM}.

\textit{Critical amplification of coherence.---}Near the transition, fluctuations reveal the effect of residual coherence. At the fully dephased threshold $K_\infty=1/(1+s)$, finite noise leaves the restoring rate
\[
 1-g(r)K_\infty=\frac{s}{(1+s)(1+r)}.
\]
This rate vanishes as $r$ grows, leaving cubic drift as the leading restoring force. Balancing that drift against magnetization diffusion of order $1/N$ gives the classical Curie--Weiss fluctuation scale $m\sim N^{-1/4}$~\cite{Ding2009}. On this scale, the residual linear drift $m/r$ competes with the cubic drift precisely when $r\sim\sqrt N$. Suppressing its effect on critical fluctuations therefore requires noise that grows with system size.

To resolve this competition, take $K_N=K_\infty+\kappa_N/\sqrt N$ with bounded $\kappa_N$, and $r_N\ge\rho_0\sqrt N$ for any fixed $\rho_0>0$. The drift expansion has cubic coefficient $c_\infty=(s+1/3)/(1+s)^2$, which sets the scales
\[
 X_N=(Nc_\infty)^{1/4}M_z,\qquad
 A_N=\sqrt{N/c_\infty}\,[g(r_N)K_N-1].
\]
Here $X_N$ is the rescaled magnetization, and $A_N$ measures the linear drift on the fluctuation scale. To obtain the stationary law of $X_N$, we eliminate the rapidly damped coherence from the quantum equation, retaining its leading feedback on magnetization. This feedback supplies the linear correction in $A_N$. Whenever $A_N\to A$, the drift and diffusion on the critical time scale give the following generator acting on a smooth function $f$:
\[
 (\mathcal G_Af)(X)=f''(X)+(AX-X^3)f'(X).
\]
Integrating its stationary equation yields
\begin{equation}
 X_N\Longrightarrow p_A(X)
 =\mathcal Z_A^{-1}
   \exp\!\left(\frac{AX^2}{2}-\frac{X^4}{4}\right),
 \label{eq:critical-law}
\end{equation}
where $\mathcal Z_A$ normalizes the density. The convergence is uniform over the stated window and includes the second moment, for both finite and diverging $r_N/\sqrt N$~\cite{SM}.

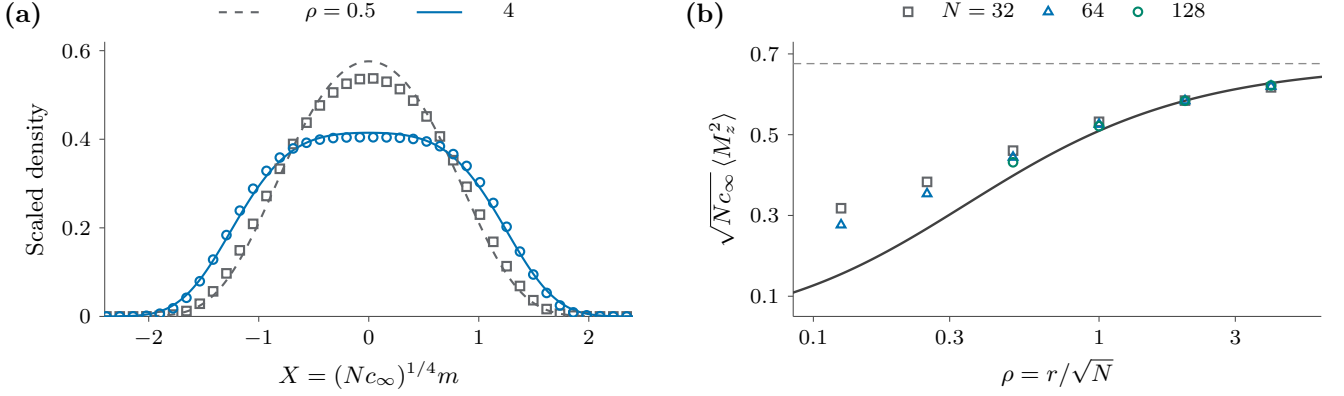
\begin{figure*}[t]
\centering
\resizebox{\textwidth}{!}{%

\begin{tikzpicture}
\path[use as bounding box] (0,0) rectangle (17.8,5.2);
\plabel{0}{5.18}{a}\plabel{9.00}{5.18}{b}
\begin{axis}[paper,at={(1.45cm,.91cm)},anchor=south west,width=6.98cm,height=3.63cm,
 xmin=-2.4,xmax=2.4,ymin=0,ymax=.62,
 xtick={-2,-1,0,1,2},ytick={0,.2,.4,.6},
 xlabel={$X=(Nc_\infty)^{1/4}m$},ylabel={Scaled density},
 legend columns=2,legend style={at={(.5,1.045)},anchor=south,column sep=13pt}]
\addplot[qgray,dashed] table[x=X,y=p]{\DataJointTheoryRhoZeroPFive};
\addlegendentry{$\rho=0.5$}
\addplot[qblue] table[x=X,y=p]{\DataJointTheoryRhoFourPZero};
\addlegendentry{$4$}
\addplot[qgray,only marks,mark=square,mark options={fill=white},mark repeat=3,forget plot]
 table[x=X,y=p]{\DataJointNOneTwoEightRhoZeroPFive};
\addplot[qblue,only marks,mark=o,mark options={fill=white},mark repeat=3,forget plot]
 table[x=X,y=p]{\DataJointNOneTwoEightRhoFourPZero};
\end{axis}
\begin{axis}[paper,at={(10.55cm,.91cm)},anchor=south west,width=6.98cm,height=3.63cm,
 xmode=log,xmin=.085,xmax=6,ymin=.05,ymax=.73,
 xtick={.1,.3,1,3},xticklabels={0.1,0.3,1,3},ytick={.1,.3,.5,.7},
 xlabel={$\rho=r/\sqrt N$},ylabel={$\sqrt{Nc_\infty}\,\langle M_z^2\rangle$},
 legend columns=3,legend style={at={(.5,1.045)},anchor=south,column sep=9pt}]
\addplot[black!45,densely dashed,line width=.5pt,forget plot] coordinates {(.085,.67597824)(6,.67597824)};
\addplot[black!75,line width=.9pt,forget plot] table[x=rho,y=Q]{\DataCrossoverTheory};
\addplot[qgray,only marks,mark=square,mark options={fill=white}] table[x=rho,y=Q]{\DataCrossoverNThreeTwo};
\addlegendentry{$N=32$}
\addplot[qblue,only marks,mark=triangle,mark size=1.9pt,mark options={fill=white}] table[x=rho,y=Q]{\DataCrossoverNSixFour};
\addlegendentry{$64$}
\addplot[qteal,only marks,mark=o,mark options={fill=white}] table[x=rho,y=Q]{\DataCrossoverNOneTwoEight};
\addlegendentry{$128$}
\end{axis}
\end{tikzpicture}
}
\caption{Residual coherence narrows the critical stationary distribution. (a) Magnetization distributions at the fully dephased critical coupling $K_\infty=1/(1+s)$, with $s=0.8$. Lines give the limiting densities $p_{A_{0,\rho}}$ for $\rho=r/\sqrt N=0.5$ and $4$; open symbols are full quantum steady-state results at $N=128$. (b) The scaled variance follows the crossover in Eq.~\eqref{eq:crossover}. Open symbols show finite-size quantum results, the solid curve is the limiting crossover, and the horizontal dashed line is the fully dephased quartic value. A finite $r/\sqrt N$ leaves a finite change in the critical fluctuations.}
\label{fig:critical}
\end{figure*}

In particular, if $\kappa_N\to\kappa$ and $r_N/\sqrt N\to\rho\in(0,\infty]$, then
\begin{equation}
 A_{\kappa,\rho}
 =\frac{(1+s)^2\kappa-s/\rho}{\sqrt{s+1/3}}.
 \label{eq:two-parameter}
\end{equation}
The coupling shift and residual coherent feedback thus change the same quadratic term in the stationary density. At the unchanged coupling $K_\infty$, every finite $\rho$ gives $A_{0,\rho}<0$ and narrows the distribution by a finite amount [Fig.~\ref{fig:critical}(a)]. Its variance obeys
\begin{equation}
 \sqrt{Nc_\infty}\,\langle M_z^2\rangle
 \longrightarrow \int X^2p_{A_{0,\rho}}(X)\,dX .
 \label{eq:crossover}
\end{equation}
Only as $\rho\to\infty$ does it reach the fully dephased quartic value $2\Gamma(3/4)/\Gamma(1/4)$ [Fig.~\ref{fig:critical}(b)]. Thus dephasing can grow arbitrarily strong on the single-spin scale while residual coherence leaves a finite imprint on the collective critical distribution.

\textit{Spatial order and probability currents.---}The enhanced local response also changes sampling with short-range interactions. Physical samplers commonly operate on bounded-degree graphs, including low-dimensional lattices and hardware connectivities such as Chimera and Pegasus. To determine the resulting spatial statistics, take $r\to\infty$ at fixed finite graph size. Equation~\eqref{eq:generator} then reduces to a classical process with single-spin flip rates, in units of $\tau$,
\begin{equation}
 w_i(z_i\to-z_i)
 =\tfrac12(1-z_i u_i)(1-sz_i u_i).
 \label{eq:rates}
\end{equation}
The extra factor $1-sz_i u_i$ suppresses flips of aligned spins and favors flips of misaligned spins, strengthening the response to the local field. The ratio of opposite flip rates corresponds to the field $F_s(a)=a+\operatorname{artanh}(s\tanh a)$. On a one-dimensional nearest-neighbor chain, $a_i=JS_i$ with $S_i=z_{i-1}+z_{i+1}\in\{-2,0,2\}$. Since $F_s$ is odd, $F_s(JS_i)=\frac12F_s(2J)S_i$, so the stationary law is exactly the one-dimensional Ising law with enhanced pair coupling $F_s(2J)/2>J$. The corresponding correlation length is enhanced; the one-dimensional chain is the lower-critical-dimensional case of the short-range problem~\cite{SM}.

On the square lattice, the local field has two nonzero magnitudes, so the nonlinear response can no longer be absorbed into a single pair coupling. Its linear part strengthens pair alignment. To find how its nonlinear part changes the sampled law $\pi_s$, we expand the stationary equation at small $J$ on a square torus. Write $E=\sum_{\langle ij\rangle}z_i z_j$ and let $T$ sum four-spin stars, each containing a site and three of its neighbors. The resulting stationary weight is
\begin{equation}
 \begin{aligned}
 \log\pi_s={}&\mathrm{const}+(1+s)JE\\
 &-\frac{s(1-s^2)}{3}J^3
       \left(10E+\tfrac32T\right)+O(J^4).
 \end{aligned}
 \label{eq:spatial-weight}
\end{equation}
The nonlinear response therefore generates a four-spin interaction in the sampled law. Substituting this law into the stationary probability flows also gives nonzero currents, with entropy-production rate per site starting at $12s^2(1-s^2)^2J^6$~\cite{Schnakenberg1976,SM}. The process remains irreversible at stationarity.

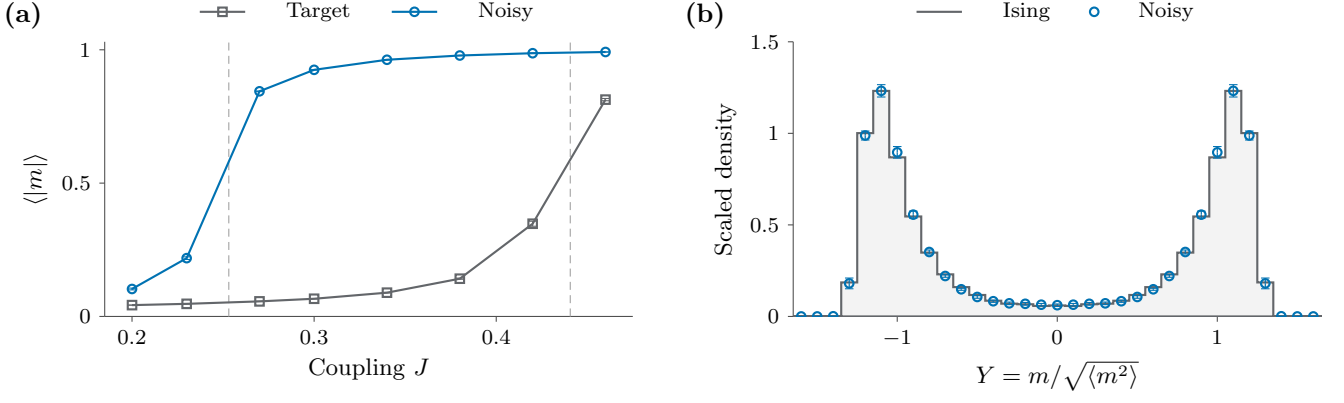
\begin{figure*}[t]
\centering
\resizebox{\textwidth}{!}{%

\begin{tikzpicture}
\path[use as bounding box] (0,0) rectangle (17.8,5.2);
\plabel{0}{5.18}{a}\plabel{9.00}{5.18}{b}
\begin{axis}[paper,at={(1.45cm,.91cm)},anchor=south west,width=6.98cm,height=3.63cm,
 xmin=.185,xmax=.475,ymin=0,ymax=1.03,xtick={.2,.3,.4},ytick={0,.5,1},
 xlabel={Coupling $J$},ylabel={$\langle |m|\rangle$},
 legend columns=2,legend style={at={(.5,1.045)},anchor=south,column sep=13pt}]
\addplot[black!30,densely dashed,line width=.5pt,forget plot] coordinates {(\CriticalJ,0)(\CriticalJ,1.03)};
\addplot[black!30,densely dashed,line width=.5pt,forget plot] coordinates {(.44068679,0)(.44068679,1.03)};
\addplot[qgray,mark=square,mark options={fill=white},
 error bars/.cd,y dir=both,y explicit,error bar style={line width=.45pt}]
 table[x=J,y=m,y error=err]{\DataSpatialTarget};
\addlegendentry{Target}
\addplot[qblue,mark=o,mark options={fill=white},
 error bars/.cd,y dir=both,y explicit,error bar style={line width=.45pt}]
 table[x=J,y=m,y error=err]{\DataSpatialNoisy};
\addlegendentry{Noisy}
\end{axis}
\begin{axis}[paper,at={(10.55cm,.91cm)},anchor=south west,width=6.98cm,height=3.63cm,
 xmin=-1.65,xmax=1.65,ymin=0,ymax=1.5,xtick={-1,0,1},ytick={0,.5,1,1.5},
 xlabel={$Y=m/\sqrt{\langle m^2\rangle}$},ylabel={Scaled density},
 legend columns=2,legend style={at={(.5,1.045)},anchor=south,column sep=13pt}]
\addplot[qgray,fill=qgray!8,mark=none] table[x=Y,y=p]{\DataHistIsingSteps}\closedcycle;
\addlegendentry{Ising}
\addplot[qblue,only marks,mark=o,mark size=1.6pt,mark options={fill=white},
 error bars/.cd,y dir=both,y explicit,error bar style={line width=.45pt}]
 table[x=Y,y=p,y error=err]{\DataHistNoisy};
\addlegendentry{Noisy}
\end{axis}
\end{tikzpicture}
}
\caption{Spatial order in the strong-dephasing process of Eq.~\eqref{eq:rates}. (a) Magnetization magnitudes for the noisy sampler and target Ising process on the same $32\times32$ lattice. Vertical lines mark their thermodynamic critical couplings; between them, the sampler is ordered and the target is disordered. (b) Standardized magnetization distributions on a $64\times64$ torus, comparing the noisy sampler at $J=0.2530$ with the equilibrium Ising model at its own critical coupling. The sampler has $s=0.8$. Bars are independent-run bootstrap errors, including uncertainty in the normalization. The SM gives critical exponents, moment ratios, and finite-size analysis.}
\label{fig:spatial}
\end{figure*}

On larger square lattices at $s=0.8$, the ordering transition shifts from the target Ising value $J_{\rm I}=\frac12\log(1+\sqrt2)=0.44069\ldots$~\cite{Onsager1944} to $J_c=\Jcritical$. Figure~\ref{fig:spatial}(a) displays the ordered sampler and disordered target at the same couplings. We characterize this transition through its scaling exponents and magnetization distribution. Finite-size fits give the correlation-length exponent $\nu=\NuExponent$ and the magnetization and susceptibility ratios $\beta/\nu=\BetaExponent$ and $\gamma/\nu=\GammaExponent$, consistent with Ising critical behavior. Torus moment ratios and the standardized distribution in Fig.~\ref{fig:spatial}(b) provide independent evidence~\cite{SalasSokal2000,SM}. At long wavelengths, the nonconserved scalar order parameter, inversion symmetry, and nonzero local noise lead to scalar relaxational dynamics. The detailed-balance-breaking terms enter at higher field or gradient order~\cite{Grinstein1985}, explaining how the observed critical statistics coexist with microscopic probability currents.

\textit{Restoring the target probability process.---}The local response calculation also shows how to make the population drift follow the target conditional update independently of coherence. Dissipative preparation permits changes of a jump's output that preserve its dark state~\cite{Fiorelli2019}. We use this freedom to redirect the orthogonal component in Eq.~\eqref{eq:kraus}, replacing the ket in $A_{i1}$ by a normalized $\ket{\chi_i}$. Both $A_{i0}$ and the Kraus effects $A_{i\mu}^\dagger A_{i\mu}$ remain unchanged, and the channel still fixes $\ket{\Psi_\theta}$. Choosing the output polarization $b_i=\langle\chi_i|Z_i|\chi_i\rangle$ to cancel the transverse term gives the longitudinal update
\begin{equation}
 b_i=\frac{1-s}{1+s}u_i,\qquad
 \Phi_i^\dagger(Z_i)=sZ_i+(1-s)u_i\openone.
 \label{eq:calibration}
\end{equation}
For fixed neighbors, the entire longitudinal evolution is now $dm/d\tau=u-m$, independently of the dephasing rate. A pure recycling state with this polarization can be prepared by rotations conditioned on the neighboring spins~\cite{Mottonen2005,Bergholm2005}. The calibrated channel uses the same local, neighbor-conditioned control structure as the original update. On bounded-degree graphs, the cost of each local update is independent of the total system size $N$, so a sweep over all spins requires $O(N)$ local operations. The calibration changes the conditional recycling rotation without changing this asymptotic scaling; circuit constructions and implementation errors are discussed in the Supplemental Material~\cite{SM}.

For a binary spin, this longitudinal update fixes both conditional outcome probabilities. Since the cancellation holds in every neighboring configuration, it also fixes the network's probability evolution. Let $\dd$ extract computational-basis probabilities and let $C_i$ replace spin $i$ by a draw from its target conditional distribution. On any graph and for every input state,
\begin{equation}
 \dd\Phi_i=[sI+(1-s)C_i]\dd ,
 \label{eq:closure}
\end{equation}
where $I$ is the identity on probability vectors. Writing $p_N=\dd\rho_N$ and summing the contributions in Eq.~\eqref{eq:generator} gives
\[
 \frac{dp_N}{d\tau}=\sum_i(C_i-I)p_N.
\]
Each $C_i$ preserves $p_\theta$ by resampling its own conditional distribution. Positive single-spin rates then make $p_\theta$ the unique stationary law on a finite graph. Equation~\eqref{eq:closure} also holds through successive updates, intervening population-preserving noise, and computational-basis readouts, outcome by outcome. With the same initial probabilities and parameter-update rule, learning driven by these readouts has the same joint law of measurements and parameter histories as the classical sampler~\cite{Theurer2019,SM}.

\textit{Discussion.---}Criticality makes the accuracy of a classical description depend on the observable and on system size. At $K_\infty$, recovering the fully dephased magnetization distribution on its natural fluctuation scale requires $r/\sqrt N\to\infty$. For $r$ of order $\sqrt N$, the local coherent correction tends to zero, yet the collective distribution retains a finite shift. In physical implementations, the update and dephasing rates may be jointly constrained by microscopic controls, restricting the accessible values of $r$. Finite environmental memory can also modify the conditional response beyond the Markovian dynamics of Eq.~\eqref{eq:generator}.

On the lattice, the four-spin term in Eq.~\eqref{eq:spatial-weight} changes the statistical model realized by the sampler. It cannot be absorbed into a change of temperature in the programmed pairwise model. The stationary currents establish irreversibility, even as the observed critical statistics remain Ising-like. Agreement in critical scaling can thus coexist with a different microscopic probability law. The latter controls the correlations used in learning.

To recover those correlations, the calibration fixes how each local update acts on measurement probabilities. This condition determines the full probability evolution and the records used for learning under repeated readout. The same target quantum state admits different dissipative updates; choosing their action on measured probabilities provides direct control over the statistical model realized during operation.

\begin{acknowledgments}
This work is supported by the Quantum Science and Technology-National Science and Technology Major Project (Grant No. 2024ZD0301000), and the National Natural Science Foundation of China (Grant No. 12275136). This work is also supported by the Key Research and Development Support Program of the Chengdu Municipal Science and Technology Bureau through the project ``Research, Development, and Application Demonstration Platform for an Independently Controllable Room-Temperature Quantum--Supercomputing Integrated System'' (Grant No. 2025-XT00-00012-GX). KLC acknowledges funding from the National Research Foundation, Singapore and the Ministry of Education, Singapore.
\end{acknowledgments}

\interlinepenalty=10000

\end{document}

\title{Supplemental Material for ``Dephasing-Enhanced Response and Phase Transitions in Quantum Boltzmann Samplers''}
\author{Yu-Xuan Zhang}
\affiliation{School of Physics, Nankai University, Tianjin 300071, People's Republic of China}
\affiliation{Centre for Quantum Technologies, National University of Singapore, 3 Science Drive 2, Singapore 117543, Singapore}

\author{Jing-Ling Chen}
\email{chenjl@nankai.edu.cn}
\affiliation{Theoretical Physics Division, Chern Institute of Mathematics, Nankai University, Tianjin 300071, People's Republic of China}

\author{Leong-Chuan Kwek}
\email{kwekleongchuan@nus.edu.sg}
\affiliation{Centre for Quantum Technologies, National University of Singapore, 3 Science Drive 2, Singapore 117543, Singapore}
\affiliation{National Institute of Education, Nanyang Technological University, 1 Nanyang Walk, Singapore 637616, Singapore}

\author{Peng Wang}
\email{wp002005@163.com}
\affiliation{Centre for Quantum Technologies, National University of Singapore, 3 Science Drive 2, Singapore 117543, Singapore}
\affiliation{School of Computer and Artificial Intelligence, Southwest Minzu University, Chengdu 610225, China}

\date{September 13, 2026}

\maketitle

\section{Local sampling dynamics and the network master equation}
\label{sec:system}

Boltzmann-machine training requires samples from the model at its current parameters. For $N$ binary variables $z_i=\pm1$, the programmed fields $h_i$ and symmetric couplings $J_{ij}=J_{ji}$ define the target distribution and its positive-amplitude encoding,
\begin{equation}
 \begin{aligned}
 p_\theta(z)&=\frac1{Z_\theta}\exp\!\left(\sum_i h_i z_i+\sum_{i<j}J_{ij}z_i z_j\right),\\
 \ket{\Psi_\theta}&=\sum_z\sqrt{p_\theta(z)}\ket z.
 \end{aligned}
 \label{eq:S-target}
\end{equation}
The parameters are real and finite, $J_{ii}=0$, and $Z_\theta$ normalizes the probabilities. Each variable is represented by a qubit with Pauli operators $X_i,Y_i,Z_i$; the computational states $\ket0_i,\ket1_i$ have $Z_i$ eigenvalues $1,-1$. Measuring $\ket{\Psi_\theta}$ in this basis samples $p_\theta$. We construct dissipative training toward this encoding using conditional dark states~\cite{Somma2007,Kraus2008,Fiorelli2019}.

To determine what an update at site $i$ should do, group the amplitudes by the configuration $\zeta=(z_j)_{j\ne i}$ of the other spins. Write $p_\theta(\zeta)=p_\theta(\zeta,1)+p_\theta(\zeta,-1)$ for its marginal probability and $p_\theta(z_i\mid\zeta)=p_\theta(\zeta,z_i)/p_\theta(\zeta)$ for the conditional probability. The corresponding conditional qubit $\ket{\nu_i(\zeta)}$ appears directly in the tensor decomposition:
\[
 \begin{aligned}
 \ket{\Psi_\theta}
 &=\sum_\zeta\ket\zeta_{\ne i}\otimes
   \left[\sqrt{p_\theta(\zeta,1)}\ket0_i+\sqrt{p_\theta(\zeta,-1)}\ket1_i\right]\\
 &=\sum_\zeta\sqrt{p_\theta(\zeta)}\,\ket\zeta_{\ne i}\otimes
   \left[\sqrt{p_\theta(1\mid\zeta)}\ket0_i+
         \sqrt{p_\theta(-1\mid\zeta)}\ket1_i\right]\\
 &=\sum_\zeta\sqrt{p_\theta(\zeta)}\,\ket\zeta_{\ne i}\otimes\ket{\nu_i(\zeta)}.
 \end{aligned}
\]
We build the local update from this conditional target. Its amplitudes follow from Eq.~\eqref{eq:S-target}: the terms containing $z_i$ combine into $z_i a_i(\zeta)$, with
\[
 a_i(\zeta)=h_i+\sum_{j\ne i}J_{ij}z_j.
\]
All other terms are common to the two values of $z_i$ and cancel when conditioning. With $u_i(\zeta)=\tanh a_i(\zeta)$,
\[
 \begin{aligned}
 p_\theta(z_i\mid\zeta)
 &=\frac{p_\theta(\zeta,z_i)}{p_\theta(\zeta,1)+p_\theta(\zeta,-1)}\\
 &=\frac{e^{z_i a_i(\zeta)}}{e^{a_i(\zeta)}+e^{-a_i(\zeta)}}
 =\frac{1+z_i u_i(\zeta)}2.
 \end{aligned}
\]
The conditional target in the tensor decomposition is therefore
\begin{equation}
 \ket{\nu_i(\zeta)}=
 \sqrt{\frac{1+u_i(\zeta)}2}\ket0_i+
 \sqrt{\frac{1-u_i(\zeta)}2}\ket1_i.
 \label{eq:S-states}
\end{equation}

We first construct the update at a fixed field $a=a_i(\zeta)$, writing $u=\tanh a$, $q=\sech a=\sqrt{1-u^2}$, and suppressing the site and configuration labels. Choose the normalized orthogonal state
\[
 \ket{\nu^\perp}=
 \sqrt{\frac{1-u}2}\ket0-\sqrt{\frac{1+u}2}\ket1.
\]
The target and orthogonal projections are
\[
 P=\ket\nu\bra\nu
 =\frac12\begin{pmatrix}1+u&q\\q&1-u\end{pmatrix}
 =\frac12(\openone+qX+uZ),
 \qquad Q=\ket{\nu^\perp}\bra{\nu^\perp}=\openone-P.
\]
We keep the target component and retain a real amplitude $0\le s<1$ in the orthogonal direction. This specifies the first Kraus operator:
\[
 A_0=A_0(P+Q)=P+sQ.
\]
Trace preservation determines the contribution of the second Kraus operator. Since $PQ=0$, $A_1$ must satisfy
\[
 \begin{aligned}
 A_1^\dagger A_1
 &=\openone-A_0^\dagger A_0\\
 &=\openone-(P+sQ)^2\\
 &=(1-s^2)Q.
 \end{aligned}
\]
Consequently, $A_1\ket\nu=0$ and $\|A_1\ket{\nu^\perp}\|^2=1-s^2$. We choose this branch to return the orthogonal component to the conditional target $\ket\nu$. The standard sampler therefore has Kraus operators and channel
\begin{equation}
 \begin{aligned}
 A_0&=P+sQ,\qquad A_1=\sqrt{1-s^2}\ket\nu\bra{\nu^\perp},\\
 \Phi(\rho)&=\sum_{\mu=0,1}A_\mu\rho A_\mu^\dagger.
 \end{aligned}
 \label{eq:S-kraus}
\end{equation}
Here $\rho$ is a single-qubit density matrix. The target is dark, $A_0\ket\nu=\ket\nu$ and $A_1\ket\nu=0$, while the $A_1$ branch recycles the orthogonal, bright component into the target. The same Kraus effects and dark-state relations hold if this branch emits any normalized qubit state; the standard choice in Eq.~\eqref{eq:S-kraus} returns it to $\ket\nu$.

The population remaining in the bright direction after applying the channel is
\[
 \begin{aligned}
 \Tr[Q\Phi(\rho)]
 &=\Tr\!\left[(A_0^\dagger QA_0+A_1^\dagger QA_1)\rho\right]\\
 &=s^2\Tr(Q\rho).
 \end{aligned}
\]
Since the trace is preserved, the population in $P$ increases by $(1-s^2)\Tr(Q\rho)$. This transfer toward the conditional target is the local dissipative training step.

We describe the continuous relaxation by a Lindblad generator. For an operator $A$, define the dissipator and anticommutator by
\[
 \begin{aligned}
 \D[A](\rho)
 &=A\rho A^\dagger-\frac12\{A^\dagger A,\rho\}\\
 &=A\rho A^\dagger-\frac12A^\dagger A\rho-\frac12\rho A^\dagger A,
 \qquad \{A,B\}=AB+BA.
 \end{aligned}
\]
Since $\sum_\mu A_\mu^\dagger A_\mu=\openone$, the dissipators sum to the channel minus the identity map $\id$:
\begin{equation}
 \begin{aligned}
 \sum_{\mu=0,1}\D[A_\mu](\rho)
 &=\sum_{\mu=0,1}A_\mu\rho A_\mu^\dagger
   -\frac12\left\{\sum_{\mu=0,1}A_\mu^\dagger A_\mu,\rho\right\}\\
 &=\Phi(\rho)-\frac12\{\openone,\rho\}\\
 &=\Phi(\rho)-\rho=(\Phi-\id)(\rho).
 \end{aligned}
 \label{eq:S-channel-generator}
\end{equation}
With training rate $\nu>0$, the local training generator is therefore $\nu(\Phi-\id)$. We add phase-flip noise with dephasing rate $\gamma\ge0$. Since $Z^2=\openone$, its dissipator acts as
\[
 \begin{aligned}
 \frac\gamma2\D[Z](\rho)
 &=\frac\gamma2\left(Z\rho Z-\frac12\{Z^2,\rho\}\right)
 =\frac\gamma2(Z\rho Z-\rho)\\
 &=\begin{pmatrix}0&-\gamma\rho_{01}\\-\gamma\rho_{10}&0\end{pmatrix}.
 \end{aligned}
\]
Thus this noise damps the computational-basis coherence while preserving the instantaneous populations. In the target basis, the same flip is
\[
 Z\ket\nu=u\ket\nu+q\ket{\nu^\perp}.
\]
Applied to the target, the flip puts population $q^2$ into the bright direction, which the training channel returns toward the target. The simultaneous evolution at fixed $a$ is
\begin{equation}
 \frac{d\rho}{dt}=\mathcal L_a(\rho)
 =\nu[\Phi(\rho)-\rho]+\frac\gamma2(Z\rho Z-\rho).
 \label{eq:S-local-master}
\end{equation}
The evolution is $\rho(t)=e^{t\mathcal L_a}\rho(0)$. Setting $\gamma=0$ gives training alone; we vary $\gamma$ at fixed $a,s,\nu$ to determine the noise response.

To determine that response, write the evolving state in terms of its computational-basis coherence and polarization,
\[
 \rho=\frac12(\openone+xX+yY+mZ)
 =\frac12\begin{pmatrix}1+m&x-iy\\x+iy&1-m\end{pmatrix},
 \qquad x=\Tr(X\rho),\quad y=\Tr(Y\rho),\quad m=\Tr(Z\rho).
\]
The overlap with $P$ fixes the weight entering the recycling branch:
\[
 \begin{aligned}
 \Tr(P\rho)&=\frac{1+qx+um}{2},
 &\Tr(Q\rho)&=1-\Tr(P\rho)=\frac{1-qx-um}{2},\\
 P\rho P&=\Tr(P\rho)P,
 &A_1\rho A_1^\dagger&=(1-s^2)\Tr(Q\rho)P.
 \end{aligned}
\]
Also, $A_0=s\openone+(1-s)P$. Multiplying its two factors and adding the recycling contribution gives
\[
 \begin{aligned}
 \Phi(\rho)
 &=[s\openone+(1-s)P]\rho[s\openone+(1-s)P]
   +(1-s^2)\Tr(Q\rho)P\\
 &=s^2\rho+s(1-s)(P\rho+\rho P)
   +\bigl[(1-s)^2\Tr(P\rho)+(1-s^2)\Tr(Q\rho)\bigr]P\\
 &=s^2\rho+s(1-s)(P\rho+\rho P)
   +(1-s)[1-s(qx+um)]P.
 \end{aligned}
\]
The anticommutator in this expression follows from the Pauli products:
\[
 \begin{aligned}
 P\rho+\rho P
 &=\frac14\bigl\{\openone+qX+uZ,\openone+xX+yY+mZ\bigr\}\\
 &=\frac12\bigl[(1+qx+um)\openone+(x+q)X+yY+(m+u)Z\bigr].
 \end{aligned}
\]
Substitution therefore puts the channel action in the same coordinates as $\rho$:
\begin{equation}
 \begin{aligned}
 \Phi(\rho)-\rho
 &=\frac{1-s}{2}\Bigl\{-xX-yY-mZ
   +[1+s-s(qx+um)](qX+uZ)\Bigr\}.
 \end{aligned}
 \label{eq:S-local-channel}
\end{equation}
The phase-flip contribution is equally explicit:
\[
 Z\rho Z=\frac12(\openone-xX-yY+mZ),
 \qquad
 \frac\gamma2(Z\rho Z-\rho)=-\frac\gamma2(xX+yY).
\]
The bright population relaxes at rate $\nu(1-s^2)$ under training alone. We express the dephasing strength relative to this rate and absorb the common factor $\nu(1-s)$ into time:
\begin{equation}
 r=\frac\gamma{\nu(1-s^2)},\qquad
 \tau=\nu(1-s)t,\qquad
 \frac\gamma{\nu(1-s)}=(1+s)r.
 \label{eq:S-time-units}
\end{equation}
Comparing the $X,Y,Z$ coefficients in Eq.~\eqref{eq:S-local-master} now gives
\begin{equation}
 \begin{aligned}
 \frac{dx}{d\tau}&=(1+s)q-[1+sq^2+(1+s)r]x-squ\,m,\\
 \frac{dy}{d\tau}&=-[1+(1+s)r]y,\\
 \frac{dm}{d\tau}&=(1+s)u-squ\,x-(1+su^2)m.
 \end{aligned}
 \label{eq:S-clamped-flow}
\end{equation}
Dephasing damps $x$ and $y$ directly, and the term $-squ\,x$ transmits the change in coherence to the polarization. Setting these derivatives to zero determines the stationary components $x_*(a),y_*(a),m_*(a)$ at the fixed field.

The second equation gives $y(\tau)=y(0)e^{-[1+(1+s)r]\tau}$ and hence $y_*=0$. The remaining stationary equations are
\begin{equation}
 M\begin{pmatrix}x_*\\m_*\end{pmatrix}
 =(1+s)\begin{pmatrix}q\\u\end{pmatrix},\qquad
 M=\begin{pmatrix}
 1+sq^2+(1+s)r&squ\\
 squ&1+su^2
 \end{pmatrix}.
 \label{eq:S-stationarymatrix}
\end{equation}
Using $q^2+u^2=1$, the determinant reduces to
\[
 \begin{aligned}
 \det M
 &=[1+sq^2+(1+s)r](1+su^2)-s^2q^2u^2\\
 &=1+s(q^2+u^2)+(1+s)r(1+su^2)\\
 &=(1+s)(1+r+sru^2)>0.
 \end{aligned}
\]
The two numerators simplify by the same identity:
\[
 \begin{aligned}
 (\det M)x_*
 &=(1+s)\bigl[q(1+su^2)-u(squ)\bigr]
 =(1+s)q,\\
 (\det M)m_*
 &=(1+s)\bigl[u(1+sq^2+(1+s)r)-q(squ)\bigr]
 =(1+s)[1+(1+s)r]u.
 \end{aligned}
\]
Thus the conditional stationary state and its components are
\begin{equation}
 \begin{aligned}
 \rho_*(a)&=\frac12\bigl[\openone+x_*(a)X+m_*(a)Z\bigr],\\
 x_*(a)&=\frac{q}{1+r+sru^2},\qquad
 y_*(a)=0,\qquad
 m_*(a)=\frac{[1+(1+s)r]u}{1+r+sru^2}.
 \end{aligned}
 \label{eq:S-response}
\end{equation}
The matrix $M$ also gives convergence to this state. Indeed,
\[
 \begin{aligned}
 M
 &=\openone_2+s\begin{pmatrix}q\\u\end{pmatrix}
                \begin{pmatrix}q&u\end{pmatrix}
   +(1+s)r\begin{pmatrix}1&0\\0&0\end{pmatrix}
 \ge\openone_2,\\
 \begin{pmatrix}x(\tau)-x_*\\m(\tau)-m_*\end{pmatrix}
 &=e^{-M\tau}\begin{pmatrix}x(0)-x_*\\m(0)-m_*\end{pmatrix}.
 \end{aligned}
\]
Together with the decay of $y$, this shows that every initial state of the qubit approaches $\rho_*(a)$ at fixed $a$.

Computational-basis measurement of this state gives probabilities $(1\pm m_*)/2$, whereas the target conditional probabilities are $(1\pm u)/2$. Dividing the numerator and denominator of $m_*$ by $1+r$ makes their difference explicit:
\begin{equation}
 \begin{aligned}
 m_*(a)
 &=\frac{(1+\alpha)u}{1+\alpha u^2},
 \qquad \alpha=\frac{sr}{1+r},\\
 m_*(a)-u
 &=\frac{(1+\alpha)u-u(1+\alpha u^2)}{1+\alpha u^2}
 =\frac{\alpha u(1-u^2)}{1+\alpha u^2}.
 \end{aligned}
 \label{eq:S-response-shift}
\end{equation}
At $r=0$, Eq.~\eqref{eq:S-response} gives $x_*=q$ and $m_*=u$, so the stationary state is the pure conditional target. For $0<s<1$ and any nonzero finite field, increasing $r$ suppresses $x_*$ and increases the magnitude of $m_*$. At full reset, $s=0$, the longitudinal equation reduces to $dm/d\tau=u-m$, and the stationary polarization remains $u$ at every noise strength.

The small-field gain follows from the same response function. Since $du/da=1-u^2$,
\[
 \frac{dm_*}{da}
 =\frac{(1+\alpha)(1-\alpha u^2)(1-u^2)}{(1+\alpha u^2)^2}.
\]
At zero field this gives
\begin{equation}
 \begin{aligned}
 x_*(0)&=\frac1{1+r},\\
 g(r)=\left.\frac{dm_*}{da}\right|_{a=0}
 &=1+\alpha
 =1+\frac{sr}{1+r}
 =1+s[1-x_*(0)].
 \end{aligned}
 \label{eq:S-gain}
\end{equation}
Thus the zero-field coherence loss gives the increase in local gain. We now restore the surrounding spins in the channel, so that their configuration supplies the field during the joint evolution.

Let $\rho_N$ be the network density matrix. The tensor decomposition of $\ket{\Psi_\theta}$ at the start of this section specifies how to embed the local operators: on each background branch $\ket\zeta_{\ne i}$, apply $A_\mu(a_i(\zeta))$ to neuron $i$. Thus
\[
 \begin{aligned}
 A_{i\mu}&=\sum_\zeta\bigl(\ket\zeta\bra\zeta\bigr)_{\ne i}\otimes
                  A_\mu\bigl(a_i(\zeta)\bigr),\\
 \Phi_i(\rho_N)&=\sum_{\mu=0,1}A_{i\mu}\rho_NA_{i\mu}^\dagger.
 \end{aligned}
\]
The same embedding sends the local phase flip $Z$ to $Z_i$. For an arbitrary initial network density matrix $\rho_{\rm in}$, summing the complete site contributions in Eq.~\eqref{eq:S-local-master} gives
\begin{equation}
 \begin{aligned}
 \frac{d\rho_N}{dt}&=\mathcal L_N(\rho_N),\qquad
 \rho_N(0)=\rho_{\rm in},\\
 \mathcal L_N
 &=\sum_i\left[\nu\sum_{\mu=0,1}\D[A_{i\mu}]
                  +\frac\gamma2\D[Z_i]\right]\\
 &=\nu\sum_i(\Phi_i-\id)+\frac\gamma2\sum_i\D[Z_i].
 \end{aligned}
 \label{eq:S-L}
\end{equation}
Every site contribution acts on the same network density matrix, with evolution $\rho_N(t)=e^{t\mathcal L_N}\rho_{\rm in}$. The programmed parameters and training channel remain fixed as $\gamma$ is varied.

The conditional fields can also be written as diagonal network operators,
\[
 a_i=h_i\openone+\sum_{j\ne i}J_{ij}Z_j,
 \qquad u_i=\tanh a_i,\qquad q_i=\sech a_i.
\]
The products below include the implicit identity factors on all other sites. The controlled target projection then has the same form as its local counterpart:
\[
 \begin{aligned}
 P_i&=\sum_\zeta\bigl(\ket\zeta\bra\zeta\bigr)_{\ne i}\otimes P(a_i(\zeta))\\
 &=\frac12(\openone+q_iX_i+u_iZ_i),\\
 Q_i&=\openone-P_i,\qquad A_{i0}=P_i+sQ_i.
 \end{aligned}
\]
Orthogonality of the control configurations lifts local trace preservation to the network:
\[
 \begin{aligned}
 \sum_{\mu=0,1}A_{i\mu}^\dagger A_{i\mu}
 &=\sum_\zeta\bigl(\ket\zeta\bra\zeta\bigr)_{\ne i}\otimes
   \left[\sum_{\mu=0,1}A_\mu(a_i(\zeta))^\dagger A_\mu(a_i(\zeta))\right]\\
 &=\sum_\zeta\bigl(\ket\zeta\bra\zeta\bigr)_{\ne i}\otimes\openone_i
 =\openone.
 \end{aligned}
\]
At zero dephasing, the target state is preserved. Applying the controlled Kraus operators to its tensor decomposition gives
\[
 \begin{aligned}
 A_{i0}\ket{\Psi_\theta}
 &=\sum_\zeta\sqrt{p_\theta(\zeta)}\ket\zeta_{\ne i}\otimes
   A_0(a_i(\zeta))\ket{\nu_i(\zeta)}
 =\ket{\Psi_\theta},\\
 A_{i1}\ket{\Psi_\theta}
 &=\sum_\zeta\sqrt{p_\theta(\zeta)}\ket\zeta_{\ne i}\otimes
   A_1(a_i(\zeta))\ket{\nu_i(\zeta)}
 =0.
 \end{aligned}
\]
The two Kraus relations imply
\[
 \Phi_i(\ket{\Psi_\theta}\bra{\Psi_\theta})
 =\ket{\Psi_\theta}\bra{\Psi_\theta},
 \qquad
 \left.\mathcal L_N(\ket{\Psi_\theta}\bra{\Psi_\theta})\right|_{\gamma=0}=0.
\]
Thus the coherent Boltzmann encoding is a stationary state of Eq.~\eqref{eq:S-L} at $\gamma=0$.

In the time $\tau$ of Eq.~\eqref{eq:S-time-units}, we extract the sampling probabilities $p_N(z,\tau)=\bra z\rho_N(\tau)\ket z$ by taking the diagonal of Eq.~\eqref{eq:S-L}. For the contribution of site $i$, the same background configuration $\zeta$ occurs on both sides. Its qubit block is
\[
 \rho_\zeta^{(i)}={}_{\ne i}\!\bra\zeta\rho_N\ket\zeta_{\ne i}
 =\begin{pmatrix}p_+&c\\c^*&p_-\end{pmatrix},
 \qquad
 p_\pm=p_N(\zeta,\pm1),\quad
 c=\langle\zeta,1|\rho_N|\zeta,-1\rangle.
\]
The trace $p_++p_-$ is the probability of $\zeta$. Let $\Phi_i(\zeta)$ denote the local channel at the field $a_i(\zeta)$. The controlled Kraus operators give its action on this block directly:
\[
 \begin{aligned}
 {}_{\ne i}\!\bra\zeta\Phi_i(\rho_N)\ket\zeta_{\ne i}
 &=\sum_{\mu=0,1}A_\mu(a_i(\zeta))\rho_\zeta^{(i)}A_\mu(a_i(\zeta))^\dagger\\
 &=\Phi_i(\zeta)(\rho_\zeta^{(i)}).
 \end{aligned}
\]
We can therefore reuse Eq.~\eqref{eq:S-local-channel}. Linearity extends that calculation to this unnormalized block through the replacements
\[
 \begin{aligned}
 1&\longmapsto\Tr\rho_\zeta^{(i)}=p_++p_-,\\
 m&\longmapsto\Tr(Z\rho_\zeta^{(i)})=p_+-p_-,\\
 x&\longmapsto\Tr(X\rho_\zeta^{(i)})=2\operatorname{Re}c.
 \end{aligned}
\]
Suppressing the argument $\zeta$ in $u_i,q_i$, the diagonal contribution at $z=(\zeta,z_i)$ is
\[
 \begin{aligned}
 \frac{\bra z[\Phi_i(\rho_N)-\rho_N]\ket z}{1-s}
 &=\frac{z_i}{2}\left[(1+s)u_i(p_++p_-)
             -(1+su_i^2)(p_+-p_-)-2su_iq_i\operatorname{Re}c\right]\\
 &=\frac{1+su_i^2+(1+s)z_i u_i}{2}\,p_N(\zeta,-z_i)
       -\frac{1+su_i^2-(1+s)z_i u_i}{2}\,p_N(\zeta,z_i)
       -sz_i u_iq_i\operatorname{Re}c.
 \end{aligned}
\]
Write $a_i(z)=a_i(z_{\ne i})$ for the scalar field at a full configuration $z$, with $u_i(z)=\tanh a_i(z)$ and $q_i(z)=\sech a_i(z)$, and let $z^i$ denote $z$ with its $i$th spin reversed. The coefficient of the outgoing population defines the flip rate
\begin{equation}
 \begin{aligned}
 w_i(z)
 &=\frac12[1+su_i(z)^2-(1+s)z_i u_i(z)]\\
 &=\frac12[1-z_i u_i(z)][1-sz_i u_i(z)].
 \end{aligned}
 \label{eq:S-population-rate}
\end{equation}
The field excludes the updated spin, so $u_i(z^i)=u_i(z)$ and $q_i(z^i)=q_i(z)$. Every phase-flip dissipator has zero diagonal. Summing the site contributions consequently gives the exact probability equation
\begin{equation}
 \partial_\tau p_N(z)
 =\sum_i\Bigl[w_i(z^i)p_N(z^i)-w_i(z)p_N(z)
       -s z_i u_i(z)q_i(z)\operatorname{Re}\langle z|\rho_N|z^i\rangle\Bigr].
 \label{eq:S-network-population}
\end{equation}
For each site, the first two terms transfer probability between configurations differing at that spin. The last term couples this transfer to the coherence between those configurations. Its two contributions on an edge $z\leftrightarrow z^i$ have opposite signs, so the equation preserves total probability. Dephasing changes the coherence in this equation, and the training channel converts that change into a population response.

The probability equation therefore requires the coherence between neighboring configurations. In Sec.~\ref{sec:collective}, we take its magnetization moment and obtain the required coherence evolution from the same network master equation.

\section{Collective dynamics and stationary order}
\label{sec:collective}
\label{sec:finite}

We use the network equation of Sec.~\ref{sec:system} to determine the magnetization distribution sampled at stationarity. Take zero field, Curie--Weiss couplings $J_{ij}=K/N$ for $i\ne j$, standard recycling, uniform $0<s<1$ and $r>0$, and $0<K<1$, where the programmed Boltzmann model is disordered. The measured magnetization and the conditional field obey
\[
 M_z=\frac1N\sum_iZ_i,\qquad
 a_i=K\left(M_z-\frac{Z_i}{N}\right),\qquad
 u_i=\tanh a_i,\quad q_i=\sech a_i.
\]
For a stationary network density $\rho_N$, its magnetization law $P_{N,r}$ is determined by
\begin{equation}
 \int f(m)\,dP_{N,r}(m)=\Tr[\rho_Nf(M_z)].
 \label{eq:S-magnetization-law}
\end{equation}
We first calculate how these magnetization functions evolve; their drift locates the possible ordered states, and their stationary equation determines the weights of those states. Retain $\tau=\nu(1-s)t$, put $\mathcal L_{\tau,N}=\mathcal L_N/[\nu(1-s)]$, and write $\langle O\rangle_\tau=\Tr[\rho_N(\tau)O]$. Moving the Kraus operators across the trace defines $\Phi_i^\dagger(O)=\sum_\mu A_{i\mu}^\dagger O A_{i\mu}$.

At fixed background, $M_z=a_i/K+Z_i/N$ has two values. The coefficient of $Z_i$ in $f(M_z)$ is their half difference,
\[
 b_{i,f}=\frac12\left[f\left(\frac{a_i}{K}+\frac1N\right)-f\left(\frac{a_i}{K}-\frac1N\right)\right].
\]
The remaining part depends only on the background and is preserved by completeness. Reusing the local $Z$ response from Sec.~\ref{sec:system} therefore gives
\begin{equation}
 \begin{aligned}
 \Phi_i^\dagger f(M_z)-f(M_z)
 &=b_{i,f}[\Phi_i^\dagger(Z_i)-Z_i]\\
 &=(1-s)b_{i,f}[(1+s)u_i-(1+su_i^2)Z_i-su_iq_iX_i].
 \end{aligned}
 \label{eq:S-diagonal-action}
\end{equation}
Dephasing commutes with $f(M_z)$, so summing the full network generator gives the exact finite-difference action
\begin{equation}
 \mathcal L_{\tau,N}^\dagger f(M_z)
 =\sum_i b_{i,f}[(1+s)u_i-(1+su_i^2)Z_i-su_iq_iX_i].
 \label{eq:S-exact-difference}
\end{equation}
In particular, $f(m)=m$ has $b_{i,f}=1/N$. Its expectation $m_N=\langle M_z\rangle_\tau$ therefore obeys
\begin{equation}
 \frac{dm_N}{d\tau}=\frac1N\sum_i
 \langle(1+s)u_i-(1+su_i^2)Z_i-su_iq_iX_i\rangle_\tau.
 \label{eq:S-exact-magnetization}
\end{equation}
The coherence in this equation requires the transverse spin averages $M_x=N^{-1}\sum_iX_i$ and $M_y=N^{-1}\sum_iY_i$, with means $x_N,y_N$. The same master equation gives
\[
 \frac{dx_N}{d\tau}
 =\frac1{N(1-s)}\sum_{i,j}\langle\Phi_j^\dagger(X_i)-X_i\rangle_\tau-(1+s)r x_N.
\]
The term $j=i$ is the local response. For $j\ne i$, neuron $i$ controls the update at $j$: writing $a_j=a_{ij}+(K/N)Z_i$, where $a_{ij}=K N^{-1}\sum_{\ell\ne i,j}Z_\ell$, shows that $X_i,Y_i$ connect the fields $a_{ij}\pm K/N$. Their contribution follows by expanding the corresponding Kraus overlap. At a local field $a$, the Pauli forms are
\[
 A_0(a)=\frac{1+s}{2}\openone+\frac{1-s}{2}(qX+uZ),\qquad
 A_1(a)=\frac{\sqrt{1-s^2}}2(qZ-uX-\mathrm iY),
 \quad u=\tanh a,\quad q=\sech a.
\]
Using $u'=q^2$, $q'=-qu$, and completeness gives
\begin{equation}
 \begin{gathered}
 \sum_\mu(A_\mu'{}^\dagger A_\mu-A_\mu^\dagger A_\mu')
 =\frac{\mathrm i q}{2}[(1-s)^2+(1-s^2)]Y=\mathrm i(1-s)qY,\\
 \sum_\mu A_\mu(a_{ij}+K/N)^\dagger A_\mu(a_{ij}-K/N)
 =\openone+\mathrm i\frac KN(1-s)q_{ij}Y_j+O(N^{-2}),
 \quad q_{ij}=\sech a_{ij}.
 \end{gathered}
 \label{eq:S-overlap}
\end{equation}
This overlap and its adjoint are the two off-diagonal blocks in the $Z_i$ basis. Consequently,
\[
 \begin{aligned}
 \frac{\Phi_j^\dagger(X_i)-X_i}{1-s}&=-\frac KNq_{ij}Y_iY_j+O(N^{-2}),\\
 \frac{\Phi_j^\dagger(Y_i)-Y_i}{1-s}&=\frac KNq_{ij}X_iY_j+O(N^{-2}).
 \end{aligned}
\]
Each control contribution is $O(N^{-1})$, and the sum over the other neurons remains finite. The local terms are those of Sec.~\ref{sec:system}; phase flips add $-(1+s)r x_N$ and $-(1+s)r y_N$. All field spectra lie in $[-K,K]$, so averaging the summed overlap remainders leaves $O(N^{-1})$ errors uniformly in $N$.

For a product preparation with single-site density $\rho(\vec v)=(\openone+xX+yY+mZ)/2$, $\vec v=(x,y,m)$, the conditional field has mean $K(N-1)m/N$ and variance $K^2(N-1)(1-m^2)/N^2$. It therefore concentrates at $Km$. Since $a_i$ excludes $i$ and $a_{ij}$ excludes both marked sites, the local terms factor in this component and the two control sums tend to $qy^2$ and $qxy$. With $u=\tanh(Km)$ and $q=\sech(Km)$, the resulting collective motion is
\begin{equation}
 \begin{aligned}
 \frac{d\vec v}{d\tau}&=\vec F(\vec v),\\
 F_x&=(1+s)q-[1+sq^2+(1+s)r]x-squm-Kqy^2,\\
 F_y&=[-1-(1+s)r+Kqx]y,\\
 F_z&=(1+s)u-(1+su^2)m-suqx.
 \end{aligned}
 \label{eq:S-mf}
\end{equation}
Each component of a mixture supplies its own magnetization and hence its own field. To follow a concentrated preparation as correlations develop, use its mean-square displacement $E_N(\tau)=\sum_{\alpha=x,y,z}\langle[M_\alpha-v_\alpha(\tau)]^2\rangle_\tau$ from this trajectory.
Excluded-spin shifts and the overlap remainders are $O(N^{-1})$; ordering collective products has the same cost because $\|[M_\alpha,M_\beta]\|\le2/N$. Thus, with symmetrized products in $F_\alpha(\vec M)$,
\begin{equation}
 \|\mathcal L_{\tau,N}^\dagger M_\alpha-F_\alpha(\vec M)\|\le C/N.
 \label{eq:S-driftbound}
\end{equation}
To control the fluctuations generated by the updates, use the product identity for a dissipative generator $\mathcal L=\sum_\ell\D[L_\ell]$:
\begin{equation}
 \begin{aligned}
 \Gamma_{\mathcal L}(O)
 &:=\mathcal L^\dagger(O^\dagger O)
 -(\mathcal L^\dagger O^\dagger)O-O^\dagger\mathcal L^\dagger O\\
 &=\sum_\ell[L_\ell,O]^\dagger[L_\ell,O].
 \end{aligned}
 \label{eq:S-product}
\end{equation}
Each commutator with a spin average is $O(N^{-1})$, including the conditional controls. Summing its square over sites gives
\begin{equation}
 \begin{aligned}
 \Gamma_{\mathcal L_{\tau,N}}(M_\alpha)
 &=\frac1{1-s}\sum_{i,\mu}[A_{i\mu},M_\alpha]^\dagger[A_{i\mu},M_\alpha]
 +\frac{(1+s)r}{2}\sum_i[Z_i,M_\alpha]^\dagger[Z_i,M_\alpha],\\
 \|\Gamma_{\mathcal L_{\tau,N}}(M_\alpha)\|&\le C'/N.
 \end{aligned}
 \label{eq:S-varianceproduction}
\end{equation}
The smooth drift and Cauchy--Schwarz now bound the derivative of $E_N$~\cite{Fiorelli2023,Carollo2024}:
\begin{equation}
 \dot E_N\le C_1E_N+C_2/N,\qquad
 E_N(\tau)\le e^{C_1\tau}E_N(0)+\frac{C_2}{C_1N}(e^{C_1\tau}-1).
 \label{eq:S-gronwall}
\end{equation}
The constants depend on fixed $K,s,r$. A product preparation has $E_N(0)=[3-|\vec v(0)|^2]/N$; every sequence with $E_N(0)\to0$ follows Eq.~\eqref{eq:S-mf} on fixed time intervals. The motion remains in the Bloch ball, since at $|\vec v|=1$,
\[
 \vec v\cdot\vec F
 =-[1-qx-um][1-s(qx+um)]-(1+s)r(x^2+y^2)\le0.
\]

We use this motion to locate the possible stationary magnetizations. Since $qx\le1$,
\[
 \frac{d}{d\tau}y^2\le-2[1-K+(1+s)r]y^2,
\]
so stationary motion lies on $y=0$. On this plane, the coefficient of $-x$ is $D(m)=(1+s)(1+r)-su^2>0$. Solving the remaining transverse equation determines its stationary value:
\begin{equation}
 \begin{gathered}
 F_x(x,0,m)=q[(1+s)-sum]-D(m)x
            =-D(m)[x-\bar x(m)],\\
 \bar x(m)=\frac{q[(1+s)-sum]}{D(m)}.
 \end{gathered}
 \label{eq:S-nullcline}
\end{equation}
Substituting this value in the magnetization equation gives
\begin{equation}
 F_{\rm eff}(m):=F_z(\bar x(m),0,m)
 =\frac{(1+s)(1+r+sru^2)}{D(m)}[m_*(Km)-m].
 \label{eq:S-effective-drift}
\end{equation}
Its prefactor is positive. The stationary magnetization must therefore reproduce the single-neuron response to its own field. With $\alpha=sr/(1+r)$ as in Sec.~\ref{sec:system},
\begin{equation}
 m=m_*(Km)=\frac{(1+\alpha)\tanh(Km)}{1+\alpha\tanh^2(Km)},
 \qquad x=\bar x(m)=x_*(Km).
 \label{eq:S-fixed}
\end{equation}
At the unpolarized solution $x=x_0=(1+r)^{-1}$, $y=m=0$, the three linear directions decouple:
\begin{equation}
 \begin{aligned}
 \lambda_x&=-(1+s)(1+r),\qquad
 \lambda_y=-1-(1+s)r+Kx_0,\\
 \lambda_m&=K[(1+s)-sx_0]-1=g(r)K-1.
 \end{aligned}
 \label{eq:S-linear}
\end{equation}
The transverse rates stay negative for $K<1$, while the feedback through the local gain changes the sign of the magnetization rate at
\begin{equation}
 K_c(r)=\frac1{g(r)}=\frac{1+r}{1+(1+s)r}<1.
 \label{eq:S-threshold}
\end{equation}
The shape of the same response determines the new stationary roots. For a local input $a>0$, put $u=\tanh a$. Then
\begin{equation}
 m_*'(a)=\frac{(1+\alpha)(1-u^2)(1-\alpha u^2)}{(1+\alpha u^2)^2}.
 \label{eq:S-concavity}
\end{equation}
All three positive factors $(1-u^2)$, $(1-\alpha u^2)$, and $(1+\alpha u^2)^{-2}$ decrease with $u$, since $0<\alpha<1$. Thus $m_*''(a)<0$ for $a>0$. The curve $m_*(Km)$ starts at zero with slope $gK$, where $g=g(r)$, and lies below one at $m=1$. Strict concavity gives only the zero root for $K\le K_c$, and exactly one positive root $m_r$ for $K>K_c$; odd symmetry supplies $-m_r$. Its onset follows by expanding this response once:
\begin{equation}
 \begin{aligned}
 m_*(a)&=ga-g(\alpha+1/3)a^3+O(a^5),\\
 m_r^2&=\frac{gK-1}{g(\alpha+1/3)K^3}+O[(gK-1)^2].
 \end{aligned}
 \label{eq:S-pitchfork}
\end{equation}
At either ordered root, concavity gives $F_{\rm eff}'<0$. Differentiating the curve $F_x(\bar x(m),0,m)=0$ shows that the planar Jacobian has determinant $-D(m)F_{\rm eff}'(m)>0$. Its trace satisfies
\begin{equation}
 \begin{aligned}
 \partial_xF_x+\partial_mF_z
 &=(1+s)Kq^2-(1+s)(1+r)-1-2sKmuq^2-sKq(1-2u^2)x\\
 &\le(1+s)(K-1-r)-1+sK<0.
 \end{aligned}
 \label{eq:S-trace}
\end{equation}
Here $u=\tanh(Km)$ again; $mu\ge0$ and $|q(1-2u^2)x|\le1$ give the bound throughout the physical disk. The ordered roots attract nearby trajectories. This negative divergence also excludes planar periodic orbits; at $x\le0$, $F_x\ge q\ge\sech K>0$, so recurrent points are precisely the stationary roots.

To determine their sampled weights, take the stationary expectation of the finite-difference equation~\eqref{eq:S-exact-difference}. The finite dynamics first selects a unique stationary density. The following argument also covers the partially noisy populations used below. On a finite graph allow positive conditional training rates $\nu_i(a)$, $0\le s_i(a)<1$, normalized recycling outputs, and $\gamma_i\ge0$ with a noisy site in every connected component:
\begin{equation}
 \mathcal L=\sum_{i,\mu}\D[L_{i\mu}]+\frac12\sum_i\gamma_i\D[Z_i],\qquad
 L_{i\mu}=\sqrt{\nu_i(a_i)}A_{i\mu}.
 \label{eq:S-general-generator}
\end{equation}
Time averaging gives a stationary density $\rho_*$: $\bar\rho_T=T^{-1}\int_0^Te^{t\mathcal L}\rho_{\rm in}\,dt$ obeys $\|\mathcal L\bar\rho_T\|_1\le2/T$, where $\|\cdot\|_1$ is the trace norm; finite-dimensional compactness gives a limit point. If $\Pi_*$ projects onto its support, positivity and stationarity imply
\begin{equation}
 0=\Tr[(\openone-\Pi_*)\mathcal L\rho_*]
 =\sum_\ell\|(\openone-\Pi_*)L_\ell\rho_*^{1/2}\|_2^2,
 \label{eq:S-support}
\end{equation}
where the sum includes all training and noise jumps and $\|B\|_2^2=\Tr B^\dagger B$. Each summand vanishes, so every jump preserves the support. The Hermitian jumps $L_{i0}$ and the noise operators $Z_i$ therefore generate a unital algebra that preserves it as well. We recover the local spin operators from this algebra, starting at a noisy site.

Write its conditional field as $a_i=h_i\openone+\sum_{j\ne i}J_{ij}Z_j$. Conjugating $L_{i0}$ by the available $Z_i$ reverses only its $X_i$ part. Since $\nu_i>0$, $s_i<1$, and $q_i>0$, the resulting coefficient is positive. Inverse square roots on a finite positive spectrum are polynomials, so normalization isolates $X_i$ within the same algebra:
\[
 C_i=\frac{L_{i0}-Z_iL_{i0}Z_i}{2}
     =\frac{1-s_i}{2}\sqrt{\nu_i(a_i)}\,q_iX_i,
 \qquad X_i=C_i(C_i^2)^{-1/2}.
\]
Conjugation by the recovered $X_i$ now isolates the $Z_i$ part, and division by the same coefficient gives
\[
 (C_i^2)^{-1/2}\frac{L_{i0}-X_iL_{i0}X_i}{2}Z_i
 =\frac{u_i}{q_i}=\sinh a_i.
\]
Applying $\operatorname{arsinh}$ by the same finite-spectrum calculus recovers $a_i$. Its commutator with the training jump at a neighbor $j$ then isolates the next spin. For $J_{ij}\ne0$,
\[
 \begin{gathered}
 H_j=\frac{[a_i,L_{j0}]}{2\mathrm iJ_{ij}}
     =\frac{1-s_j}{2}\sqrt{\nu_j(a_j)}\,q_jY_j,
 \qquad Y_j=H_j(H_j^2)^{-1/2},\\
 X_j=-\frac{[a_i,Y_j]}{2\mathrm iJ_{ij}},\qquad
 Z_j=\frac{[X_j,Y_j]}{2\mathrm i}.
 \end{gathered}
\]
We can now repeat the local calculation at $j$ to recover $a_j$ and continue along its edges. A noisy site in every connected component thus generates every local Pauli operator, hence the full matrix algebra~\cite{ZhangBarthel2024}. It has no proper invariant subspace, so $\rho_*>0$.

To obtain relaxation, apply the product identity~\eqref{eq:S-product} to a nondecaying adjoint mode $\mathcal L^\dagger O=\mathrm i\omega O$. Stationarity gives
\[
 0=\Tr[\rho_*\Gamma_{\mathcal L}(O)]
   =\sum_\ell\|[L_\ell,O]\rho_*^{1/2}\|_2^2.
\]
Full rank makes every commutator vanish. Thus $O$ commutes with the full matrix algebra, so it is scalar and $\omega=0$. The bounded finite-dimensional semigroup has no imaginary-axis Jordan blocks. Its zero eigenspace is therefore one-dimensional, and every initial density converges to the same stationary density:
\begin{equation}
 \lim_{t\to\infty}e^{t\mathcal L}\rho_{\rm in}=\rho_*,\qquad \rho_*>0.
 \label{eq:S-finite-relaxation}
\end{equation}

Write $\langle O\rangle_*=\Tr(\rho_NO)$ for stationary expectations. For the uniform sampler, this unique density $\rho_N$ inherits site permutations and global spin inversion $\mathcal F=\bigotimes_iX_i$, giving $\Tr(\rho_NM_z)=0$. Its full distribution distinguishes a central peak from two oppositely polarized peaks. Take a subsequence on which every fixed-$k$ reduced density $\rho_N^{(k)}$ converges. The limits are consistent and permutation invariant, so the quantum de Finetti theorem gives~\cite{HudsonMoody1976}
\begin{equation}
 \rho^{(k)}=\int_{|\vec v|\le1}\rho(\vec v)^{\otimes k}\,d\mu(\vec v).
 \label{eq:S-definetti}
\end{equation}
In a fixed-degree collective moment, repeated site indices occupy only $O(N^{-1})$ of the sum; the distinct indices are determined by these reduced states. Its expectation therefore tends to the corresponding moment of $\mu$. The drift and product bounds similarly give, for a fixed collective polynomial $f$,
\begin{equation}
 \mathcal L_{\tau,N}^\dagger f(\vec M)
 =(\vec F\cdot\nabla f)(\vec M)+O(N^{-1}).
 \label{eq:S-polynomial-generator}
\end{equation}
Its stationary expectation yields $\int\vec F\cdot\nabla f\,d\mu=0$; smooth approximation proves that $\mu$ is invariant under the collective motion. The decay of $y^2$ and planar recurrence restrict its support to the fixed points found above. For $K\le K_c$ only $(x_0,0,0)$ remains. For $K>K_c$, the support can contain the center and the two ordered roots. Every such point satisfies $x=\bar x(m)$. The residual coherence $\delta X=M_x-\bar x(M_z)$ therefore has vanishing stationary mean square:
\begin{equation}
 e_N=\Tr[\rho_N(\delta X)^2]\longrightarrow
 \int[x-\bar x(m)]^2\,d\mu=0.
 \label{eq:S-slaving}
\end{equation}

It remains to determine the possible weight at the center. Above threshold, $F_{\rm eff}(m)=(gK-1)m+O(m^3)$ points outward. To retain its positive contribution near zero, the derivative of a magnetization test should behave as $1/m$. The finite difference also produces a second derivative at order $1/N$, so regularize on the scale $m^2\sim N^{-1}$ and integrate:
\begin{equation}
 f_N'(m)=\frac{m}{m^2+\eta/N},\qquad
 f_N(m)=\frac12\log(m^2+\eta/N),\qquad\eta>0.
 \label{eq:S-log-test}
\end{equation}
The half difference in Eq.~\eqref{eq:S-exact-difference} is
\[
 b_{i,f_N}=N^{-1}f_N'(M_z)-N^{-2}Z_if_N''(M_z)
 +O(N^{-3}\|f_N^{(3)}\|_\infty).
\]
With $u(m)=\tanh(Km)$ and $q(m)=\sech(Km)$, the second term gives the fluctuation coefficient $B(m)=1+su(m)^2-(1+s)u(m)m$. Pairing the two off-diagonal entries of the coherence term preserves its Hermitian form:
\begin{equation}
 \begin{aligned}
 \mathcal L_{\tau,N}^\dagger f_N(M_z)
 &=f_N'(M_z)[(1+s)u(M_z)-(1+su(M_z)^2)M_z]\\
 &\quad-\frac s2\{(f_N'uq)(M_z),M_x\}
 +\frac{B(M_z)}Nf_N''(M_z)+R_N.
 \end{aligned}
 \label{eq:S-log-adjoint}
\end{equation}
The remainder obeys $\|R_N\|\le C_\eta N^{-1/2}$: $\|f_N^{(3)}\|_\infty=O(N^{3/2})$ makes the summed Taylor remainder $O(N^{-1/2})$. The anticommutator averages $f_N'uq$ at $a_i/K\pm N^{-1}$; $\|(f_N'uq)''\|_\infty=O(N)$ makes that replacement error $O(N^{-1})$. Moreover $|f_N'uq|\le K$, so Cauchy--Schwarz bounds the effect of $\delta X$ in the stationary expectation by $sK\sqrt{e_N}\to0$. Substituting $M_x=\bar x(M_z)+\delta X$ now leaves an equation for the magnetization law alone:
\begin{equation}
 0=\int\left[
 \frac{F_{\rm eff}(m)}m\frac{Nm^2}{Nm^2+\eta}
 +B(m)\frac{\eta-Nm^2}{(Nm^2+\eta)^2}
 \right]dP_{N,r}(m)+o(1).
 \label{eq:S-repeller-test}
\end{equation}
The ratio $F_{\rm eff}(m)/m$ has continuous value $gK-1$ at zero. Choose a small central neighborhood where it is at least $(gK-1)/2$ and $B$ has positive upper and lower bounds. For $\eta>8\sup B/(gK-1)$, the integrand has a positive lower bound throughout this neighborhood: if $Nm^2\ge\eta$, the first term is at least $(gK-1)/4$ and dominates the negative part of the second; if $Nm^2\le\eta$, both terms are nonnegative and cannot vanish together. Outside the neighborhood the integrand converges uniformly to $F_{\rm eff}(m)/m$, which vanishes at the only remaining support points $\pm m_r$. Stationarity therefore removes all central weight. Spin inversion fixes equal weights at the ordered roots, and every subsequence has the same limit:
\begin{equation}
 P_{N,r}\Longrightarrow
 \begin{cases}
 \delta_0,&K\le K_c,\\
 \tfrac12\delta_{m_r}+\tfrac12\delta_{-m_r},&K_c<K<1.
 \end{cases}
 \label{eq:S-stationary-theorem}
\end{equation}
Thus the sampled first moment is zero while the second tends to $m_r^2$ in the ordered interval. The same support points determine all fixed-$k$ quantum marginals. Put $\rho_0=(\openone+x_0X)/2$ and $\rho_\pm=[\openone+\bar x(m_r)X\pm m_rZ]/2$ in Eq.~\eqref{eq:S-definetti} to obtain
\begin{equation}
 \rho_N^{(k)}\longrightarrow
 \begin{cases}
 \rho_0^{\otimes k},&K\le K_c,\\
 \tfrac12\rho_+^{\otimes k}+\tfrac12\rho_-^{\otimes k},&K_c<K<1.
 \end{cases}
 \label{eq:S-local-quantum-limit}
\end{equation}
These fixed-$k$ limits are separable, and their ordered components retain $\bar x(m_r)>0$ at finite noise.

\begin{figure}[htbp]
\centering
\resizebox{.94\linewidth}{!}{%
\begin{tikzpicture}
\path[use as bounding box] (0,0) rectangle (17.8,5.5);
\plabel{0}{5.4}{a}\plabel{9.0}{5.4}{b}
\begin{axis}[paper,at={(1.5cm,1.02cm)},anchor=south west,width=6.95cm,height=3.7cm,xmin=0,xmax=134,ymin=.325,ymax=.435,xlabel={Number of spins $N$},ylabel={$\langle M_z^2\rangle$},xtick={0,32,64,96,128},ytick={.34,.38,.42},legend columns=2,legend style={at={(.5,1.06)},anchor=south,column sep=14pt}]

\addplot[qgray,dashed] coordinates {(0,.400824)(134,.400824)};\addlegendentry{Stationary limit}
\addplot[qblue,only marks,mark=o,mark options={fill=white},mark size=2pt] table[x=N,y=m2]{\DataOrderedSizes};\addlegendentry{Finite $N$}

\end{axis}
\begin{axis}[paper,at={(10.5cm,1.02cm)},anchor=south west,width=6.95cm,height=3.7cm,xmode=log,ymode=log,xmin=10,xmax=145,ymin=.005,ymax=.11,xlabel={Number of spins $N$},ylabel={$\langle(\delta X)^2\rangle$},xtick={12,32,64,128},xticklabels={12,32,64,128},ytick={.01,.03,.1},yticklabels={0.01,0.03,0.1}]

\addplot[qgray,dashed,domain=10:140,samples=80] {.94/x};
\addplot[qblue,only marks,mark=o,mark options={fill=white},mark size=2pt] table[x=N,y=slaving_variance]{\DataSlaving};

\end{axis}
\end{tikzpicture}}
\caption{Two readouts of the finite quantum stationary state at $K=0.75$, $s=0.8$, and $r=3$. (a) The magnetization second moment from Eq.~\eqref{eq:S-observables}; the dashed line is the limit $m_r^2\simeq0.400824$ determined by Eq.~\eqref{eq:S-fixed}. (b) The transverse deviation $e_N=\langle[M_x-\bar x(M_z)]^2\rangle_*$ in the same states. The line $0.94/N$ is a reference for the displayed sizes.}
\label{fig:S-finite-memory}
\end{figure}
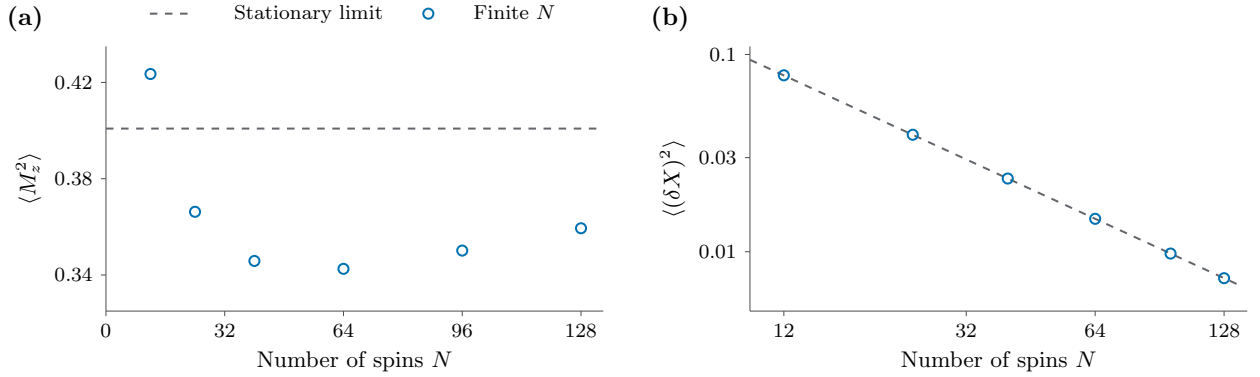

Figure~\ref{fig:S-finite-memory} evaluates this stationary distribution directly at finite $N$. To solve for the same density, collect computational matrix units $|\xi\rangle\langle\zeta|$ by their local ket--bra counts $\vec n=(n_{00},n_{01},n_{10},n_{11})$. Simultaneous permutations produce an orbit $\mathcal O_{\vec n}$ of size $w_{\vec n}=N!/\prod_{a,b}n_{ab}!$. Normalizing its sum gives a Hilbert--Schmidt orthonormal basis for the density:
\[
 \begin{aligned}
 E_{\vec n}&=w_{\vec n}^{-1/2}\sum_{(\xi,\zeta)\in\mathcal O_{\vec n}}|\xi\rangle\langle\zeta|,\\
 \rho_N&=\sum_{\vec n}R_{\vec n}E_{\vec n}.
 \end{aligned}
\] Replacing an input pair $cd$ by $ab$ sends $\vec n$ to $\vec n-\vec e_{cd}+\vec e_{ab}$, where $\vec e_{ab}$ is the corresponding unit count vector. The ket and bra down-spin counts give their magnetizations; removing the updated spin then gives their conditional fields:
\[
 \begin{aligned}
 m_\xi&=1-\frac{2(n_{10}+n_{11})}{N},
 &m_\zeta&=1-\frac{2(n_{01}+n_{11})}{N},\\
 a_\xi&=K[m_\xi-(1-2c)/N],
 &a_\zeta&=K[m_\zeta-(1-2d)/N].
 \end{aligned}
\]
There are $n_{cd}$ possible updated sites. Counting them and including the orbit normalization gives the generator matrix (input column $\vec n$, output row $\vec n'$)
\begin{equation}
 \begin{aligned}
 L_{\vec n',\vec n}
 &=\frac1{1-s}\sqrt{\frac{w_{\vec n}}{w_{\vec n'}}}
 \sum_{\substack{c,d\in\{0,1\}\\n_{cd}>0}}n_{cd}
 \sum_{a,b\in\{0,1\}}\delta_{\vec n',\,\vec n-\vec e_{cd}+\vec e_{ab}}
 \sum_{\mu=0,1}[A_\mu(a_\xi)]_{ac}[A_\mu(a_\zeta)]_{bd}^*\\
 &\quad-\left[\frac N{1-s}+(1+s)r(n_{01}+n_{10})\right]\delta_{\vec n',\vec n},
 \end{aligned}
 \label{eq:S-orbit-matrix}
\end{equation}
where the losses are the identity subtractions and the decay of each differing ket--bra bit. The stationary coefficients solve
\begin{equation}
 L\vec R=0,\qquad
 \sum_{k=0}^N\sqrt{\binom Nk}R_{(N-k,0,0,k)}=1.
 \label{eq:S-orbit}
\end{equation}
The $\binom{N+3}{3}=\sum_j(2j+1)^2$ coordinates retain every total-spin sector and its redistribution by local dephasing~\cite{Shammah2018}; here $j$ runs from $N/2$ down to $0$ or $1/2$ in unit steps. For $k$ down spins, $m_k=1-2k/N$. The diagonal coefficients give their probability, and hence the magnetization law and second moment:
\begin{equation}
 \begin{gathered}
 \Pi_N(k)=\sqrt{\binom Nk}R_{(N-k,0,0,k)},\\
 P_{N,r}=\sum_{k=0}^N\Pi_N(k)\delta_{m_k},\qquad
 \langle M_z^2\rangle_*=\sum_{k=0}^Nm_k^2\Pi_N(k).
 \end{gathered}
 \label{eq:S-observables}
\end{equation}
For the transverse deviation, put $d_{\vec n}=n_{01}+n_{10}$. The matrix elements of $M_x$ and the off-diagonal elements of $M_x^2$ are $1/N$ for one differing bit and $2/N^2$ for two, respectively. Hence
\begin{equation}
 \begin{gathered}
 \langle M_x^2\rangle_*=\frac1N+\frac2{N^2}\sum_{d_{\vec n}=2}\sqrt{w_{\vec n}}R_{\vec n},\\
 \langle\{M_x,\bar x(M_z)\}\rangle_*=\frac1N\sum_{d_{\vec n}=1}\sqrt{w_{\vec n}}R_{\vec n}[\bar x(m_\xi)+\bar x(m_\zeta)],\\
 e_N=\langle M_x^2\rangle_*-\langle\{M_x,\bar x(M_z)\}\rangle_*
 +\sum_{k=0}^N\Pi_N(k)\bar x(m_k)^2.
 \end{gathered}
 \label{eq:S-orbit-transverse}
\end{equation}
At $K=0.75$, $s=0.8$, and $r=3$, the second moment approaches $m_r^2\simeq0.400824$, while the transverse deviation decreases toward zero, as shown in the figure.

The feedback calculation also identifies the ordering pattern when only some populations are noisy. Let population $a$ occupy fraction $f_a>0$, $\sum_af_a=1$, with symmetric pair coupling $kK_{ab}^{(0)}/N$, local parameters $\nu_a>0$, $0\le s_a<1$, $\gamma_a\ge0$, and Bloch components $m_a,x_a,y_a$. Each population has noise ratio $r_a=\gamma_a/[\nu_a(1-s_a^2)]$, local gain $g_a=1+s_ar_a/(1+r_a)$, and population relaxation rate $d_a=\nu_a(1-s_a)$.

The field and the linearized response in physical time are
\begin{equation}
 \begin{aligned}
 a_a&=k\sum_bK_{ab}^{(0)}f_bm_b,\\
 \frac{dm_a}{dt}&=d_a(-m_a+g_aa_a)
 =d_a\left[-m_a+kg_a\sum_bK_{ab}^{(0)}f_bm_b\right].
 \end{aligned}
 \label{eq:S-heterolinear}
\end{equation}
To find the first unstable pattern, rescale $z_a=\sqrt{f_a/g_a}\,m_a$. The interaction coefficients become symmetric:
\[
 \frac{\dot z_a}{d_a}
 =-z_a+k\sum_b
 \underbrace{\sqrt{f_ag_a}\,K_{ab}^{(0)}\sqrt{f_bg_b}}_{H_{ab}=H_{ba}}\,z_b.
\]
The coordinates $z_a/\sqrt{d_a}$ have symmetric linear matrix $\sqrt{d_ad_b}(kH_{ab}-\delta_{ab})$, whose inertia is that of $k\mathsf H-\openone$. Positive training rates set the time scales at fixed gains; the first magnetization instability is
\begin{equation}
 k\lambda_{\max}(\mathsf H)=1.
 \label{eq:S-spectral}
\end{equation}
For $K_{ab}^{(0)}=1$, $\lambda_{\max}=\sum_af_ag_a$. A noisy fraction $f$ coupled to a gain-one population therefore has $k_c=[1+fsr/(1+r)]^{-1}$.

Setting all gains to one gives the target matrix $(\mathsf J_0)_{ab}=\sqrt{f_af_b}K_{ab}^{(0)}$ and its disordered interval $k\lambda_{\max}(\mathsf J_0)<1$. At the unpolarized state $x_{0,a}=(1+r_a)^{-1}$, the control calculation gives
\[
 \dot y_a=-(d_a+\gamma_a)y_a+kx_{0,a}\sum_bK_{ab}^{(0)}f_bd_by_b.
\]
Multiplying by $f_ad_ay_a/x_{0,a}$ and using the symmetric interaction yields
\[
 \begin{aligned}
 \frac12\frac{d}{dt}\sum_a\frac{f_ad_a}{x_{0,a}}y_a^2
 &=-\sum_a\frac{f_ad_a(d_a+\gamma_a)}{x_{0,a}}y_a^2
 +k\sum_{a,b}K_{ab}^{(0)}f_af_bd_ad_by_ay_b\\
 &\le-[1-k\lambda_{\max}(\mathsf J_0)]\sum_af_ad_a^2y_a^2.
 \end{aligned}
\]
Thus $y$ decays throughout that interval; the $x$ rates are already $-d_a(1+s_a)(1+r_a)<0$. The unstable direction is a magnetization pattern.

For an onset inside this target-disordered interval, its nonlinear amplitude follows from the same local saturation as in the uniform system:
\begin{equation}
 m_a=\frac{g_a\tanh a_a}{1+(g_a-1)\tanh^2a_a}.
 \label{eq:S-pop-stationary}
\end{equation}
Using $a_a=k(\mathsf H\vec z)_a/\sqrt{f_ag_a}$ in its cubic expansion gives
\begin{equation}
 z_a=k(\mathsf H\vec z)_a-
 \frac{g_a-2/3}{f_ag_a}[k(\mathsf H\vec z)_a]^3+O(\|\vec z\|^5).
 \label{eq:S-mode-expansion}
\end{equation}
Suppose the largest eigenvalue $\lambda>0$ is simple, with real unit eigenvector $\vec e$. Decompose $\vec z=A\vec e+\vec w$, where $A=\vec e^{\,T}\vec z$ and $\vec e^{\,T}\vec w=0$. The stationary equation on the complementary space is invertible at $k_c=1/\lambda$, giving $\vec w=O(A^3)$. Its projection onto $\vec e$ gives
\[
 0=(k\lambda-1)A-(k\lambda)^3\beta A^3+O(A^5),\qquad
 \beta=\sum_a\frac{g_a-2/3}{f_ag_a}e_a^4>0,
\]
and hence the emerging pattern
\begin{equation}
 \begin{aligned}
 A^2&=\frac{k\lambda-1}{\beta}+O[(k\lambda-1)^2],\\
 m_a&=\sqrt{g_a/f_a}\,Ae_a+O(A^3).
 \end{aligned}
 \label{eq:S-cubic}
\end{equation}
For its dynamical attraction, put $\epsilon=k\lambda-1$. On $\vec y=0$, the transverse stationary curve is
\[
 \bar x_a(\vec m)=\frac{q_a[(1+s_a)-s_au_am_a]}{1+s_aq_a^2+(1+s_a)r_a},\qquad
 u_a=\tanh a_a,\quad q_a=\sech a_a.
\]
Near onset $\bar x_a-x_{0,a}=O(A^2)$ and $\dot A=O(\epsilon A+A^3)$, so the center-manifold equation gives $x_a-\bar x_a=O(|\epsilon|A^2+A^4)$. Its contribution to $\dot m_a$ has the extra factor $u_a=O(A)$ and leaves the cubic coefficient unchanged. The left zero vector has components $e_a/d_a$; projecting the dynamics therefore gives
\begin{equation}
 \left(\sum_a\frac{e_a^2}{d_a}\right)\frac{dA}{dt}
 =\epsilon A-\beta A^3+O(\epsilon^2A+|\epsilon|A^3+A^5).
 \label{eq:S-mode-dynamics}
\end{equation}
Its positive temporal coefficient and $\beta>0$ make the emerging branch stable while the other modes remain damped.

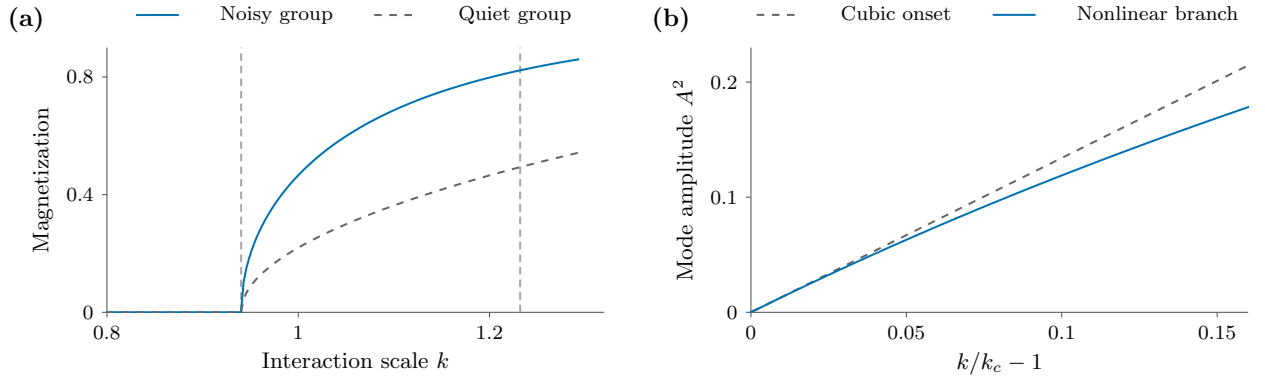
\begin{figure}[htbp]
\centering
\resizebox{.94\linewidth}{!}{%
\begin{tikzpicture}
\path[use as bounding box] (0,0) rectangle (17.8,5.5);
\plabel{0}{5.4}{a}\plabel{9.0}{5.4}{b}
\begin{axis}[paper,at={(1.5cm,1.02cm)},anchor=south west,width=6.95cm,height=3.7cm,xmin=.8,xmax=1.32,ymin=0,ymax=.9,xlabel={Interaction scale $k$},ylabel={Magnetization},xtick={.8,1,1.2},ytick={0,.4,.8},legend columns=2,legend style={at={(.5,1.06)},anchor=south,column sep=14pt}]

\addplot[black!35,densely dashed,forget plot] coordinates {(.94033,0)(.94033,.9)};
\addplot[black!35,densely dashed,forget plot] coordinates {(1.23206,0)(1.23206,.9)};
\addplot[qblue] table[x=coupling,y=m_noisy]{\DataHeterogeneous};\addlegendentry{Noisy group}
\addplot[qgray,dashed] table[x=coupling,y=m_quiet]{\DataHeterogeneous};\addlegendentry{Quiet group}

\end{axis}
\begin{axis}[paper,at={(10.5cm,1.02cm)},anchor=south west,width=6.95cm,height=3.7cm,xmin=0,xmax=.16,ymin=0,ymax=.23,xlabel={$k/k_c-1$},ylabel={Mode amplitude $A^2$},xtick={0,.05,.1,.15},ytick={0,.1,.2},legend columns=2,legend style={at={(.5,1.06)},anchor=south,column sep=14pt}]

\addplot[qgray,dashed] table[x=distance,y=normal_form]{\DataHeterogeneous};\addlegendentry{Cubic onset}
\addplot[qblue] table[x=distance,y=amplitude2]{\DataHeterogeneous};\addlegendentry{Nonlinear branch}

\end{axis}
\end{tikzpicture}}
\caption{The ordering mode for the two populations specified in Eq.~\eqref{eq:S-pop-example}. (a) Their magnetizations on the positive stationary branch. The vertical lines mark the noisy and target thresholds. (b) The squared projection of the same branch onto the critical eigenvector, compared with the cubic onset $(k/k_c-1)/\beta$. Both panels follow from Eq.~\eqref{eq:S-pop-stationary}.}
\label{fig:S-mode}
\end{figure}

Figure~\ref{fig:S-mode} shows the resulting two-population pattern for
\begin{equation}
 (f_1,f_2)=(0.35,0.65),\quad(r_1,r_2)=(3,0),\quad(s_1,s_2)=(0.8,0.6),\qquad
 K^{(0)}=\begin{pmatrix}1.25&0.8\\0.8&0.65\end{pmatrix}.
 \label{eq:S-pop-example}
\end{equation}
The gains $(1.6,1)$ give $k_c=0.94033\ldots$ and $k_{c,\rm target}=1.23206\ldots$. Solving Eq.~\eqref{eq:S-pop-stationary} gives the two magnetizations; projection onto $\vec e$ gives the amplitude and its slope fixed by Eq.~\eqref{eq:S-cubic}. Both populations order, with unequal amplitudes, before the target instability.

The same mode quantifies a small additional noise change. At fixed couplings, differentiating $\mathsf H$ and its simple eigenvalue gives
\[
 \begin{aligned}
 \mathrm dH_{ab}&=\frac12\left(\frac{\mathrm dg_a}{g_a}+\frac{\mathrm dg_b}{g_b}\right)H_{ab},\\
 \mathrm d\lambda&=\vec e^{\,T}(\mathrm d\mathsf H)\vec e
 =\lambda\sum_ae_a^2\frac{\mathrm dg_a}{g_a},\\
 \frac{\mathrm dk_c}{k_c}&=-\frac{\mathrm d\lambda}{\lambda}
 =-\sum_ae_a^2\frac{\mathrm dg_a}{g_a}.
 \end{aligned}
\]
A single site in a delocalized mode has weight $O(N^{-1})$, although noise there can already ensure finite-graph relaxation through the algebra constructed above. Smooth even $\nu(a)$ and $s(a)$ have vanishing first derivatives at zero, so their onset uses only $\nu(0),s(0)$.

The zero-field gain also compares continuous training conventions. At fixed bright-state relaxation rate $\kappa=\nu(1-s^2)$, put $V(a)=|\nu\rangle\langle\nu^\perp|$. Then
\[
 \nu(\D[A_0]+\D[A_1])
 =\kappa\frac{1-s}{1+s}\D[Q]+\kappa\D[V]
 \longrightarrow\kappa\D[V]\qquad(s\uparrow1).
\]
The field-tilted zero-field jump has $|\pm\rangle=(|0\rangle\pm|1\rangle)/\sqrt2$ and $D_a=\operatorname{diag}(1,e^{-a})$. Restoring the conditional-state normalizations gives
\begin{equation}
 D_a|+\rangle\langle-|D_a^{-1}
 =\frac12\begin{pmatrix}1&-e^a\\e^{-a}&-1\end{pmatrix}
 =\cosh(a)V(a).
 \label{eq:S-normalization}
\end{equation}
Its controlled version $\cosh(a_i)V_i$ has an extra rate $\cosh^2a$, even and equal to one at zero. Both normalizations therefore have the limiting onset
\begin{equation}
 r=\gamma/\kappa,\qquad
 K_c=\frac1{1+sr/(1+r)}\longrightarrow\frac{\kappa+\gamma}{\kappa+2\gamma}.
 \label{eq:S-continuous}
\end{equation}
The magnetization law is now determined. Its finite-size width and the time needed to form or leave an ordered peak follow by retaining the same finite differences in Eq.~\eqref{eq:S-exact-difference}, as we do next.
\FloatBarrier

\section{Magnetization fluctuations and relaxation}
\label{sec:critical}
\label{sec:fluctuations}
\label{sec:times}

The stationary peaks found in Sec.~\ref{sec:collective} acquire a finite width at finite $N$. We determine that width, its coherent correction near criticality, and the formation and sign-loss times from the same magnetization generator, Eq.~\eqref{eq:S-exact-difference}. We retain uniform Curie--Weiss coupling, standard recycling, $0<s<1$, and $\tau=\nu(1-s)t$.

Writing the dephasing rate as $\lambda=(1+s)r$, separate the two actions on observables and then compose the backward generator:
\begin{equation}
 \begin{gathered}
 \mathcal C_NO=\frac1{1-s}\sum_i[\Phi_i^\dagger(O)-O],\qquad
 \mathcal D_NO=\frac12\sum_i(Z_iOZ_i-O),\\
 \mathcal L_{\tau,N}^\dagger=\mathcal C_N+\lambda\mathcal D_N.
 \end{gathered}
 \label{eq:S-split-generator}
\end{equation}
Let $\mathcal PO=\sum_z\langle z|O|z\rangle|z\rangle\langle z|$ retain measured, diagonal entries and $\mathcal Q=\id-\mathcal P$ retain coherences. On an entry differing at $d$ spins, $\mathcal D_N(|z\rangle\langle z'|)=-d|z\rangle\langle z'|$. For $F=f(M_z)$, the exact half difference $b_{i,f}$ of Eq.~\eqref{eq:S-exact-difference} writes $\mathcal C_NF=B_F+T_F$, where
\begin{equation}
 \begin{aligned}
 B_F=\mathcal P\mathcal C_NF
 &=\sum_i b_{i,f}[(1+s)u_i-(1+su_i^2)Z_i],\\
 T_F=\mathcal Q\mathcal C_NF
 &=-s\sum_i b_{i,f}u_iq_iX_i.
 \end{aligned}
 \label{eq:S-uniform-BT}
\end{equation}
Thus $B_F$ changes the measured magnetization, while every entry of $T_F$ differs at one spin and satisfies $\mathcal D_NT_F=-T_F$. Eliminating this damped coherence gives the leading measured process; retaining its feedback gives the first correction.

At fixed $N$, write $\rho_{\rm d}=\mathcal P\rho$, $\rho_{\rm o}=\mathcal Q\rho$, and $c_N=\|\mathcal C_N^\dagger\|_{2\to2}$. The forward equation gives
\[
 \begin{aligned}
 \dot\rho_{\rm d}&=\mathcal P\mathcal C_N^\dagger\rho_{\rm d}
                  +\mathcal P\mathcal C_N^\dagger\rho_{\rm o},\\
 \dot\rho_{\rm o}&=\mathcal Q\mathcal C_N^\dagger\rho_{\rm d}
     +(\lambda\mathcal D_N+\mathcal Q\mathcal C_N^\dagger\mathcal Q)\rho_{\rm o}.
 \end{aligned}
\]
On the off-diagonal subspace $\mathcal D_N\le-\id$. Integrating the second equation for $\lambda>c_N$ and inserting it into the first yields
\[
 \begin{aligned}
 \|\rho_{\rm o}(\tau)\|_2
 &\le e^{-(\lambda-c_N)\tau}\|\rho_{\rm o}(0)\|_2+\frac{c_N}{\lambda-c_N},\\
 \rho_{\rm d}(\tau)
 &=e^{\tau\mathcal P\mathcal C_N^\dagger\mathcal P}\rho_{\rm d}(0)
      +O_{N,T}(\lambda^{-1})\qquad(0\le\tau\le T).
 \end{aligned}
\]
The decaying integral includes any initial coherence. At stationarity the same off-diagonal inverse, whose norm is $O_N(\lambda^{-1})$, gives $\rho_{\rm o}=O_N(\lambda^{-1})$ and $\mathcal P\mathcal C_N^\dagger\rho_{\rm d}=O_N(\lambda^{-1})$. Thus $\mathcal P\mathcal C_N\mathcal P$ governs both the leading measured evolution and its stationary limit. Reading its rates from the diagonal equation of Sec.~\ref{sec:system} gives, with $u_i(z)=\tanh a_i(z)$,
\begin{align}
 w_i(z)&=\frac12[1-z_i u_i(z)][1-sz_i u_i(z)],
 \label{eq:S-strongrate}\\
 (W_\infty f)(z)&=\sum_iw_i(z)[f(z^i)-f(z)],\notag\\
 \partial_\tau p(z,\tau)&=(W_\infty^*p)(z,\tau)
 =\sum_i[w_i(z^i)p(z^i,\tau)-w_i(z)p(z,\tau)].
 \label{eq:S-projected-master}
\end{align}
All rates are positive at finite fields, so their unique stationary law is the fixed-size strong-dephasing limit of the quantum measurement law. Since a reversal leaves its own conditional field unchanged, the directional ratio is
\begin{equation}
 \begin{aligned}
 \frac{w_i(z)}{w_i(z^i)}
 &=\frac{1-z_i u_i}{1+z_i u_i}\frac{1-sz_i u_i}{1+sz_i u_i}\\
 &=\exp\{-2z_i[a_i+\artanh(s\tanh a_i)]\}.
 \end{aligned}
 \label{eq:S-effectivefield}
\end{equation}
This identifies $F_s(a)=a+\artanh(s\tanh a)$ as the effective field in the ratio $e^{-2z_iF_s(a_i)}$.

For Curie--Weiss coupling, let $k$ count down spins, $m_k=1-2k/N$, and include a uniform field $h$. The $N-k$ up spins each see $K(m_k-1/N)+h$ and increase $k$ on reversal; the $k$ down spins see $K(m_k+1/N)+h$ and decrease it. Write the corresponding polarizations as $u_k^\pm=\tanh[K(m_k\pm1/N)+h]$. Summing the single-spin rates closes the magnetization process:
\begin{equation}
 \begin{gathered}
 B_k=\frac{N-k}{2}(1-u_k^-)(1-su_k^-),\qquad
 D_k=\frac{k}{2}(1+u_k^+)(1+su_k^+),\\
 (W_\infty f)(k)=B_k[f(k+1)-f(k)]+D_k[f(k-1)-f(k)].
 \end{gathered}
 \label{eq:S-birthdeath}
\end{equation}
The endpoint rates satisfy $B_N=D_0=0$. For the stationary law $\pi_{N,h}$, the current $B_k\pi_{N,h}(k)-D_{k+1}\pi_{N,h}(k+1)$ is constant and vanishes at the endpoints. Because $u_k^-=u_{k+1}^+$, Eq.~\eqref{eq:S-effectivefield} gives
\[
 \frac{\pi_{N,h}(k+1)}{\pi_{N,h}(k)}
 =\frac{B_k}{D_{k+1}}
 =\frac{N-k}{k+1}
   e^{-2F_s(K[1-(2k+1)/N]+h)}.
\]
Multiplying these adjacent ratios yields the complete stationary law,
\begin{equation}
 \pi_{N,h}(k)=\frac1{\mathcal Z_{N,h}}\binom Nk
 \exp\left[-2\sum_{\ell=0}^{k-1}
 F_s\left(K\left[1-\frac{2\ell+1}{N}\right]+h\right)\right].
 \label{eq:S-exact-macro}
\end{equation}
Its large-$N$ exponent determines the peak locations and widths. Stirling's formula gives the entropy per spin,
\[
 S(m)=-\frac{1+m}{2}\log\frac{1+m}{2}
       -\frac{1-m}{2}\log\frac{1-m}{2}.
\]
The remaining exponent is a midpoint sum with mesh $2/N$:
\[
 \begin{aligned}
 -\frac2N\sum_{\ell=0}^{k-1}
 F_s\left(K\left[1-\frac{2\ell+1}{N}\right]+h\right)
 &\longrightarrow-\int_m^1 F_s(Kx+h)\,dx\\
 &=\int_0^m F_s(Kx+h)\,dx-\int_0^1 F_s(Kx+h)\,dx.
 \end{aligned}
\]
Absorbing the last term into normalization gives $\pi_{N,h}(k)=\exp[-NI_{s,h}(m_k)+\mathrm{const}+o(N)]$, where
\begin{equation}
 I_{s,h}(m)=-S(m)-\int_0^mF_s(Kx+h)\,dx.
 \label{eq:S-memory-potential}
\end{equation}
At $h=0$, write $I_s=I_{s,0}$. Differentiating the exponent recovers the strong-noise response equation:
\[
 I_s'(m)=\artanh m-F_s(Km)=0
 \quad\Longleftrightarrow\quad
 m=\frac{(1+s)\tanh(Km)}{1+s\tanh^2(Km)}.
\]
Response concavity from Sec.~\ref{sec:collective} gives one minimum at zero for $(1+s)K\le1$, and two at $\pm m_\infty$ for $(1+s)K>1$. In the latter case the stationary suppression between a peak and the center is $e^{-N\Delta I}$, where $\Delta I=I_s(0)-I_s(m_\infty)>0$.

The potential also determines which ordered peak is selected by a small field. At $h=0$ the weights are equal; differentiating the two minimum values gives
\begin{equation}
 \begin{aligned}
 \left.\partial_hI_{s,h}(\pm m_\infty)\right|_{h=0}
 &=-\int_0^{\pm m_\infty}F_s'(Kx)\,dx\\
 &=\mp\frac{F_s(Km_\infty)}K
 =\mp\frac{\artanh m_\infty}K.
 \end{aligned}
 \label{eq:S-field-chi}
\end{equation}
Denote the positive coefficient by $\chi_s=\artanh(m_\infty)/K$. Since $I_s'(\pm m_\infty)=0$, motion of the minima contributes no linear term. Their exponent difference $2Nh\chi_s$ selects $h_N=\eta/N$; equal zero-field prefactors then give the limiting weight ratio $e^{2\eta\chi_s}$:
\begin{equation}
 \begin{aligned}
 \sum_k\pi_{N,h_N}(k)\delta_{m_k}
 &\Longrightarrow
 \frac{e^{\eta\chi_s}\delta_{m_\infty}
       +e^{-\eta\chi_s}\delta_{-m_\infty}}{2\cosh(\eta\chi_s)},\\
 \lim_{N\to\infty}\langle m\rangle_{h_N}
 &=m_\infty\tanh(\eta\chi_s).
 \end{aligned}
 \label{eq:S-field-selection}
\end{equation}
For $h_N=o(N^{-1})$ the weights remain equal. If $h_N\to0$ and $Nh_N\to\pm\infty$, one sign is selected. In the interval $K<1$, the programmed Gibbs law has finite susceptibility per spin, so its mean still tends to zero under $h_N=\eta/N$. This field resolves the sampler's ordered peaks while leaving the target macroscopically unpolarized.

At zero field, expanding about a nondegenerate minimum $m_0$ gives a Gaussian peak of width $[NI_s''(m_0)]^{-1/2}$. At threshold the curvature vanishes, so the next term determines the width:
\[
 \begin{aligned}
 F_s(a)&=(1+s)a-\frac{s(1-s^2)}3a^3+O(a^5),\\
 I_s'(m)&=[1-(1+s)K]m+\frac{1+s(1-s^2)K^3}{3}m^3+O(m^5).
 \end{aligned}
\]
The curvature vanishes at $K_\infty=1/(1+s)$. There the cubic coefficient of $I_s'$ is $c_\infty=(s+1/3)/(1+s)^2$, so balancing the quartic exponent $Nc_\infty m^4$ with one fixes the fluctuation coordinate:
\begin{equation}
 X_N=(Nc_\infty)^{1/4}M_z.
 \label{eq:S-uniform-scales}
\end{equation}
The critical width is $N^{-1/4}$; write $X=(Nc_\infty)^{1/4}m$ for the measured coordinate.

With $\epsilon_\infty=(1+s)K-1$, the same expansion gives the nearby barrier,
\begin{equation}
 \begin{gathered}
 I_s(m)-I_s(0)=-\frac{\epsilon_\infty}{2}m^2+\frac{c_\infty}{4}m^4
                  +O(\epsilon_\infty m^4,m^6),\\
 \Delta I=I_s(0)-I_s(m_\infty)
 =\frac{\epsilon_\infty^2}{4c_\infty}+O(\epsilon_\infty^3).
 \end{gathered}
 \label{eq:S-critical-memory}
\end{equation}
To follow fluctuations on this scale, use the first two increments of the same birth--death process. At zero field,
\begin{equation}
 \begin{aligned}
 \frac{2(D_k-B_k)}N
 &=\epsilon_\infty m_k+O(m_k^3,N^{-1}),\\
 \frac{4(B_k+D_k)}{N^2}
 &=\frac2N+O(m_k^2/N,N^{-2}).
 \end{aligned}
 \label{eq:S-formation}
\end{equation}
Retaining the cubic drift and a small field gives $\epsilon_\infty m-c_\infty m^3+(1+s)h$, with diffusion coefficient $1/N$ to leading order. In $X$, both act at rate $\sqrt{c_\infty/N}$. The same rescaling gives the detuning and field parameters
\[
 A=\epsilon_\infty\sqrt{N/c_\infty},\qquad
 H_c=(1+s)hN^{3/4}c_\infty^{-1/4}.
\]
The backward generator in time $\tau\sqrt{c_\infty/N}$ is therefore
\begin{equation}
 (\mathcal G_{A,H_c}f)(X)
 =f''(X)+(AX-X^3+H_c)f'(X).
 \label{eq:S-critical-generator}
\end{equation}
Its zero-current density is $p_{A,H_c}(X)=\mathcal Z_{A,H_c}^{-1}
 e^{AX^2/2-X^4/4+H_cX}$; write $\mathcal G_A=\mathcal G_{A,0}$ and $p_A=p_{A,0}$ at zero field. This identifies fluctuation, field, and relaxation scales $N^{-1/4}$, $N^{-3/4}$, and $\sqrt N$, respectively~\cite{Ding2009}.

At zero field, finite noise changes the detuning through the gain of Sec.~\ref{sec:collective}:
\[
 g(r)K-1=(1+s)(K-K_\infty)-\frac{sK}{1+r}.
\]
The coherent restoring term competes with critical detuning when $r\sim\sqrt N$. We therefore retain the feedback of $T_F$ into $B_F$ in Eq.~\eqref{eq:S-uniform-BT}. Fix $s\in(0,1)$, $0\le\kappa_0<\infty$, and $\rho_{\min}>0$. For sequences with $|\kappa_N|\le\kappa_0$ and $r_N\ge\rho_{\min}\sqrt N$, set the coupling and then its resulting detuning to
\begin{equation}
 \begin{aligned}
 K_N&=K_\infty+\frac{\kappa_N}{\sqrt N},\\
 A_N&=\sqrt{\frac N{c_\infty}}[g(r_N)K_N-1].
 \end{aligned}
 \label{eq:S-triangular-array}
\end{equation}
We determine the stationary quantum law in this window by taking $F=f(X_N)=f_N(M_z)$, where $f\in C_c^8(\mathbb R)$ and $f_N(m)=f((Nc_\infty)^{1/4}m)$. Set $v_i=M_z-Z_i/N$; the conditional field then gives $u_i=\tanh(K_Nv_i)$ and $q_i=\sech(K_Nv_i)$. The half difference in Eq.~\eqref{eq:S-uniform-BT} is $b_{i,f_N}=[f_N(v_i+1/N)-f_N(v_i-1/N)]/2$.
With $\lambda_N=(1+s)r_N$, the identity $\mathcal D_NT_F=-T_F$ shows how to cancel the first off-diagonal term: add $T_F/\lambda_N$ to the observable to obtain
\[
 \mathcal L_{\tau,N}^\dagger(F+T_F/\lambda_N)
 =B_F+\frac{\mathcal P\mathcal C_NT_F}{\lambda_N}
       +\frac{\mathcal Q\mathcal C_NT_F}{\lambda_N}.
\]
Only an update at the flipped site returns $T_F$ to the diagonal. The local $X_i$ equation therefore gives its feedback,
\begin{equation}
 \mathcal P\mathcal C_NT_F
 =-s\sum_i b_{i,f_N}u_iq_i^2[(1+s)-su_iZ_i].
 \label{eq:S-uniform-PCT}
\end{equation}
The remaining off-diagonal term differs at one or two spins. Set $S_F=(-\mathcal D_N)^{-1}\mathcal Q\mathcal C_NT_F$, dividing each entry by its number of differing spins. A second correction then leaves
\begin{equation}
 \begin{aligned}
 \widehat F&=F+\frac{T_F}{\lambda_N}+\frac{S_F}{\lambda_N^2},\\
 \mathcal L_{\tau,N}^\dagger\widehat F
 &=B_F+\frac{\mathcal P\mathcal C_NT_F}{\lambda_N}
       +\frac{\mathcal C_NS_F}{\lambda_N^2}.
 \end{aligned}
 \label{eq:S-exact-corrector}
\end{equation}
The first two terms are the measured process and its coherent correction. The critical time rescales the last by $O(\sqrt N/r_N^2)$, so the joint limit requires an $N$-independent bound on $\mathcal C_NS_F$.

On the support of $f_N$, the factors in $T_F$ obey $b_{i,f_N}=O(N^{-3/4})$ and $u_i=O(N^{-1/4})$, so their product remains bounded after the site sum. Each further generator action can mark at most one additional spin. In local matrix units $E_i^{ab}=|a\rangle_i\langle b|$, the terms needed for $T_F$, $S_F$, and $\mathcal C_NS_F$ consequently have the form
\[
 N^{-d}\sum_{i_1,\ldots,i_d}^{\ne}
 \psi_N\left(\frac1N\sum_{j\notin\{i_1,\ldots,i_d\}}Z_j\right)
 E_{i_1}^{a_1b_1}\cdots E_{i_d}^{a_db_d},\qquad d\le3.
\]
The indices are distinct, and the prefactor offsets the at most $N^d$ summands, giving norm at most $\|\psi_N\|_\infty$. The initial coefficient functions are supported on $|m|\le RN^{-1/4}+2/N$, with $R$ fixed by the test, and obey
\[
 \|\psi_N^{(j)}\|_\infty\le C_jN^{j/4}\qquad(0\le j\le6).
\]

A generator action either updates a displayed site, changes the coefficient through another spin, or changes a coherent control. The first has boundedly many local terms. For the second, pair the two values of the unmarked spin before summing. Their first difference multiplies an odd drift bounded by $C(|m|+d/N)$, while the second difference has total rate at most $CN$. For the third, ket and bra fields differ by at most $2d|K_N|/N$. Kraus completeness cancels the zeroth-order overlap, so summing over the remaining sites again gives a bounded coefficient. The mean-value theorem on these paired differences bounds each resulting coefficient by
\begin{equation}
 C_d\left(\|\psi_N\|_\infty+\|m \psi_N'\|_\infty
       +\frac{\|\psi_N'\|_\infty+\|\psi_N''\|_\infty}{N}\right).
 \label{eq:S-coefficient-estimate}
\end{equation}

Differentiating the paired differences gives the same estimate for the scaled derivative norms. Each action uses at most two derivatives and enlarges the support by $O(N^{-1})$, so two actions preserve these bounds. Removing diagonal entries with $\mathcal Q$ and dividing the remaining entries by one or two with $(-\mathcal D_N)^{-1}$ preserves the coefficient bounds as well. Thus
\begin{equation}
 \begin{gathered}
 \|T_F\|+\|S_F\|+\|\mathcal C_NS_F\|\le C_f,\\
 \|\widehat F-F\|
 \le\frac{\|T_F\|}{\lambda_N}+\frac{\|S_F\|}{\lambda_N^2}
 \le C_f(r_N^{-1}+r_N^{-2}).
 \end{gathered}
 \label{eq:S-uniform-corrector-bound}
\end{equation}
The constants are independent of $N$ and $r_N$. The same construction applies to $F=M_z^2$: its coefficients and their derivatives are bounded on $[-1,1]$, so Eq.~\eqref{eq:S-uniform-corrector-bound} again holds with a constant bound.

Use this second-moment observable to determine how much stationary probability lies outside the critical scale. At $K_\infty$, the diagonal generator has drift
\[
 F_\infty(m)=(1+s)\tanh(K_\infty m)
       -[1+s\tanh^2(K_\infty m)]m.
\]
Response concavity gives $mF_\infty(m)<0$ off zero and $-F_\infty(m)/m^3\to c_\infty>0$ at zero. Compactness gives $2mF_\infty(m)\le-bm^4$, $b>0$; adding detuning and finite-size differences yields
\[
 B_{M_z^2}\le-bM_z^4+\frac{C\kappa_0}{\sqrt N}M_z^2+\frac CN.
\]
Its half difference $2v_i/N$ satisfies $v_i\tanh(K_Nv_i)\ge0$ for large $N$. Together with $(1+s)-su_iZ_i>0$, this makes the feedback~\eqref{eq:S-uniform-PCT} nonpositive. The stationary expectation of Eq.~\eqref{eq:S-exact-corrector} then gives
\[
 \begin{aligned}
 b\langle M_z^4\rangle
 &\le\frac{C\kappa_0}{\sqrt N}\langle M_z^2\rangle+\frac CN+\frac C{r_N^2}\\
 &\le\frac b2\langle M_z^4\rangle+\frac{C(1+\kappa_0^2)}N+\frac C{r_N^2}.
 \end{aligned}
\]
Hence
\begin{equation}
 \begin{gathered}
 \langle M_z^4\rangle\le C(N^{-1}+r_N^{-2}),\\
 \langle X_N^4\rangle=Nc_\infty\langle M_z^4\rangle
 \le C(1+N/r_N^2)\le C(1+\rho_{\min}^{-2}).
 \end{gathered}
 \label{eq:S-uniform-moment}
\end{equation}
Thus the critical coordinate is tight and its second moment is uniformly integrable throughout the joint window.

With probability concentrated on this scale, return to the tests $F=f(X_N)$. Expanding the two diagonal terms in Eq.~\eqref{eq:S-exact-corrector} gives
\[
 \frac{f_N''(m)}N+
 \left[\left((1+s)K_N-1-\frac{sK_N}{r_N}\right)m
               -c_\infty m^3\right]f_N'(m).
\]
Since $1/r_N=1/(1+r_N)+O(r_N^{-2})$, the linear term becomes $g(r_N)K_N-1$ to the retained order. On the critical time scale, excluded-self shifts contribute $O(N^{-1/4})$, higher drift powers $O(N^{-1/2})$, and cubic feedback $O(r_N^{-1})$. The linear replacement and second-corrector remainder contribute $O(\sqrt N/r_N^2)$. Combining these bounds gives
\begin{equation}
 \left\|\sqrt{\frac N{c_\infty}}\,
       \mathcal L_{\tau,N}^\dagger\widehat F
       -(\mathcal G_{A_N}f)(X_N)\right\|
 \le C_f\left(N^{-1/4}+\frac1{r_N}+\frac{\sqrt N}{r_N^2}\right).
 \label{eq:S-uniform-residual}
\end{equation}
The coefficient estimates make the constant uniform throughout the window~\eqref{eq:S-triangular-array}.

Taking stationary expectations in Eq.~\eqref{eq:S-uniform-residual}, every subsequence with $A_N\to A$ and $X_N\Rightarrow\mu$ satisfies
\[
 \int[f''(X)+(AX-X^3)f'(X)]\,d\mu(X)=0
 \qquad(f\in C_c^8).
\]
The unique integrable stationary density is $p_A$, and the fourth-moment bound allows its second moment to pass to the limit. Thus
\begin{equation}
 \begin{aligned}
 X_N&\Longrightarrow p_A,\\
 \langle X_N^2\rangle&\longrightarrow\int_{\mathbb R}X^2p_A(X)\,dX.
 \end{aligned}
 \label{eq:S-critical-distribution}
\end{equation}
Continuity in bounded $A$ and the same subsequence argument make both conclusions uniform relative to $p_{A_N}$ throughout the joint window, for every fixed bounded continuous test and for the second moment.

When $\kappa_N\to\kappa$ and $r_N/\sqrt N\to\rho\in(0,\infty]$, separating the coupling and noise contributions gives
\begin{equation}
 \begin{aligned}
 A_N&=\frac{(1+s)^2\kappa_N-s\sqrt N/(1+r_N)
                -s(1+s)\kappa_N/(1+r_N)}{\sqrt{s+1/3}},\\
 &\longrightarrow\frac{(1+s)^2\kappa-s/\rho}{\sqrt{s+1/3}}
 =:A_{\kappa,\rho}.
 \end{aligned}
 \label{eq:joint-two-controls}
\end{equation}
A coupling shift $\kappa=s/[(1+s)^2\rho]$ compensates the residual coherent restoring force and gives the pure quartic density $p_0$. At the unchanged coupling $K_N=K_\infty$, this detuning becomes $A_\rho=-s/[\rho\sqrt{s+1/3}]$, so
\begin{equation}
 \sqrt{Nc_\infty}\langle M_z^2\rangle
 \longrightarrow\int X^2p_{A_\rho}(X)\,dX.
 \label{eq:S-noise-crossover}
\end{equation}
For $\rho=\infty$, the substitution $t=X^4/4$ gives the pure quartic second moment $2\Gamma(3/4)/\Gamma(1/4)=0.675978\ldots$, with $\Gamma$ Euler's gamma function. Every finite positive $\rho$ retains a quadratic restoring term. As $\rho\downarrow0$ in this limiting family, the coordinate $Y=\sqrt{-A_\rho}X$ has density proportional to $\exp[-Y^2/2-Y^4/(4A_\rho^2)]$ and approaches a unit Gaussian.

The same quartic family gives the leading finite-noise expansion near $K_c=1/g(r)$. Only the longitudinal restoring rate vanishes there. Write $\epsilon_r=g(r)K-1$ for the linear growth rate. Expanding the transverse stationary curve~\eqref{eq:S-nullcline} in the drift~\eqref{eq:S-effective-drift} gives the cubic coefficient and the resulting longitudinal motion:
\[
 \begin{aligned}
 c_r&=\frac{\alpha+1/3}{(1+\alpha)^2},\\
 F_{\rm eff}(m)&=\epsilon_rm-c_rm^3+O(\epsilon_rm^3,m^5).
 \end{aligned}
\]
At threshold $\bar x(m)-x_0=O(m^2)$ and $\dot m=O(m^3)$, so following this curve changes $x$ only at order $m^4$ and supplies the cubic drift. The product identity of Sec.~\ref{sec:collective} gives its fluctuation strength directly:
\[
 \mathcal L_{\tau,N}^\dagger(M_z^2)-\{M_z,\mathcal L_{\tau,N}^\dagger M_z\}
 =\frac2{N^2}\sum_i[1+su_i^2-(1+s)u_iZ_i].
\]
The leading variance rate $2/N$ gives the same density $p_A$ for the coordinate $(Nc_r)^{1/4}m$, with $A=\epsilon_r\sqrt{N/c_r}$. At $A=0$, finite quantum stationary states with $s=0.8$, $N=96$ give scaled second moments $0.671811$ at $r=0.5$ and $0.667049$ at $r=3$, compared with $0.675978\ldots$ for $p_0$. At the unchanged coupling $K_\infty$, the linear restoring rate instead gives the leading Gaussian variance $\langle M_z^2\rangle\simeq(1+s)(1+r)/(sN)$. It reaches the quartic scale when $r\sim\sqrt N$, where Eq.~\eqref{eq:S-noise-crossover} supplies the joint law; $c_r\to c_\infty$ and the Gaussian limit of $p_{A_\rho}$ join the two regimes.

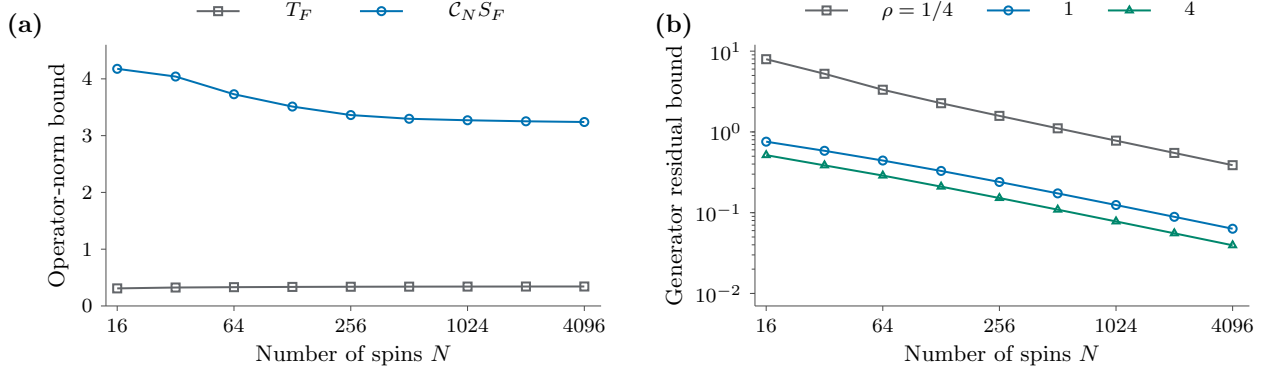
\begin{figure}[htbp]
\centering
\resizebox{.94\linewidth}{!}{%
\begin{tikzpicture}
\path[use as bounding box] (0,0) rectangle (17.8,5.2);
\plabel{0}{5.15}{a}\plabel{9.05}{5.15}{b}
\begin{axis}[paper,at={(1.5cm,.94cm)},anchor=south west,
 width=6.92cm,height=3.65cm,xmode=log,
 xmin=14,xmax=5000,ymin=0,ymax=4.6,
 xtick={16,64,256,1024,4096},xticklabels={16,64,256,1024,4096},
 ytick={0,1,2,3,4},xlabel={Number of spins $N$},
 ylabel={Operator-norm bound},
 legend columns=2,legend style={at={(.5,1.065)},anchor=south,column sep=16pt}]
\addplot[qgray,mark=square,mark options={fill=white}]
 table[x=N,y=T_bound]{\DataUniformRhoOnePZero};
\addlegendentry{$T_F$}
\addplot[qblue,mark=o,mark options={fill=white}]
 table[x=N,y=CS_bound]{\DataUniformRhoOnePZero};
\addlegendentry{$\mathcal C_NS_F$}
\end{axis}
\begin{axis}[paper,at={(10.57cm,.94cm)},anchor=south west,
 width=6.91cm,height=3.65cm,xmode=log,ymode=log,
 xmin=14,xmax=5000,ymin=.007,ymax=12,
 xtick={16,64,256,1024,4096},xticklabels={16,64,256,1024,4096},
 ytick={.01,.1,1,10},xlabel={Number of spins $N$},
 ylabel={Generator residual bound},
 legend columns=3,legend style={at={(.5,1.065)},anchor=south,column sep=11pt}]
\addplot[qgray,mark=square,mark options={fill=white}]
 table[x=N,y=generator_error_bound]{\DataUniformRhoZeroPTwoFive};
\addlegendentry{$\rho=1/4$}
\addplot[qblue,mark=o,mark options={fill=white}]
 table[x=N,y=generator_error_bound]{\DataUniformRhoOnePZero};
\addlegendentry{$1$}
\addplot[qteal,mark=triangle,mark options={fill=white}]
 table[x=N,y=generator_error_bound]{\DataUniformRhoFourPZero};
\addlegendentry{$4$}
\end{axis}
\end{tikzpicture}}
\caption{Coherent corrections on the critical scale at $K=K_\infty$, $s=0.8$. (a) Maximum absolute row-sum bounds for two operators in Eq.~\eqref{eq:S-uniform-corrector-bound}. (b) The row-sum bound on the full residual on the left of Eq.~\eqref{eq:S-uniform-residual}, at $r=\rho\sqrt N$. Here $f(X)=[1-(X/2.5)^2]^{10}$ for $|X|<2.5$ and zero otherwise. The matrix-entry calculations extend to $N=4096$. The lines join the calculated bounds.}
\label{fig:S-uniform}
\end{figure}

The permutation counts of Sec.~\ref{sec:collective} also evaluate the coherent correction: a row with $k$ down spins has $\binom{N-k}{n_{01}}\binom{k}{n_{10}}$ entries with $n_{01}$ up-to-down and $n_{10}$ down-to-up differences. The maximum multiplicity-weighted absolute row sum bounds each Hermitian operator norm. Figure~\ref{fig:S-uniform} shows these bounds and the residual in Eq.~\eqref{eq:S-uniform-residual}.

\FloatBarrier

Formation and sign loss are two passage problems for the same rates. For a chosen absorbing set, the mean hitting time $T_k$ satisfies the backward equation. To find sign loss at $h=0$, $K>K_\infty$, take the absorbing center $b=\lceil N/2\rceil$ and the positive-well site $k_\star$ nearest to $N(1-m_\infty)/2$. The boundary conditions $T_b=0$ and $D_0=0$ complete
\[
 -1=B_k(T_{k+1}-T_k)+D_k(T_{k-1}-T_k).
\]
Write $\pi(k)=\pi_{N,0}(k)$. Multiplication by $\pi(k)$ and use of $\pi(k)D_k=\pi(k-1)B_{k-1}$ turns the recurrence into a difference of neighboring terms. Summing first from $0$ to $n$ and then from $k_\star$ to $b-1$ gives
\begin{equation}
 \begin{gathered}
 B_n\pi(n)(T_n-T_{n+1})=\sum_{j=0}^n\pi(j),\\
 T_{k_\star\to b}
 =\sum_{n=k_\star}^{b-1}
       \frac{\sum_{j=0}^n\pi_{N,0}(j)}{B_n\pi_{N,0}(n)}.
 \end{gathered}
 \label{eq:S-passage}
\end{equation}
The numerator contains the well's mass, while the denominator is exponentially smallest at the center. The other factors are subexponential, so upper and lower Laplace bounds give
\begin{equation}
 \lim_{N\to\infty}\frac1N\log T_{k_\star\to b}
 =I_s(0)-I_s(m_\infty)=\Delta I.
 \label{eq:S-memory-exponent}
\end{equation}
Either well is reached from the center with probability tending to one half, so passage to the opposite well has the same exponent. Sign retention also follows in probability: the initial stationary law conditioned on the positive well is bounded by $(2+o(1))\pi_{N,0}$, and evolution preserves this bound. The expected central-edge crossings up to $T$ are at most $\operatorname{poly}(N)T e^{-N\Delta I}$; for $T=e^{Na}$, $a<\Delta I$, sign loss therefore has vanishing probability.

For formation, the initial point is the unstable center and the absorbing set is $|m|\ge a$, with fixed $0<a<m_\infty$. The linear drift and variance in Eq.~\eqref{eq:S-formation} make an initial $N^{-1/2}$ fluctuation grow at rate $\epsilon_\infty$.

For even $N$, identify symmetric states $k$ and $N-k$. The resulting chain starts at $c=N/2$ and moves toward smaller $k$. Combining each reflected pair gives weights $\mu_c=\pi(c)$ and $\mu_k=2\pi(k)$ for $k<c$; combining the two exits from the center gives outward rates $\widetilde D_c=2D_c$ and $\widetilde D_k=D_k$ elsewhere. With absorbing endpoint $a_N=\lfloor N(1-a)/2\rfloor$, the same telescoping calculation yields
\begin{equation}
 \mathbb E T_{\rm form}=T_{c\to a_N}
 =\sum_{n=a_N+1}^c\frac{\sum_{j=n}^c\mu_j}{\widetilde D_n\mu_n}.
 \label{eq:S-formation-sum}
\end{equation}
The inner region $m=O(N^{-1/2})$ contributes $O(1)$. For $L/\sqrt N\le m_n\le\delta$, with $L$ large and $\delta>0$ small, the rate is $D_n=N[1+O(m_n)]/4$. Endpoint summation then gives
\[
 \frac{\sum_{j=n}^c\mu_j}{\mu_n}
 =\frac1{2\epsilon_\infty m_n}
   \left[1+O\left(m_n+\frac1{Nm_n^2}\right)\right].
\]
With mesh $2/N$, the passage sum becomes $(1/\epsilon_\infty)\int dm/m$; the displayed errors contribute bounded matching terms. Hence
\begin{equation}
 \mathbb E T_{\rm form}=\frac{\log N}{2\epsilon_\infty}+O(1).
 \label{eq:S-formation-time}
\end{equation}
Motion between fixed thresholds below $m_\infty$ adds only finite time. Thus the same backward equation gives logarithmic formation and exponentially slow sign loss. Near threshold, Eq.~\eqref{eq:S-critical-memory} requires $N\epsilon_\infty^2\gg1$ for activated passage; in the window $\epsilon_\infty=O(N^{-1/2})$, Eq.~\eqref{eq:S-critical-generator} supplies the intervening $\sqrt N$ relaxation scale.

\begin{figure}[htbp]
\centering
\resizebox{.94\linewidth}{!}{%
\begin{tikzpicture}
\path[use as bounding box] (0,0) rectangle (17.8,5.5);
\plabel{0}{5.4}{a}\plabel{9.0}{5.4}{b}
\begin{axis}[paper,at={(1.5cm,1.02cm)},anchor=south west,width=6.95cm,height=3.7cm,ymode=log,xmin=0,xmax=260,ymin=1,ymax=1e9,xlabel={Number of spins $N$},ylabel={First-passage time $\tau$},xtick={0,64,128,192,256},ytick={1,1e3,1e6,1e9},legend columns=2,legend style={at={(.5,1.06)},anchor=south,column sep=14pt}]

\addplot[qgray,mark=square,mark options={fill=white},mark size=1.7pt] table[x=N,y=tau_form_exact]{\DataMemory};\addlegendentry{Formation}
\addplot[qblue,mark=o,mark options={fill=white},mark size=1.7pt] table[x=N,y=erasure]{\DataMemory};\addlegendentry{Erasure}

\end{axis}
\begin{axis}[paper,at={(10.5cm,1.02cm)},anchor=south west,width=6.95cm,height=3.7cm,xmin=-4,xmax=4,ymin=-.9,ymax=.9,xlabel={Scaled field $Nh$},ylabel={$\langle M_z\rangle$},xtick={-4,-2,0,2,4},ytick={-.8,0,.8},legend columns=3,legend style={at={(.5,1.06)},anchor=south,column sep=10pt}]

\addplot[black!40,dotted,forget plot] coordinates {(-4,0)(4,0)};
\addplot[qgray,only marks,mark=o,mark repeat=5] table[x=eta,y=mean]{\DataFieldNSixFour};\addlegendentry{$N=64$}
\addplot[qblue,only marks,mark=square,mark repeat=5] table[x=eta,y=mean]{\DataFieldNOneZeroTwoFour};\addlegendentry{$1024$}
\addplot[qred] table[x=eta,y=two_well]{\DataFieldNOneZeroTwoFour};\addlegendentry{$\infty$}

\end{axis}
\end{tikzpicture}}
\caption{Formation, sign loss, and field selection in the strong-dephasing process at $K=0.75$, $s=0.8$. (a) Exact mean first-passage times from the backward equation: formation starts at zero and stops at $|m|\ge m_\infty/2$; erasure reaches the center from the positive well. (b) Stationary sign selection by $h=\eta/N$. Symbols are the exact finite-size laws, and the curve is Eq.~\eqref{eq:S-field-selection}.}
\label{fig:S-barrier}
\end{figure}
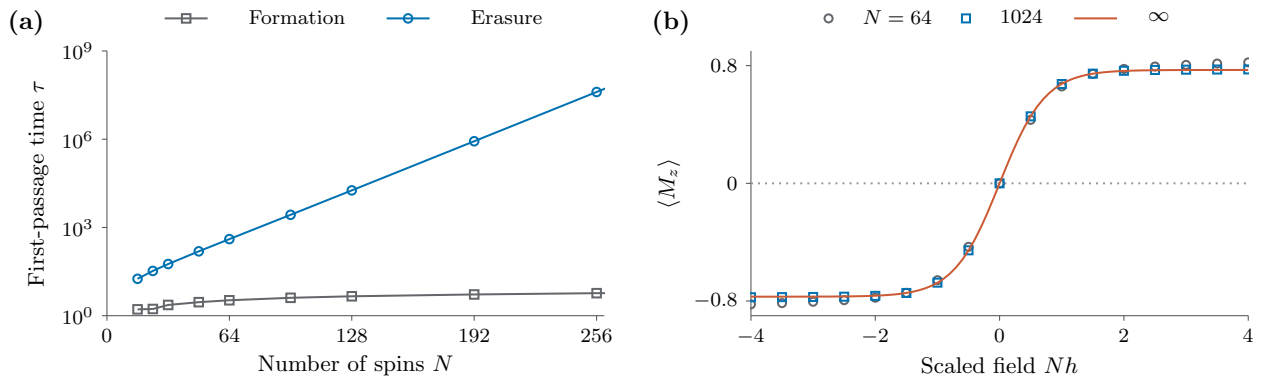

Figure~\ref{fig:S-barrier} evaluates these results at $K=0.75$, $s=0.8$: panel (a) compares formation to $|m|\ge m_\infty/2$ with erasure to the center; panel (b) compares the exact law~\eqref{eq:S-exact-macro} at $h=\eta/N$ with the two-well limit~\eqref{eq:S-field-selection}.

\FloatBarrier

At finite noise, the quantum matrix~\eqref{eq:S-orbit-matrix} resolves the same separation of times. Spin inversion separates odd and even sectors, probed by $M_z$ and $M_z^2$, respectively. Their leading nonzero eigenvalues $\zeta_{\rm o,e}$ give $\tau_{\rm o,e}=1/[-\operatorname{Re}\zeta_{\rm o,e}]$. At $K=0.75$, $s=0.8$, $r=3$, the leading odd mode overlaps with $M_z$, and $(\tau_{\rm o},\tau_{\rm e})=(49.49,4.264)$ for $N=64$ and $(223.50,5.767)$ for $N=128$: even moments settle before odd relaxation erases the sign~\cite{Macieszczak2016}. The preparation $\rho_N(0)=\varrho^{\otimes N}$ with $\varrho=(\openone+0.25X+0.5Z)/2$ excites both sectors. Its magnetization returns to zero at $N=8,16,32$, whereas the collective trajectory tends to $m_r=0.633106\ldots$. These outcomes express the two orders of limits: for a product preparation in the positive basin at $K_c<K<1$,
\begin{equation}
 \begin{aligned}
 \lim_{\tau\to\infty}\lim_{N\to\infty}
 \Tr[M_z e^{\tau\mathcal L_{\tau,N}}\varrho^{\otimes N}]&=m_r,\\
 \lim_{N\to\infty}\lim_{\tau\to\infty}
 \Tr[M_z e^{\tau\mathcal L_{\tau,N}}\varrho^{\otimes N}]&=0.
 \end{aligned}
 \label{eq:S-two-limits}
\end{equation}
For a spin-inversion-symmetric preparation the mean remains zero; formation instead builds the second moment from finite-size fluctuations near the center, with $\lim_{N\to\infty}\lim_{\tau\to\infty}\Tr[\rho_N(\tau)M_z^2]=m_r^2$.

The same elimination applies to a spatial network. Its configuration probabilities still follow Eq.~\eqref{eq:S-projected-master}, but the local fields now depend on the arrangement of neighboring spins. We keep this full configuration equation in the next section and determine the stationary interactions and probability currents generated by those fields.

\section{Spatial stationary statistics}
\label{sec:spatial}

We now use the same strong-dephasing generator $W_\infty$ to determine how the local response changes spatial order. For a periodic one-dimensional chain with nearest-neighbor coupling $J>0$ and zero field, write $S_i=z_{i-1}+z_{i+1}$, so that $a_i=JS_i$ and $S_i\in\{-2,0,2\}$. Since $F_s$ is odd,
\[
  F_s(JS_i)=\frac12F_s(2J)S_i.
\]
A flip at $i$ changes $\sum_j z_jz_{j+1}$ by $-2z_iS_i$, so Eq.~\eqref{eq:S-effectivefield} satisfies detailed balance with the one-dimensional Ising weight
\[
\pi(z)\propto \exp\left[\frac12F_s(2J)\sum_i z_i z_{i+1}\right].
\]
Since $F_s(2J)>2J$ for $s>0$, dephasing increases the effective nearest-neighbor coupling and the corresponding correlation length,
\[
 \left\{-\log\tanh\left[\frac12F_s(2J)\right]\right\}^{-1}.
\]
The one-dimensional chain gives the lower-critical-dimensional case of the short-range problem, while in two dimensions the same enhanced response can support an ordering transition.

On an even periodic $L\times L$ square lattice, $L\ge4$, write $\Lambda=L^2$ and let $\partial i$ be the four neighbors of $i$. With nearest-neighbor coupling $J>0$ and zero field, their spin sum $S_i$ supplies the field and hence the flip rate:
\[
 \begin{aligned}
 a_i(z)&=J\sum_{j\in\partial i}z_j=JS_i(z),\\
 w_i(z)&=\frac12[1-z_i\tanh(JS_i)][1-sz_i\tanh(JS_i)].
 \end{aligned}
\]
The configuration equation remains $\partial_\tau p=W_\infty^*p$ in time $\tau=\nu(1-s)t$, but its stationary law now depends on the spin arrangement as well as the magnetization. Positive single-spin-flip rates select a unique law $\pi_{s,J}$, analytic at fixed $L$ near $J=0$, where $\pi_{s,0}=2^{-\Lambda}$.

We construct this law by balancing the probability flow. Since a spin flip leaves its own conditional field unchanged, Eq.~\eqref{eq:S-effectivefield} gives
\[
 \frac12\log\frac{w_i(z^i)}{w_i(z)}=z_iF_s(JS_i).
\]
For the edge sum $E(z)=\sum_{\langle ij\rangle}z_i z_j$, with each edge counted once, $E(z^i)-E(z)=-2z_iS_i$. Thus the linear part of $F_s$ is supplied by a pair log weight. At $s=0$ and $s\to1$ this gives the Ising weights at couplings $J$ and $2J$. At intermediate strength, the cubic coefficient $c=s(1-s^2)/3$ in Sec.~\ref{sec:fluctuations} gives
\begin{equation}
 F_s(a)=(1+s)a-ca^3+O(a^5).
 \label{eq:S-F-series}
\end{equation}
Inserting the resulting leading weight into the stationary equation produces the source for its next interaction:
\begin{equation}
 \begin{aligned}
 \frac{W_\infty^*e^{(1+s)JE}}{e^{(1+s)JE}}
 &=\sum_iw_i(z)\left\{e^{2z_i[F_s(JS_i)-(1+s)JS_i]}-1\right\}\\
 &=-cJ^3\sum_i z_iS_i^3+O(J^4).
 \end{aligned}
 \label{eq:S-defect}
\end{equation}
Write the required log-weight correction as $J^3H_3$. Expanding the source with $z_j^2=1$ separates repeated from distinct neighbor indices:
\begin{equation}
 S_i^3=10\sum_{j\in\partial i}z_j
       +6\sum_{\{j,k,\ell\}\subset\partial i}z_jz_kz_\ell.
 \label{eq:S-star-identity}
\end{equation}
Multiplying by $z_i$ and summing counts each edge twice and produces four-spin stars. Writing $T$ for their total gives
\begin{equation}
 \begin{aligned}
 \sum_i z_iS_i^3
 &=20E+6\sum_i\sum_{\{j,k,\ell\}\subset\partial i}z_i z_jz_kz_\ell\\
 &=20E+6T.
 \end{aligned}
 \label{eq:S-edge-star}
\end{equation}
At cubic order, the correction acts through the symmetric zero-coupling generator $W_0$. Each spin in a monomial $z_A=\prod_{i\in A}z_i$ flips at rate $1/2$, so $W_0z_A=-|A|z_A$. Applying this to the two-spin and four-spin terms solves the source equation:
\[
 \begin{aligned}
 W_0H_3&=c(20E+6T)\\
 &=W_0\left(-10cE-\frac{3c}{2}T\right).
 \end{aligned}
\]
The remaining additive constant fixes normalization. Substituting $H_3$ gives the stationary law
\begin{equation}
 \log\pi_{s,J}(z)
 =\mathrm{const}+(1+s)JE(z)-cJ^3\left[10E(z)+\frac32T(z)\right]+O(J^4).
 \label{eq:S-stationary-expansion}
\end{equation}
The enhanced local response strengthens the pair interaction and generates a four-spin interaction that cannot be absorbed into temperature. More generally, analytic spin-inversion-symmetric rates with zero-field value $1/2$ give the same construction: the linear and cubic coefficients of their log rate ratio fix this correction, while the even part of the rate enters at higher order.

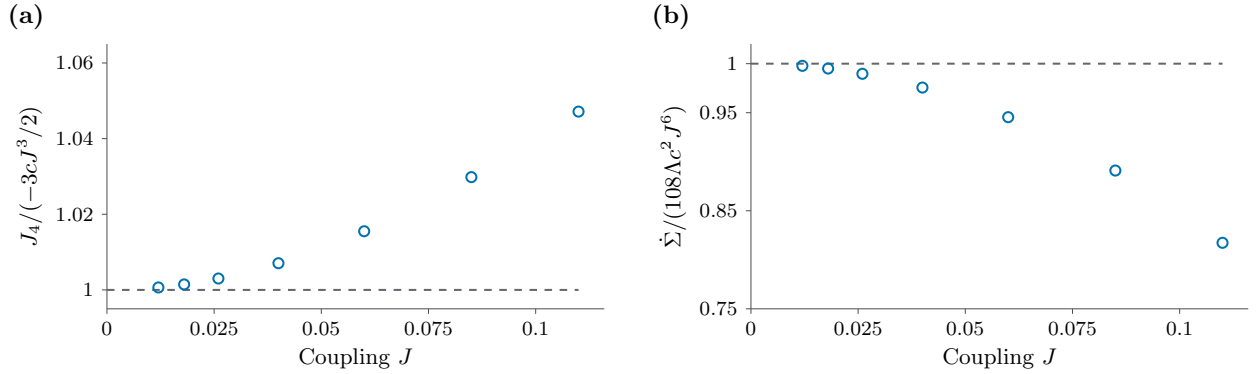
\begin{figure}[htbp]
\centering
\resizebox{.94\linewidth}{!}{%
\begin{tikzpicture}
\path[use as bounding box] (0,0) rectangle (17.8,5.5);
\plabel{0}{5.4}{a}\plabel{9.0}{5.4}{b}
\begin{axis}[paper,at={(1.5cm,1.02cm)},anchor=south west,width=6.95cm,height=3.7cm,xmin=0,xmax=.116,ymin=.995,ymax=1.065,xlabel={Coupling $J$},ylabel={$J_4/(-3cJ^3/2)$},xtick={0,.025,.05,.075,.1},ytick={1,1.02,1.04,1.06}]
\addplot[qgray,dashed] coordinates {(0,1)(.11,1)};
\addplot[qblue,only marks,mark=o,mark options={fill=white},mark size=2pt] table[x=J,y=cubic_ratio]{\DataSpatialSeries};
\end{axis}
\begin{axis}[paper,at={(10.5cm,1.02cm)},anchor=south west,width=6.95cm,height=3.7cm,xmin=0,xmax=.116,ymin=.75,ymax=1.02,xlabel={Coupling $J$},ylabel={$\dot\Sigma/(108\Lambda c^2J^6)$},xtick={0,.025,.05,.075,.1},ytick={.75,.85,.95,1}]
\addplot[qgray,dashed] coordinates {(0,1)(.11,1)};
\addplot[qblue,only marks,mark=o,mark options={fill=white},mark size=2pt] table[x=J,y=entropy_ratio]{\DataSpatialSeries};
\end{axis}
\end{tikzpicture}}
\caption{Two consequences of the stationary correction on a $4\times4$ torus at $s=0.8$. (a) The four-spin star coefficient of $\log\pi_{s,J}$ divided by $-3cJ^3/2$ from Eq.~\eqref{eq:S-stationary-expansion}. (b) The entropy-production rate evaluated with the same stationary probabilities, divided by $108\Lambda c^2J^6$ from Eq.~\eqref{eq:S-epr-series}. The horizontal lines give the common small-$J$ limit of one.}
\label{fig:S-spatial-expansion}
\end{figure}

This weight balances the total flow into each configuration. Its individual edge flows $Q_i(z)=\pi_{s,J}(z)w_i(z)$ follow from the same correction. Flipping $i$ reverses the four stars centered at $i$, whose sum is $T_i$, and the twelve stars centered at its neighbors and containing $i$, whose sum is $R_i$. Thus $T(z)-T(z^i)=2(T_i+R_i)$ and $z_iS_i^3=10z_iS_i+6T_i$. In the logarithmic flux ratio the pair terms cancel:
\begin{equation}
 \begin{aligned}
 \log\frac{Q_i(z)}{Q_i(z^i)}
 &=\log\frac{\pi_{s,J}(z)}{\pi_{s,J}(z^i)}-2z_iF_s(JS_i)\\
 &=3cJ^3[3T_i(z)-R_i(z)]+O(J^4).
 \end{aligned}
 \label{eq:S-edge-force}
\end{equation}
Along a closed sequence of flips, the stationary weights cancel altogether. Choose adjacent up spins $i,j$ whose other neighbors sum to three and one. The sequence $i,j,i,j$ has fields $4J,0,2J,2J$ and signs $+,+,-,-$, giving the exact cycle affinity
\begin{equation}
 \mathcal A=2[2F_s(2J)-F_s(4J)]>0
 \qquad(0<s<1,\ J>0).
 \label{eq:S-squareaffinity}
\end{equation}
Indeed, $F_s(0)=0$ and
\[
 F_s''(a)=-\frac{2s(1-s^2)\tanh a\,\sech^2a}
 {[1-s^2\tanh^2a]^2}<0\qquad(a>0).
\]
The nonzero affinity requires a stationary current on at least one cycle edge at every intermediate strength. Weighting each flux ratio by its net current measures the resulting Markov-path entropy production~\cite{Schnakenberg1976}:
\begin{equation}
 \dot\Sigma=\frac{d\Sigma}{d\tau}
 =\frac12\sum_{z,i}[Q_i(z)-Q_i(z^i)]
       \log\frac{Q_i(z)}{Q_i(z^i)}.
 \label{eq:S-epr-definition}
\end{equation}
Each undirected edge is counted once. To cubic order, its current is $Q_i(z)-Q_i(z^i)=(2^{-\Lambda}/2)\log[Q_i(z)/Q_i(z^i)]+O(J^4)$. Thus the leading entropy production is the uniform mean square of the cubic flux bias. Denoting uniform averages by $\langle\cdot\rangle_0$ and using orthogonality of the four center-star and twelve neighbor-star monomials gives
\[
 \begin{aligned}
 \frac{\dot\Sigma}{\Lambda}
 &=\frac{9c^2J^6}{4}\langle(3T_i-R_i)^2\rangle_0+O(J^7)\\
 &=\frac{9c^2J^6}{4}(4\cdot3^2+12)+O(J^7).
 \end{aligned}
\]
Flipping one sublattice maps $J$ to $-J$ and preserves entropy production, so odd powers vanish:
\begin{equation}
 \frac{\dot\Sigma}{\Lambda}
 =108c^2J^6+O(J^8)
 =12s^2(1-s^2)^2J^6+O(J^8).
 \label{eq:S-epr-series}
\end{equation}
The physical-time rate is $\nu(1-s)\dot\Sigma$. For a star with site set $A$, orthogonality extracts its coefficient as $J_4=\langle z_A\log\pi_{s,J}\rangle_0=-3cJ^3/2+O(J^4)$. Figure~\ref{fig:S-spatial-expansion} shows this interaction and the dissipation from the same stationary law.

For this $4\times4$ torus calculation, $s=0.8$ and $0.012\le J\le0.11$. Probability-preserving uniformization solves $W_\infty^*\pi_{s,J}=0$. The same stationary vector gives $J_4$ and $\dot\Sigma$, each divided by its leading expression in the figure.

To follow how this stationary order extends over large distances, apply the same generator to the block magnetization $m_B=|B|^{-1}\sum_{i\in B}z_i$. A flip in $B$ changes it by $-2z_i/|B|$; weighting this increment and its square by the rates gives the drift and noise variance. With $u_i=\tanh(JS_i)$,
\begin{equation}
 \begin{gathered}
 W_\infty m_B
 =\frac1{|B|}\sum_{i\in B}[(1+s)u_i-z_i(1+su_i^2)],\\
 W_\infty(m_B^2)-2m_BW_\infty m_B
 =\frac4{|B|^2}\sum_{i\in B}w_i.
 \end{gathered}
 \label{eq:S-block-magnetization}
\end{equation}
Magnetization is nonconserved, its noise variance is positive and of order $|B|^{-1}$, and disjoint blocks have zero instantaneous cross increment. We describe slowly varying block magnetization by a scalar field $\varphi(\vec x,\tau)$, odd under spin inversion.

To obtain its drift, expand the neighbor functions exactly into spin monomials. Odd symmetry allows one-spin and three-spin terms in $u_i$; even symmetry allows zero-spin, two-spin, and four-spin terms in $u_i^2$. Their values at $S_i=0,2,4$ fix all coefficients:
\begin{equation}
 \begin{aligned}
 u_i&=A_1\sum_{j\in\partial i}z_j+
 A_3\sum_{\{j,k,\ell\}\subset\partial i}z_jz_kz_\ell,\\
 u_i^2&=B_0+B_2\sum_{\{j,k\}\subset\partial i}z_jz_k+
 B_4\prod_{j\in\partial i}z_j,\\
 A_1&=\frac{\tanh(4J)+2\tanh(2J)}8,\qquad
 A_3=\frac{\tanh(4J)-2\tanh(2J)}8,\\
 B_2&=\frac{\tanh^2(4J)}8,\\
 B_0&=B_2+\frac{\tanh^2(2J)}2,\qquad
 B_4=B_2-\frac{\tanh^2(2J)}2.
 \end{aligned}
 \label{eq:S-walsh-rate}
\end{equation}
Local factorization replaces spins in these products by their local magnetizations. At uniform $\varphi$, counting the four neighbors, four triples, and six pairs gives
\[
 \begin{aligned}
 &(1+s)(4A_1\varphi+4A_3\varphi^3)
   -\varphi[1+s(B_0+6B_2\varphi^2+B_4\varphi^4)]\\
 &\quad=[4(1+s)A_1-1-sB_0]\varphi
   -[6sB_2-4(1+s)A_3]\varphi^3-sB_4\varphi^5.
 \end{aligned}
\]
The cubic restoring coefficient is positive because $B_2>0$ and $A_3<0$. Spatial variation enters through the same neighbor sum: with lattice spacing one, $\sum_{\vec\delta}\varphi(\vec x+\vec\delta)=4\varphi+\nabla^2\varphi+O(\nabla^4\varphi)$, yielding diffusion coefficient $(1+s)A_1>0$. Write $D,a,b$ for the diffusion and reaction coefficients after spatial fluctuations are included, and $\zeta$ for zero-mean Gaussian noise of strength $D_0$. Retaining these drift terms and the block noise gives the leading coarse-grained equation
\begin{equation}
 \begin{gathered}
 \partial_\tau\varphi
 =D\nabla^2\varphi+a\varphi-b\varphi^3+\zeta+\text{higher-order terms},\\
 \langle\zeta(\vec x,\tau)\zeta(\vec x',\tau')\rangle
 =2D_0\delta(\vec x-\vec x')\delta(\tau-\tau'),\qquad D,D_0>0.
 \end{gathered}
 \label{eq:S-model-A}
\end{equation}
For a continuous transition $b>0$; the retained equation has stationary profile weight in spatial dimension $d$
\[
 \mathcal P_{\rm st}[\varphi]\propto
 \exp\!\left[-\frac1{D_0}\int d^dx\left(
 \frac D2|\nabla\varphi|^2-\frac a2\varphi^2+\frac b4\varphi^4\right)\right].
\]
The remaining lattice terms give higher field powers, nonlinear gradients, and field-dependent noise. Near the upper critical dimension, the leading nonlinear-gradient and multiplicative-noise couplings have dimension $2-d$, compared with $4-d$ for the quartic interaction. The scalar symmetry, nonconservation, and positive noise therefore lead to the renormalization-group prediction of Ising critical statistics~\cite{Grinstein1985}, while the microscopic process retains the currents found above.

We resolve the two-dimensional transition by following the distribution of $m=\Lambda^{-1}\sum_jz_j$ as $L$ grows. Its shape is measured by normalized even moments $V_{2n}$ and the Binder ratio $U_L$; the variance of $|m|$ measures fluctuations after combining the two signs:
\begin{equation}
 \begin{gathered}
 V_{2n}=\frac{\langle m^{2n}\rangle}{\langle m^2\rangle^n}\quad(n\ge2),\\
 U_L=1-\frac{V_4}{3}
     =1-\frac{\langle m^4\rangle}{3\langle m^2\rangle^2},\qquad
 \chi_L=\Lambda\operatorname{Var}(|m|)
        =L^2\bigl(\langle m^2\rangle-\langle|m|\rangle^2\bigr).
 \end{gathered}
 \label{eq:S-criticalobservables}
\end{equation}
All averages are stationary. To determine the spatial extent of these fluctuations, use the structure factor $S(\vec k)=\Lambda^{-1}\langle|\sum_jz_je^{\mathrm i\vec k\cdot\vec r_j}|^2\rangle$, with site positions $\vec r_j$. Comparing its uniform value $S(\vec0)=\Lambda\langle m^2\rangle$ with the smallest modes, $\vec k_x=(2\pi/L,0)$ and $\vec k_y=(0,2\pi/L)$, gives the second-moment correlation length
\[
 \frac{\xi_L}{L}
 =\frac1{2L\sin(\pi/L)}
   \sqrt{\frac{S(\vec0)}{[S(\vec k_x)+S(\vec k_y)]/2}-1}.
\]
The size dependence locates $J_c$ and gives the correlation-length, magnetization, and fluctuation exponents $\nu,\beta,\gamma$. The two dimensionless ratios $R_L\in\{U_L,\xi_L/L\}$ depend on the same scaled distance $x=(J-J_c)L^{1/\nu}$. Expanding about their critical intercepts $R_*$ and allowing a correction exponent $\omega>0$ gives the joint fit
\begin{equation}
 R_L=R_*+\alpha_{R,1}x+\alpha_{R,2}x^2
       +\lambda_R(L/20)^{-\omega},
 \label{eq:S-fit}
\end{equation}
with common $J_c,\nu$ and separate coefficients and intercepts. At these $J_c,\nu$, the powers of the magnetization magnitude and its fluctuation follow from
\begin{equation}
 \begin{aligned}
 \log\langle|m|\rangle
 &=\alpha_{m,0}+\alpha_{m,1}x+\alpha_{m,2}x^2
   +\lambda_m(L/20)^{-\omega}-(\beta/\nu)\log L,\\
 \log\chi_L
 &=\alpha_{\chi,0}+\alpha_{\chi,1}x+\alpha_{\chi,2}x^2
   +\lambda_\chi(L/20)^{-\omega}+(\gamma/\nu)\log L.
 \end{aligned}
 \label{eq:S-exponent-fit}
\end{equation}
The two lines have separate coefficients and freely fitted powers; Eq.~\eqref{eq:S-fit} applied to $V_6,V_8$ gives the higher critical moment ratios. Figure~\ref{fig:S-spatial} and Tables~\ref{tab:fits}--\ref{tab:universality} collect the resulting spatial statistics.

\begin{figure}[htbp]
\centering
\resizebox{.94\linewidth}{!}{%
\begin{tikzpicture}
\path[use as bounding box] (0,0) rectangle (17.8,5.5);
\plabel{0}{5.4}{a}\plabel{9.0}{5.4}{b}
\begin{axis}[paper,at={(1.5cm,1.02cm)},anchor=south west,width=6.95cm,height=3.7cm,xmin=.2475,xmax=.2585,ymin=.47,ymax=.68,xlabel={Coupling $J$},ylabel={Binder ratio $U_L$},xtick={.248,.253,.258},ytick={.5,.55,.6,.65},legend columns=4,legend style={at={(.5,1.06)},anchor=south,column sep=12pt}]
\addplot[qgray,mark=o,mark options={fill=white},mark size=1.65pt,error bars/.cd,y dir=both,y explicit,error bar style={line width=.5pt}] table[x=J,y=U,y error=U_err]{\DataCrossingsLTwoZero};\addlegendentry{20}
\addplot[qpurple,mark=square,mark options={fill=white},mark size=1.65pt,error bars/.cd,y dir=both,y explicit,error bar style={line width=.5pt}] table[x=J,y=U,y error=U_err]{\DataCrossingsLThreeSix};\addlegendentry{36}
\addplot[qblue,mark=triangle,mark options={fill=white},mark size=1.65pt,error bars/.cd,y dir=both,y explicit,error bar style={line width=.5pt}] table[x=J,y=U,y error=U_err]{\DataCrossingsLSixFour};\addlegendentry{64}
\addplot[qred,mark=diamond,mark options={fill=white},mark size=1.65pt,error bars/.cd,y dir=both,y explicit,error bar style={line width=.5pt}] table[x=J,y=U,y error=U_err]{\DataCrossingsLNineSix};\addlegendentry{96}
\end{axis}
\begin{axis}[paper,at={(10.5cm,1.02cm)},anchor=south west,width=6.95cm,height=3.7cm,xmin=.2475,xmax=.2585,ymin=.45,ymax=1.7,xlabel={Coupling $J$},ylabel={$\xi_L/L$},xtick={.248,.253,.258},ytick={.5,1,1.5},legend columns=4,legend style={at={(.5,1.06)},anchor=south,column sep=12pt}]
\addplot[qgray,mark=o,mark options={fill=white},mark size=1.65pt,error bars/.cd,y dir=both,y explicit,error bar style={line width=.5pt}] table[x=J,y=xi,y error=xi_err]{\DataCrossingsLTwoZero};\addlegendentry{20}
\addplot[qpurple,mark=square,mark options={fill=white},mark size=1.65pt,error bars/.cd,y dir=both,y explicit,error bar style={line width=.5pt}] table[x=J,y=xi,y error=xi_err]{\DataCrossingsLThreeSix};\addlegendentry{36}
\addplot[qblue,mark=triangle,mark options={fill=white},mark size=1.65pt,error bars/.cd,y dir=both,y explicit,error bar style={line width=.5pt}] table[x=J,y=xi,y error=xi_err]{\DataCrossingsLSixFour};\addlegendentry{64}
\addplot[qred,mark=diamond,mark options={fill=white},mark size=1.65pt,error bars/.cd,y dir=both,y explicit,error bar style={line width=.5pt}] table[x=J,y=xi,y error=xi_err]{\DataCrossingsLNineSix};\addlegendentry{96}
\end{axis}
\end{tikzpicture}}
\caption{Spatial correlations near the stationary transition at $s=0.8$. (a) Binder ratios obtained from the second and fourth magnetization moments. (b) The second-moment correlation length divided by $L$, obtained from the uniform and smallest nonzero Fourier modes. Curves join estimates for the displayed sizes; error bars are independent-run jackknife standard errors. The two quantities share the fitted $J_c,\nu$ in Eq.~\eqref{eq:S-fit}; the fit uses all sizes in the specified window, as summarized in Table~\ref{tab:fits}.}
\label{fig:S-spatial}
\end{figure}
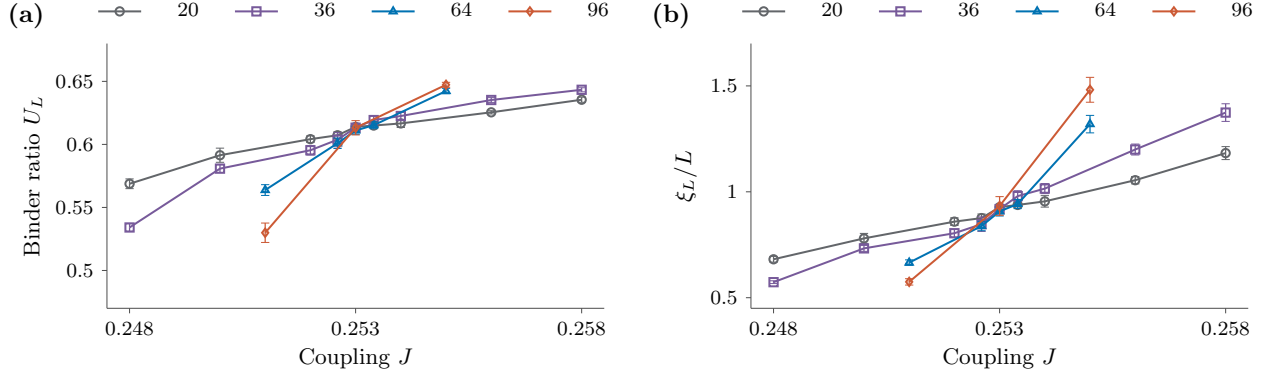
\begin{table}[!ht]
\caption{Size and coupling dependence of the spatial critical fits. The first row is the primary fit. The critical exponents $\nu$, $\beta/\nu$, and $\gamma/\nu$ are fitted freely in each row; the specified correction exponent $\omega$ enters Eqs.~\eqref{eq:S-fit} and~\eqref{eq:S-exponent-fit}.}
\label{tab:fits}
\centering\small\setlength{\tabcolsep}{7pt}\renewcommand{\arraystretch}{1.15}\begin{tabular}{rrrrrrr}\toprule
$L_{\min}$ & $J$ interval & $\omega$ & $J_c$ & $\nu$ & $\beta/\nu$ & $\gamma/\nu$\\\midrule
20 & 0.250--0.256 & 2 & 0.25312 & 0.957 & 0.1234 & 1.753\\
28 & 0.250--0.256 & 2 & 0.25320 & 0.962 & 0.1225 & 1.814\\
20 & 0.251--0.255 & 2 & 0.25301 & 1.130 & 0.1276 & 1.749\\
20 & 0.250--0.256 & 1 & 0.25312 & 0.957 & 0.1243 & 1.757\\
12 & 0.250--0.256 & 2 & 0.25315 & 0.990 & 0.1222 & 1.773\\
\bottomrule\end{tabular}

\end{table}
\begin{table}[!ht]
\caption{Critical powers and distribution ratios from the same stationary moment fits. Intervals are 95\% independent-run bootstrap intervals for the primary window. The Ising values of the torus ratios are from Ref.~\cite{SalasSokal2000}, with $U_*=1-V_4/3$.}
\label{tab:universality}
\centering\small\setlength{\tabcolsep}{8pt}\renewcommand{\arraystretch}{1.12}\begin{tabular}{lrrr}\toprule
Quantity & Estimate & 95\% interval & Ising\\\midrule
$\nu$ & 0.9574 & [0.8676, 1.0742] & 1\\
$\beta/\nu$ & 0.1234 & [0.1172, 0.1296] & 0.125\\
$\gamma/\nu$ & 1.7533 & [1.6783, 1.8307] & 1.75\\
$U_*$ & 0.6127 & [0.6072, 0.6176] & 0.6106924\\
$(\xi/L)_*$ & 0.9215 & [0.8780, 0.9662] & 0.90504883\\
$V_6$ & 1.4382 & [1.3949, 1.4863] & 1.4556491\\
$V_8$ & 1.8575 & [1.7636, 1.9609] & 1.89252\\
\bottomrule\end{tabular}

\end{table}

\FloatBarrier
The primary window $L\ge20$, $0.250\le J\le0.256$, $\omega=2$ gives $J_c=\Jcritical$; the other windows vary the size cutoff, coupling interval, and correction exponent.

The spatial-process samples use uniform site selection followed by a flip with probability
\begin{equation}
 p_{\mathrm{flip}}(z,i)=\frac{w_i(z)}{1+s}
 =\frac{(1-z_i u_i)(1-sz_i u_i)}{2(1+s)}.
 \label{eq:S-uniformization}
\end{equation}
Since $w_i\le1+s$, an attempt applies $\openone+W_\infty^*/[\Lambda(1+s)]$ to probabilities and preserves $\pi_{s,J}$; $\Lambda$ attempts make one sweep. At $s=0$ and $s\to1$, the thresholds are $J_{\rm I}=\tfrac12\log(1+\sqrt2)$ and $J_{\rm I}/2$. The Letter's broad comparison uses $L=32$, $s=0,0.8$, and four runs at each $J=0.20,0.23,0.27,0.30,0.34,0.38,0.42,0.46$, with $10000+32000$ equilibration and measurement sweeps.

At $s=0.8$, the critical grid covers $L=12,20,28,36,48,64,96$ and $0.248\le J\le0.258$, with eight runs per point. For $L=12,20,28,36,48$ and coupling spacing $0.002$, lengths are $12000+48000$ sweeps for $L\le28$ and $32000+192000$ for $L\ge36$. The finer grid $L=20,28,36,48,64$, $0.2526\le J\le0.2534$, spacing $0.0004$, uses $24000+128000$ for $L<48$ and $64000+384000$ for $L\ge48$; $L=64,96$ at $J=0.251,0.253,0.255$ uses $80000+512000$. Each group has equal numbers of ordered and random preparations and records every four sweeps.

Each run contributes $|m|,m^2,m^4,m^6,m^8$ moments and the average of the two nonzero-mode structure factors. Independent-run jackknife errors give fit weights and plotted errors; 600 whole-run bootstrap samples repeat the joint ratio, exponent, and higher-moment fits, retaining time correlations within runs and propagating the shared $J_c,\nu$ uncertainty. Table~\ref{tab:universality} gives 95\% intervals and Ising reference values.

For the Letter's distribution comparison, $Y=m/\sqrt{\langle m^2\rangle}$ standardizes each law by its own measured width. The noisy sample uses $L=64$, $J=0.2530$. The equilibrium reference uses eight independent Wolff runs at $J=J_{\rm I}$ for each $L=32,64$, with 2000 equilibration and 50000 recorded clusters. Opposite histogram bins are combined by spin-inversion symmetry; whole-run bootstrap resampling recomputes the width and histogram together.

The fitted powers, torus ratios, and standardized distribution support Ising critical statistics at a continuous transition below $J_{\rm I}$. The stationary equation has thus connected the enhanced local response to stronger spatial order, a new four-spin interaction, and persistent currents. We next choose the local recycling output to recover the programmed Boltzmann law.
\FloatBarrier

\section{Calibrating the local channel to recover Boltzmann sampling}
\label{sec:calibration}

The probability equation~\eqref{eq:S-network-population} identifies how noise changes the sampler: coherence enters the population drift. We now choose the recycling output so that this drift follows the target conditional update independently of coherence. The resulting local identity will determine the network's sampled law and its successive training records.

Retain $P=\ket\nu\bra\nu=(\openone+qX+uZ)/2$, $Q=\openone-P$, and $A_0=P+sQ$, where $u=\tanh a$, $q=\sech a$, and $0\le s<1$. Completeness fixes the remaining Kraus effect to $(1-s^2)Q$, leaving its output free. A recycling state $\tau_{\rm rec}=\sum_\ell t_\ell\ket{\chi_\ell}\bra{\chi_\ell}$ is realized by $A_{1\ell}=\sqrt{(1-s^2)t_\ell}\ket{\chi_\ell}\bra{\nu^\perp}$, giving
\begin{equation}
 \Phi(\rho)=(P+sQ)\rho(P+sQ)
                  +(1-s^2)\tau_{\rm rec}\Tr(Q\rho).
 \label{eq:S-generalrecycling}
\end{equation}
Every choice preserves $\ket\nu$. To find the output polarization $b=\Tr(Z\tau_{\rm rec})$ that removes coherence from the measured response, use $PZ+ZP=Z+u\openone$ and $PZP=uP$:
\begin{equation}
 \begin{aligned}
 \Phi^\dagger(Z)
 &=(P+sQ)Z(P+sQ)+(1-s^2)bQ\\
 &=sZ+s(1-s)u\openone+(1-s)^2uP+(1-s^2)bQ\\
 &=sZ+\frac{(1-s^2)(u+b)}2\openone
   +\frac{(1-s)^2u-(1-s^2)b}{2}(qX+uZ).
 \end{aligned}
 \label{eq:S-adjointgeneral}
\end{equation}
At finite fields $q>0$, cancellation of the $X$ coefficient fixes $b=(1-s)u/(1+s)$ and reduces the response to
\begin{equation}
 \Phi^\dagger(Z)=sZ+(1-s)u\openone.
 \label{eq:S-calibration}
\end{equation}
Phase flips have zero $Z$ drift, so the complete local master equation becomes $dm/d\tau=u-m$. Its solution $m(\tau)=u+[m(0)-u]e^{-\tau}$ holds for every dephasing strength. The same choice calibrates the target conditional populations: for $\rho_u=(\openone+uZ)/2$,
\begin{equation}
 \Tr[Z\Phi(\rho_u)]-u
 =\frac{(1-u^2)(1-s^2)}2
       \left[b-\frac{1-s}{1+s}u\right].
 \label{eq:S-driftcal}
\end{equation}
The prefactor is positive, making stationary calibration equivalent to cancellation of input coherence. Any output with this $b$ works; a pure realization is $\ket{\chi_b}=\sqrt{(1+b)/2}\ket0+\sqrt{(1-b)/2}\ket1$.

Apply this choice on each conditional branch of the network, allowing $0\le s_i<1$ and $b_i(\zeta)=(1-s_i)u_i(\zeta)/(1+s_i)$. The conditional block $\rho_\zeta^{(i)}=\langle\zeta|_{\ne i}\rho_N|\zeta\rangle_{\ne i}$ is unnormalized. Its trace sums the two populations, while its $Z$ expectation subtracts them. Thus Eq.~\eqref{eq:S-calibration} gives
\begin{equation}
 \begin{aligned}
 \langle\zeta,z_i|\Phi_i(\rho_N)|\zeta,z_i\rangle
 &=\tfrac12\Tr\!\left[
  \bigl(\openone+z_i[s_iZ+(1-s_i)u_i(\zeta)\openone]\bigr)\rho_\zeta^{(i)}\right]\\
 &=\tfrac{s_i}{2}\Tr[(\openone+z_iZ)\rho_\zeta^{(i)}]
   +\tfrac{1-s_i}{2}[1+z_i u_i(\zeta)]\Tr\rho_\zeta^{(i)}\\
 &=s_i p_N(\zeta,z_i)+(1-s_i)p_\theta(z_i\mid\zeta)
                  \sum_{z_i'=\pm1}p_N(\zeta,z_i').
 \end{aligned}
 \label{eq:S-calibrated-probability-step}
\end{equation}
The right-hand side retains the old spin with weight $s_i$ and draws from the target conditional law with weight $1-s_i$. Denote this classical update by $T_i^*$, with the star indicating action on probabilities, and write $(\dd\rho)(z)=\langle z|\rho|z\rangle$. Equation~\eqref{eq:S-calibrated-probability-step} is then the identity
\begin{equation}
 \dd\Phi_i=T_i^*\dd,
 \label{eq:S-popclosure}
\end{equation}
valid for every input state. It is the detection-incoherent property of Ref.~\cite{Theurer2019}.

For site rates $\nu_i>0$ and population-preserving noise $\dd\mathcal N=0$, the full network generator is $\mathcal L_{\rm cal}=\sum_i\nu_i(\Phi_i-\id)+\mathcal N$. Applying Eq.~\eqref{eq:S-popclosure} first to its evolution equation and then to its solution gives
\begin{equation}
 \begin{gathered}
 \frac{dp_N}{dt}=\dd\mathcal L_{\rm cal}(\rho_N)
 =\sum_i\nu_i(T_i^*-\openone)p_N=:W^*p_N,\\
 \dd e^{t\mathcal L_{\rm cal}}=e^{tW^*}\dd.
 \end{gathered}
 \label{eq:S-calibrated-evolution}
\end{equation}
Reading the single-spin changes in $T_i^*$ gives the rates and hence the action of $W^*$:
\[
 \begin{aligned}
 \widetilde w_i(z)&=\frac{\nu_i(1-s_i)}2[1-z_i u_i(z_{\ne i})],\\
 (W^*p)(z)&=\sum_i[\widetilde w_i(z^i)p(z^i)-\widetilde w_i(z)p(z)].
 \end{aligned}
\]
These are the programmed heat-bath rates; for common $\nu,s$, the prefactor is absorbed in $\tau=\nu(1-s)t$. Conditional factorization now gives detailed balance:
\[
 \begin{aligned}
 p_\theta(\zeta,z_i)\widetilde w_i(\zeta,z_i)
 &=\nu_i(1-s_i)p_\theta(\zeta)
      p_\theta(z_i\mid\zeta)p_\theta(-z_i\mid\zeta)\\
 &=p_\theta(\zeta,-z_i)\widetilde w_i(\zeta,-z_i).
 \end{aligned}
\]
All single-spin rates are positive at finite fields, making this finite classical chain irreducible. Its unique stationary law therefore determines the sampled output:
\begin{equation}
 \lim_{t\to\infty}\dd e^{t\mathcal L_{\rm cal}}\rho_{\rm in}=p_\theta.
 \label{eq:S-calibrated-stationary}
\end{equation}
In particular, every quantum stationary state $\mathcal L_{\rm cal}(\rho_*)=0$ has $\dd\rho_*=p_\theta$.

Training composes updates and readouts. For prescribed applications and intervening noise channels $E_n$ with $\dd E_n=\dd$, Eq.~\eqref{eq:S-popclosure} composes to
\begin{equation}
 \dd(E_T\Phi_{i_T}\cdots E_1\Phi_{i_1}\rho)
 =T_{i_T}^*\cdots T_{i_1}^*\dd\rho.
 \label{eq:S-sequence}
\end{equation}
Sites, strengths, and parameters may vary between updates. At fixed parameters every $T_i^*$ preserves $p_\theta$; random scan with positive site weights is irreducible and has holding probability, while a complete cyclic sweep has positive probability of every output configuration. Both repeated scans converge to $p_\theta$.

A readout outcome $y$ with diagonal Kraus operators $B_{y\ell}$ defines $\mathcal M_y(\rho)=\sum_\ell B_{y\ell}\rho B_{y\ell}^\dagger$. Taking its diagonal identifies the classical outcome matrix $D_y$:
\begin{equation}
 \begin{aligned}
 (\dd\mathcal M_y\rho)(z)
 &=\sum_\ell|\bra zB_{y\ell}\ket z|^2(\dd\rho)(z)\\
 &=(D_y\dd\rho)(z).
 \end{aligned}
 \label{eq:S-instrument}
\end{equation}
Completeness gives $\sum_yD_y=\openone$; this construction includes full and partial computational-basis readout. If a previous record $H$ selects the next update, let $\rho_H$ be its unnormalized conditional state and $E_H$ the intervening noise, with $\dd E_H=\dd$. Then $p_H=\dd\rho_H$ evolves as
\[
 p_{Hy}=\dd\mathcal M_y E_H\Phi_{i(H)}(\rho_H)
       =D_yT_{i(H)}^*p_H,
\]
and the probability of the extended record is $\Pr(Hy)=\sum_zp_{Hy}(z)$. This is the matched classical sampler's recursion. Induction gives identical finite record distributions, conditional populations, and parameter histories under any common feedback rule, including nonlinear updates using finitely many samples. Continuous intervals use the same recursion with $e^{tW^*}$.

The same closure fixes population relaxation. Let $T_i,W$ be the duals of $T_i^*,W^*$ and $O_f=\sum_z f(z)\ket z\bra z$. Since $\mathcal N^\dagger O_f=0$, duality followed by summation gives
\begin{align}
 \Phi_i^\dagger O_f&=O_{T_i f},
 \label{eq:S-diagobservable}\\
 \mathcal L_{\rm cal}^\dagger O_f
 &=\sum_i\nu_i(O_{T_i f}-O_f)\notag\\
 &=O_{Wf},
 \label{eq:S-spectrum-inclusion}
\end{align}
where $(Wf)(z)=\sum_i\widetilde w_i(z)[f(z^i)-f(z)]$. Thus $Wf=\lambda f$ implies $\mathcal L_{\rm cal}^\dagger O_f=\lambda O_f$. Detailed balance makes these eigenvalues real, establishing $\operatorname{spec}(W)\subseteq\operatorname{spec}(\mathcal L_{\rm cal})$. With $\Delta(\mathcal K)=\min_{\lambda\in\operatorname{spec}(\mathcal K),\,\lambda\ne0}[-\operatorname{Re}\lambda]$, this gives
\begin{equation}
 \Delta(\mathcal L_{\rm cal})\le\Delta(W).
 \label{eq:S-gap}
\end{equation}
Populations have exactly the heat-bath relaxation, while additional quantum modes can decay more slowly.

The pure recycling channel can be implemented by resolving the target and bright directions, transferring bright weight to an auxiliary qubit, and rotating its output. For $R_y(\varphi)=e^{-\mathrm i\varphi Y/2}$, the required angles are
\begin{equation}
 \vartheta(\zeta)=\arccos u(\zeta),\qquad
 \beta(\zeta)=\arccos\!\left[\frac{1-s}{1+s}u(\zeta)\right].
 \label{eq:S-angles}
\end{equation}
First $R_y(-\vartheta)$ sends $\ket\nu$ to $\ket0$ and $\ket{\nu^\perp}$ to $-\ket1$. Prepare the auxiliary $e$ in $\ket0_e$, apply a neuron-controlled $R_y(2\arccos s)$ to it, then a controlled-NOT from $e$ to the neuron. In the order neuron, auxiliary,
\begin{equation}
 \ket{0,0}\longmapsto\ket{0,0},
 \qquad \ket{1,0}\longmapsto s\ket{1,0}+\sqrt{1-s^2}\ket{0,1}.
 \label{eq:S-dilation}
\end{equation}
Finally rotate the neuron by $\vartheta$ for $e=0$ and $\beta$ for $e=1$. The isometry becomes $\ket\psi\ket0_e\mapsto A_0\ket\psi\ket0_e-A_1\ket\psi\ket1_e$, with $A_1=\sqrt{1-s^2}\ket{\chi_b}\bra{\nu^\perp}$; discarding $e$ implements $\Phi$. Standard recycling uses $\beta=\vartheta$. For degree $d$, direct uniformly controlled rotations require $O(2^d)$ gates and one auxiliary preparation and discard per update. Reversible field arithmetic trades depth for workspace at larger degree~\cite{Mottonen2005,Bergholm2005}; independent samples additionally require the heat-bath mixing time.

Precision of this circuit determines sampling error. Keep $A_0$ fixed and bound the recycling polarization error by $|\delta b(\zeta)|\le\eta_b$. With $Q_i(\zeta)$ the conditional bright projector, Eq.~\eqref{eq:S-generalrecycling} gives
\[
 [\dd(\Phi_{i,\rm err}-\Phi_i)\rho](\zeta,z_i)
 =\frac{z_i\delta b(\zeta)}2(1-s_i^2)\Tr[Q_i(\zeta)\rho_\zeta^{(i)}].
\]
The conditional blocks are positive and satisfy $\sum_\zeta\Tr[Q_i(\zeta)\rho_\zeta^{(i)}]\le1$. Hence the output probabilities, measured by $D_{\rm TV}(p,q)=\tfrac12\sum_z|p(z)-q(z)|$, obey
\begin{equation}
 \begin{aligned}
 D_{\rm TV}(\dd\Phi_{i,\rm err}\rho,T_i^*\dd\rho)
 &=\frac{1-s_i^2}{2}\sum_\zeta|\delta b(\zeta)|
          \Tr[Q_i(\zeta)\rho_\zeta^{(i)}]\\
 &\le\frac{1-s_i^2}{2}\eta_b.
 \end{aligned}
 \label{eq:S-tolerance}
\end{equation}
Stochastic contraction accumulates these errors along a prescribed sequence. For equal initial populations,
\[
 D_{\rm TV}(p_T,p_T^{\rm ref})
 \le\frac12\sum_{n=1}^T(1-s_n^2)\eta_{b,n}.
\]
Angle errors also perturb $A_0$. The single-rotation bound and a telescoping estimate for the ideal and implemented gate products $U,U_{\rm err}$ give, after discarding the auxiliary and measuring,
\[
 \begin{aligned}
 \|R_y(\varphi+\delta\varphi)-R_y(\varphi)\|
 &\le\frac{|\delta\varphi|}{2},\\
 D_{\rm TV}(\dd\Phi_{\rm err}\rho,\dd\Phi\rho)
 &\le\|U_{\rm err}-U\|
 \le\frac12\sum_\ell|\delta\varphi_\ell|.
 \end{aligned}
\]
The field precision sets these angle errors. The target angle has $|d\vartheta/da|=q\le1$; the recycling angle obeys
\[
 \begin{aligned}
 \left|\frac{d\beta}{da}\right|
 &=\frac{(1-s)q^2}{(1+s)\sqrt{1-[(1-s)/(1+s)]^2u^2}}\\
 &\le\frac{1-s}{1+s}q\le1.
 \end{aligned}
\]

The calibrated training channels also preserve the coherent target: $A_0\ket\nu=\ket\nu$ and $A_{1\ell}\ket\nu=0$ lift through the conditional decomposition to preserve $\ket{\Psi_\theta}$. On an isolated zero-field edge of coupling $J$, its concurrence follows from the amplitudes $c_{ab}$:
\begin{equation}
 \begin{aligned}
 \ket{\Psi_J}
 &=\frac{e^{J/2}(\ket{00}+\ket{11})+e^{-J/2}(\ket{01}+\ket{10})}
         {2\sqrt{\cosh J}},\\
 C&=2|c_{00}c_{11}-c_{01}c_{10}|=|\tanh J|.
 \end{aligned}
 \label{eq:S-two-spin-entanglement}
\end{equation}
Thus the target is entangled for $J\ne0$. Its parameter information is also retained by computational-basis sampling. For a field or coupling parameter $\theta_\mu$, let $F_\mu$ be respectively $Z_i$ or $Z_iZ_j$, with $F_\mu(z)=\bra zF_\mu\ket z$, and $\partial_\mu=\partial/\partial\theta_\mu$. Differentiating the amplitudes, with expectations in $p_\theta$, gives
\[
 \begin{aligned}
 \partial_\mu\log p_\theta(z)&=F_\mu(z)-\langle F_\mu\rangle,\\
 \partial_\mu\ket{\Psi_\theta}
 &=\tfrac12(F_\mu-\langle F_\mu\rangle)\ket{\Psi_\theta}.
 \end{aligned}
\]
The centered derivative has $\langle\Psi_\theta|\partial_\mu\Psi_\theta\rangle=0$, so the quantum Fisher information reduces to the same covariance:
\begin{equation}
 \begin{aligned}
 \mathcal I^{\rm Q}_{\mu\nu}
 &=4\operatorname{Re}\langle\partial_\mu\Psi_\theta|\partial_\nu\Psi_\theta\rangle
 =\langle F_\mu F_\nu\rangle-\langle F_\mu\rangle\langle F_\nu\rangle\\
 &=\sum_zp_\theta(z)\,
    \partial_\mu\log p_\theta(z)\,\partial_\nu\log p_\theta(z)
 =\mathcal I^{\rm classical}_{\mu\nu}.
 \end{aligned}
 \label{eq:S-fisher}
\end{equation}
Computational-basis measurement attains the pure target's quantum Fisher information for these amplitude parameters. By Eq.~\eqref{eq:S-calibrated-stationary}, stationary samples under population-preserving noise retain this classical covariance.
\FloatBarrier
\renewcommand{\bibfont}{\small}
\setlength{\bibsep}{1.5pt}